\documentclass[edbkps]{kapedbk}
\usepackage{amsmath}
\usepackage{epsf}
\usepackage{graphicx}
\normallatexbib
\input{tcilatex}
\begin{document}

\articletitle[Multi-scale Phase Modulations in Colossal Magnetoresistance Manganites]{MULTI-SCALE PHASE MODULATIONS\\IN  COLOSSAL MAGNETORESISTANCE
\\MANGANITES }

\author{K. H. Kim}
\affil{National High Magnetic Field Laboratory, MS E536, Los Alamos
National Laboratory, Los Alamos, NM 87545}
\email{khkim@lanl.gov}

\author{M. Uehara}
\affil{Department of Physics, Yokohama National University, Yokohama,
240-8501}
\email{uehara@ynu.ac.jp}
\author{V. Kiryukhin}
\affil{Department of Physics and Astronomy, Rutgers University,
Piscataway,NJ 08854}
\email{vkir@physics.rutgers.edu}
\and
\author{S-W. Cheong}
\affil{Department of Physics and Astronomy, Rutgers University,
Piscataway,NJ 08854}
\email{sangc@physics.rutgers.edu}

\begin{abstract}
Extensive experimental results are presented on the multi-scale phase
modulation phenomena observed in colossal magnetoresistance manganites. Two
key types of phase inhomogeneities directly relevant to the colossal
magnetoresistance (CMR) are discussed. The first type involves
micrometer-scale coexistence of structurally and electronically different
phases. We present extensive experimental data for a prototypical system
exhibiting such a phase coexistence, (La,Pr)$_{5/8}$Ca$_{3/8}$MnO$_{3}$.
These data reveal that percolative transport phenomena play the key role in
the metal-insulator transition in this system, and are largely responsible
for the significant magnitude of the magnetoresistance. The phase
composition of the multiphase states is governed by both the electron
correlations and the effects of martensitic accommodation strain. The second
type of an inhomogeneous state is realized in the paramagnetic state
commonly found in manganites at high-temperatures. In this state,
nanometer-scale structural correlations associated with nanoscale
charge/orbital ordered regions are observed. Experimental investigation of
numerous manganite systems indicates that these correlations are generic in
orthorhombic mixed-valent manganites, and that the correlated regions play
an essential role in the CMR effect. This chapter is organized as follows.
An introduction to the subject is given in Section 1. In Section 2, the
phase diagram of (La,Ca)MnO$_{3}$ is presented. Electronic and structural
properties of both the high- and low-temperature phases are discussed.
Section 3 is devoted to the studies of the micrometer-scale multiphase
states in (La,Pr)$_{5/8}$Ca$_{3/8}$MnO$_{3}$. In Section 4, high temperature
nanoscale correlations are discussed. Extensive studies of electronic and
thermal transport, and optical spectroscopy are presented. Section 5 focuses
on x-ray scattering studies of the nanoscale correlations. Finally, Section
6 concludes this Chapter.
\end{abstract}

%------------ article title  ------------------->>

% If you use \\'s , please supply an alternate version of the title
% in square brackets, i.e., 
%\articletitle[Communism, Sparta, and Plato]
%{COMMUNISM, SPARTA,\\ and PLATO}

%% optional, to supply a shorter version of the title for the running head:
%%\chaptitlerunninghead{}

%% multiple authors may be separated with \\
%% \author{Samuel Bostaph\\
%% and Gregor Kariotis}

%% or,

%% \author{Samuel Bostaph}
%% \author{George Lewis}
%% \and   % <=== Type in \and before the last author so that `and' will
%% \author{Cleon Jones}    % print between the last two authors
% in the table of contents.
%%

%% \begin{keywords}
%% This is a Key word.
%% \end{keywords}

% optional prologue
%\prologue{<text>}{<author, year>}

% optional abstract
% \begin{abstract}
% text...
% \end{abstract}

% optional keywords

\tableofcontents
%------------ body of article ------------------->>

\section{Introduction}

\label{sec1} The colossal magnetoresistance phenomena discovered in the
early 90's in mixed-valent manganites has stimulated world-wide research
activities on the subject during the last decade \cite{1,2,3}. Intense
research has revealed many new aspects of this complex material system,
which now turn out to be basic to the physics of correlated many-body
systems. Several key phenomena observed in manganites, such as strong
coupling among spin/charge/orbital degrees of freedom, competition among
several order parameters, and the resulting inhomogeneous ground state, now
appear to be broadly applicable to other correlated systems including high 
\textit{T}$_{c}$ cuprates. Early experimental and theoretical results that
uncovered many exciting physical properties are discussed in many excellent
review articles (Ramirez, 1997 \cite{4}; Coey, Viret, and von Molnar, 1999 %
\cite{5}; Tokura and Tomioka, 1999 \cite{6}), and in several edited books
(Rao and Raveau, 1998 \cite{7}; Tokura, 1999 \cite{8}; Kaplan and Mahanti
1999 \cite{9}). One of us (SWC) also reviewed early experimental progress in
1998 and 1999 \cite{10,11}. Early works on optical spectroscopy of the
manganites were reviewed by Noh, Jung, and Kim \cite{12}, Okimoto and Tokura %
\cite{13}, and Cooper\cite{14}. An experimental overview of structure and
transport of the manganites was prepared by Salamon and Jaime, 2001 \cite{15}
and a theoretical review by Dagotto, Hotta, and Moreo, 2001 \cite{16} on
phase separation phenomena in manganites.

In this review, we cover the extensive progress made in understanding these
interesting systems in the last about three years, with emphasis on
multi-scale phase modulation phenomena. We include our recent experimental
results that provided unambiguous evidence for multiscale
charge/lattice/orbital inhomogeneity and the concomitant phase separation
phenomena in the manganites. We begin with a discussion of the basic physics
underlying these effects, and a brief summary of earlier relevant progress
in the physics of the manganite system.

Roughly speaking, colossal magnetoresistance (CMR) results from a
magnetic-field-induced shift of the ferromagnetic Curie temperature, and the
accompanying metal-insulator transition. The initial extensive research on
the origin of CMR effect revealed two important aspects of the intriguing
system. First, it appears that charge carriers in the mixed-valent
manganites are (Jahn-Teller-type) polaronic. In other words, charge carriers
in the \textit{e}$_{g}$ band of Mn$^{3+}$ tend to be localized by
accompanying lattice distortions and to be strongly dressed by clouds of
phonons. The lattice distortions accompanying the carriers probably resemble
local Jahn-Teller distortions and the polaronic nature of charge carriers
was much emphasized. However, it is noteworthy that the exact nature (i.e.,
metallic or insulating) of the ground state of the many-polaron problem is
still largely a mystery. The second important aspect, which is somewhat
related to the Jahn-Teller polaron issue, is the unusually strong coupling
among charge/spin/orbital degrees of freedom in manganites. For example,
when charge carriers in the \textit{e}$_{g}$ orbital can hop readily from
one Mn site to another, the double exchange mechanism produces ferromagnetic
coupling between localized \textit{t}$_{2g}$ spins \cite{17}. However, when
charge carriers tend to localize, superexchange coupling becomes active. It
turns out that the magnitude as well as the sign of this superexchange
coupling depends on the orbital character of the localized charge \cite{18}.
For example, when the \textit{e}$_{g}$ orbital electron of Mn$^{3+}$ faces
the neighboring Mn$^{4+}$, the superexchange coupling between the Mn$^{3+}$
and Mn$^{4+}$ is ferromagnetic. Otherwise, the superexchange coupling is
antiferromagnetic.

The prototypical manganite system is Ca-doped LaMnO$_{3}$, where the
concentration of Mn$^{4+}$ is controlled by the doping level of Ca$^{2+}$
into La$^{3+}$ sites. The general trend of physical properties with
concentration variation of Mn$^{3+}$ (or Mn$^{4+}$) can be summarized in the
following way: (1) When there are more Mn$^{3+}$ than Mn$^{4+}$ (i.e., 
\textit{x}\TEXTsymbol{<}0.5 in La$_{1-x}$Ca$_{x}$MnO$_{3}$), the ground
state tends to be metallic and ferromagnetic, probably induced by the double
exchange mechanism. (2) When there are more Mn$^{4+}$ than Mn$^{3+}$ (i.e, 
\textit{x}\TEXTsymbol{>}0.5), the ground state often becomes
charge-localized (insulating) and, in fact, charge-ordered with intriguing
patterns. Magnetic coupling in this insulator is dominated by superexchange
coupling, which is sensitive to orbital ordering. (3) Surprisingly, high
temperature properties differ strongly from those of the ground state. At
high temperatures, systems with \textit{x}\TEXTsymbol{<}0.5 tend to be more
insulating than those with \textit{x}\TEXTsymbol{>}0.5. For example, the
ground state of La$_{5/8}$Ca$_{3/8}$MnO$_{3}$ is ferromagnetic-metallic, but
the system becomes insulating above the Curie temperature. Thus, we may
conclude that the Jahn-Teller polaron nature is a controlling effect for 
\textit{x}\TEXTsymbol{<}0.5 at high temperatures. The details of these
aspects are well discussed in various review papers \cite%
{4,5,6,7,8,9,10,11,12,13,14,15,16}.

Recently, there has been an unexpected turn of events in manganite research.
The steadily accumulating body of experimental results clearly indicates
that the mixed-valent manganite system is electronically/magnetically
inhomogeneous so that a Hamiltonian approach may have severe limitations.
This inhomogeneity appears at various length scales as well as time scales.
Trivial inhomogeneity such as chemical inhomogeneity through inappropriate
synthesis of materials is partially responsible. However, there are
non-trivial inhomogeneities not directly related to the origin of CMR
effects. For example, in the low doping range with \textit{x}\TEXTsymbol{<}$%
\sim $0.1 and \textit{x}\TEXTsymbol{>}$\sim $0.9 in La$_{1-x}$Ca$_{x}$MnO$%
_{3}$, the system is a weakly doped antiferromagnet, with a tendency to
exhibit very-small-scale (likely nano-scale) phase mixtures of an
antiferromagnetic phase and a ferromagnetic phase. In fact, this is a
general trend when carriers are introduced into antiferromagnetic
(especially, Mott-type insulating) oxides.

There are, however, two important non-trivial inhomogeneities (or phase
modulations) in mixed-valent manganites that can be relevant to CMR effects;
these CMR-related phase modulations are the main theme of this chapter. One
non-trivial phase modulation occurs at low temperatures with micro-meter
length scales. For example, in Pr$^{3+}$-substituted La$_{5/8}^{3+}$Ca$%
_{3/8}^{2+}$MnO$_{3}$, ferromagnetic-metallic regions and charge-ordered
insulating regions coexist at low temperatures with sub-micro-meter scales.
The relative volume of the electronically/magnetically distinct regions
varies with Pr substitution. Furthermore, this relative volume is sensitive
to applied magnetic field (favoring the ferromagnetic-metallic state), which
leads to huge negative magnetoresistance. It turns out that long-range
strain, resulting from the significantly-different crystallographic
structures of the metallic and insulating phases, plays an important role in
producing large-length-scale phase coexistence in the system. This tendency
resembles what generally happens in martensitic phase transformations. In
martensitic systems, crystallographically-different phases can coexist at
low temperatures to accommodate the significant long-range strain associated
with martensitic transformation. Therefore, it appears that manganites are a
wonderfully unique system in which martensitic effects and
strongly-correlated physics both play critical roles.

The other important inhomogeneity is related to the presence of nano-scale
charge/orbital correlated nano-clusters (or nano-scale charge/orbital
ordering) in the background paramagnetic state above the Curie temperature.
For example, above the Curie temperature of 275 K in La$_{5/8}$Ca$_{3/8}$%
\newline
MnO$_{3}$, nano-scale charge/orbital ordering has been observed.
Furthermore, the nano-scale charge/orbital ordering is correlated with the
insulating nature of the system above the Curie temperature. In fact,
systematics of various manganites reveal that the insulating nature above
the Curie temperature is always associated with the presence of charge/%
\newline
orbital correlated nano-clusters, existing only in the orthorhombic
crystallographic structure. The theoretical reexamination of the exact role
of Jahn-Teller polarons as well as polaron-polaron correlations in the
insulator above the Curie temperature will be essential to understand the
systematics.

To address the non-trivial phase modulations that are directly related to
CMR properties in mixed-valent manganites, we organize this chapter as
follows. A revisited phase diagram on (La,Ca)MnO$_{3}$ system covering high
temperature structural transition boundaries is discussed in Section \ref%
{sec2}. Carrier-concentration-dependence of various electronic and thermal
transport properties is also presented in this section to complete the phase
diagram of this prototypical system. In Section \ref{sec3}, new experimental
findings are discussed, especially relating to the phase separation
phenomena with sub-micro-meter scales at low temperature. We focus on the
(La,Pr,Ca)MnO$_{3}$ system produced with varying chemical pressure.
Magnetotransport, electron microscopy, and optical conductivity studies are
used to investigate extensively the nature of the low temperature phase
separation. In Section \ref{sec4}, high temperature-nanoscale phase
modulations are explored in (La,Ca)MnO$_{3}$ systems using various
experimental approaches such as electronic/thermal transport and optical
conductivity measurements. Section \ref{sec5} focuses on the nanoscale phase
coexistence problem studied with \textit{x}-ray/neutron scattering
measurements. Section \ref{sec6} concludes this chapter.

\section{Phase diagram of La$_{1-x}$Ca$_{x}$MnO$_{3}$ manganites revisited}

\label{sec2}

\subsection{Construction of the phase diagram of (La,Ca)MnO$_{3}$}

The perovskite manganite (La,Ca)MnO$_{3}$ is a prototypical system that
exhibits various ground states as a function of carrier concentration. The
Ca ionic size is almost identical to the La one, and thus a true solid
solution forms in the entire range of Ca concentrations with minimal
disorder effects, which might come from a size mismatch of the A-site ions
in the ABO$_{3}$ perovskite structure. By varying relative ratios of La and
Ca, we can explore the influence of carrier concentrations on the physical
properties of this system.

To accurately obtain desired carrier concentrations, a large number of
polycrystalline La$_{1-x}$Ca$_{x}$MnO$_{3}$ specimens with high density were
prepared with similar synthesis conditions. The prepared Ca concentrations
include \textit{x}=0.0, 0.03, 0.04, 0.05, 0.08, 0.1, 0.125, 0.15, 0.2, 0.21,
0.25, 0.3, 0.333, 0.375, 0.4, 0.45, 0.48, 0.5, 0.52, 0.55, 0.6, 0.625, 0.65,
0.666, 0.67, 0.7, 0.75, 0.8, 0.83, 0.875, 0.9, 0.95, and 1.0. The
appropriate mixtures of pre-baked La$_{2}$O$_{3}$, CaCO$_{3}$, and MnO$_{2}$
underwent solid-state reaction at 1100-1400 $^{\circ }$C in air for four
days, and the samples were furnace-cooled. Because of the different
vaporization rates of the starting constituent materials, initial low
temperature calcining with frequent regrindings was performed. In addition,
the polycrystalline samples near the phase boundary such as ones near 
\textit{x}=0.5 (\textit{x}=0.485, 0.49, 0.495, 0.5, and 0.51) and near 
\textit{x}=0.20 (\textit{x}=0.185, 0.19, 0.195, 0.20, and 0.21) were
synthesized. To get more accurate oxygen stoichiometry, the samples for 
\textit{x}\TEXTsymbol{<}$\sim $0.2 were post-annealed in a flowing nitrogen
gas, and the samples for \textit{x}\TEXTsymbol{>}$\sim $0.8 in an oxygen for
12-60 hrs.

A previous study has focused on low temperature phase diagram based on the
detailed measurements of magnetic susceptibility ($\chi $) (or
magnetization) \cite{10}. The data were still incomplete to understand
detailed phase diagram. In this study, we extended our measurements to
resistivity ($\rho $), thermal conductivity ($\kappa $) and thermoelectric
power (\textit{S}) at low temperatures to determine doping-dependent
variations of electronic (charge) and phononic (lattice) degrees of freedom
in the system. In addition, to complete the phase diagram, we also
determined structural phase transition temperatures as a function of Ca
concentrations through the high-temperature-dc resistivity measurements.

\subsection{High temperature structural phase transitions}

High temperature $\rho $ was measured with a transport stick located inside
a quartz tube in a furnace. Temperature was measured with a Pt and Pt-Rh
thermocouple from 300 to $\sim $1000 K. The thermocouple was pre-calibrated
as a function of temperature. In each measurement, we used the conventional
4-probe technique with silver or gold pastes suitable for high temperature
measurements. Depending on the carrier concentration, care was taken to flow
a mixture of nitrogen and oxygen gases so that oxygen stoichiometry of a
sample does not change while measuring resistivity at high temperatures
above $\sim $500 K. As explained above, when prepared in air, samples with 
\textit{x}\TEXTsymbol{<}$\sim $0.2 tend to absorb oxygen, while samples with 
\textit{x}\TEXTsymbol{>}$\sim $0.8 are susceptible to oxygen deficiency.
Thus, larger amount of nitrogen gas were flown for low \textit{x} samples
while larger ratio of oxygen gas was required for high \textit{x} samples.
In all the measurements, we verified that both resistivity value and
structural transition temperature of each sample are reproducible after
several cooling and heating runs. 
\begin{figure}[tb]
\epsfxsize=60mm
\centerline{\epsffile{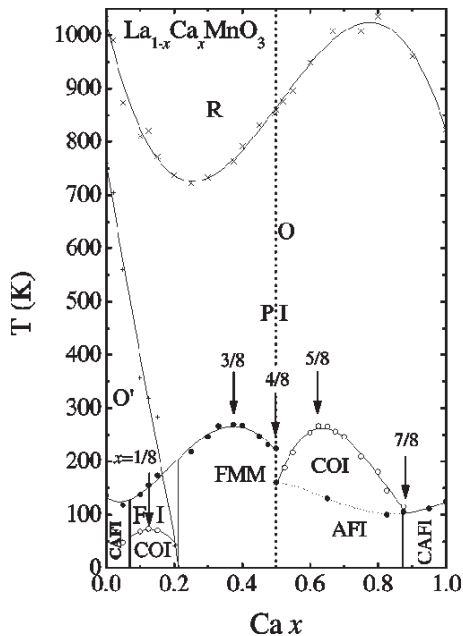}}
\caption{Phase diagram of La$_{1-x}$Ca$_{x}$MnO$_{3}$. There exist
well-defined anomalies at commensurate concentration of \textit{x}=N/8 (N=1,
3, 5, and 7) at low temperatures. PI: Paramagnetic Insulator, FMM:
Ferromagnetic Metal, COI: Charge Ordered Insulator, AFI: Antiferromagnetic
Insulator, FI: Ferromagnetic Insulator, and CAFI: Canted Antiferromagnetic
Insulator. Structural transitions occur at high temperatures. R:
Rhombohedral, $O$%
%TCIMACRO{\U{b4}}%
%BeginExpansion
\'{}%
%EndExpansion
: Orthorhombic with Jahn-Teller distorted, and $O$: Orthorhombic with
octahedron rotated. }
\label{Fig2-1}
\end{figure}

\begin{figure}[tb]
\epsfxsize=60mm
\centerline{\epsffile{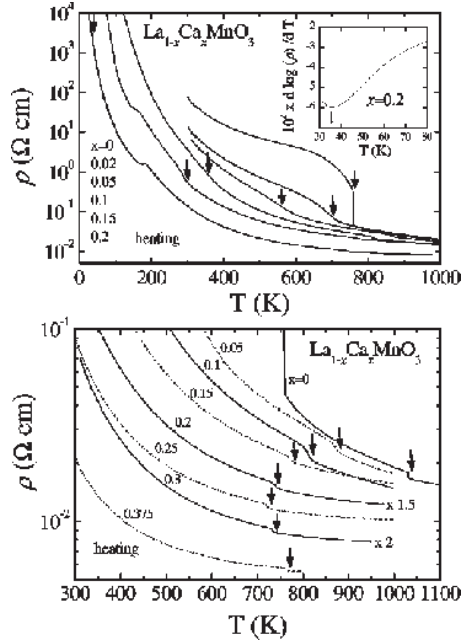}}
\caption{Upper panel: high temperature resistivity data of La$_{1-x}$Ca$_{x}$%
MnO$_{3}$ (0.0$\leq \mathit{x}\leq $20). Structural transition temperatures
from rhombohedral (high temperature) to $O$%
%TCIMACRO{\U{b4} }%
%BeginExpansion
\'{}
%EndExpansion
-orthorhombic (low temperature) are indicated as arrows. The inset indicates
that the structural transition temperature of \textit{x}=0.2 is located
around 36 K. Bottom panel: high temperature resistivity data of La$_{1-x}$Ca$%
_{x}$MnO$_{3}$ (0.0$\leq $\textit{x}$\leq $0.375) showing evolution of
another structural transition temperatures (arrows) from rhombohedral (high
temperature) to $O$-orthorhombic (low temperature). }
\label{Fig2-2}
\end{figure}

Fig. \ref{Fig2-2}(top panel) shows that the Jahn-Teller distortion of 
\textit{x}=0.0 occurring around 758 K is accompanied by a sharp increase of $%
\rho $. It is known that the Jahn-Teller distortion in a Mn$^{3+}$ ion
gives rise to a cooperative ordering of \textit{e}$_{g}$ orbitals in the
orthorhombic \textit{ab} plane. Structural analyses revealed that the
orbital ordering for the \textit{x}=0.0 sample leads to an elongation of 
\textit{a}- \& \textit{b}-axes and a shortening of \textit{c}-axis,
stabilizing the so-called $O$%
%TCIMACRO{\U{b4}}%
%BeginExpansion
\'{}%
%EndExpansion
-structure (\textit{a}\TEXTsymbol{>}\textit{b}\TEXTsymbol{>}\textit{c}/$%
\sqrt{2}$ ) below 758 K (754 K) for heating (cooling). [We are using an
orthorhombic \textit{Pbnm} notation, in which \textit{a}$\sim \sqrt{2}$%
\textit{a}$_{c}$, \textit{b}$\sim \sqrt{2}$\textit{a}$_{c}$, and \textit{c}$%
\sim $2\textit{a}$_{\mathit{c}}$, where \textit{a}$_{\mathit{c}}$ is the
cubic perovskite lattice constant.] At those temperatures, there exists a
sharp increase of $\rho $, as shown in Fig. \ref{Fig2-2} (top panel). As
doping \textit{x} increases, the increase of $\rho $ at the transition
becomes reduced and transition width becomes broad, indicating the
Jahn-Teller structural transtions become more incoherent with carrier
concentration \textit{x} increases. Furthermore, the transition temperature
(arrows in the top panel of Fig. \ref{Fig2-2}) decreases systematically. At 
\textit{x}=0.20, this transition seems to be located at 35 K as indicated by
a derivative of $\mathit{\rho }$ (the inset of Fig. \ref{Fig2-2}). The
structural distortion of \textit{x}=0.20 at such a low temperature should be
confirmed later with other techniques. Above the Jahn-Teller distortion
temperatures, it is known that a normal orthorhombic structure ($O$ type; 
\textit{a}$\approx $\textit{b}\TEXTsymbol{>}\textit{c}/$\sqrt{2}$) is
stabilized. The systematic evolution of the Jahn-Teller transition
temperatures is summarized in Fig. \ref{Fig2-1} as crosses (+). 

\begin{figure}[tbp]
\epsfxsize=60mm
\centerline{\epsffile{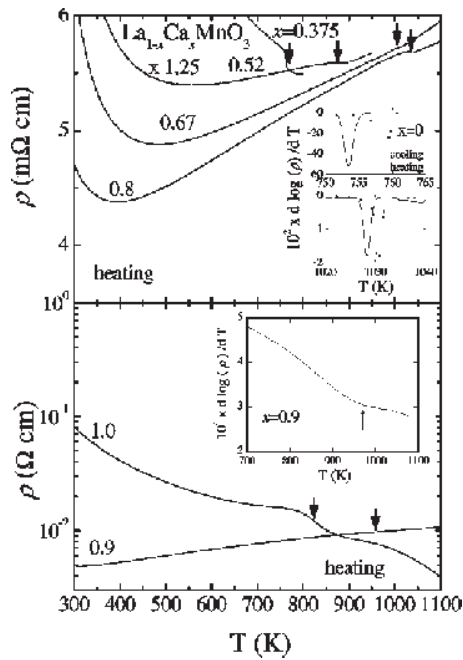}}
\caption{Upper panel: high temperature resistivity data of La$_{1-x}$Ca$_{x}$%
MnO$_{3}$ (0.375$\leq $\textit{x} $\leq $0.80). Structural transitions from
rhombohedral (high temperature) to O-orthorhombic structure are indicated as
arrows. The inset shows that two high-temperature structural transitions of 
\textit{x}=0.0 have large thermal hysteresis, showing the first-order
nature. Bottom panel: high temperature resistivity data for \textit{x}=0.9
and 1.0 exhibiting the structural transition temperatures (arrows). The
inset shows derivative of resistivity of the \textit{x}=0.9 sample,
indicating a structural transition around 980 K (arrow).}
\label{Fig2-3}
\end{figure}

On the other hand, in the bottom panel of Fig. \ref{Fig2-2} and in the inset
of Fig. \ref{Fig2-3}, another structural phase transition is clearly
identified at 1031 K (1028 K) for heating (cooling) in the \textit{x}=0.0
sample. By monitoring temperatures where a sudden drop of $\mathit{\rho }$
occurs, we determined evolutions of the structural transition temperatures
with increasing carrier concentrations. The arrows in the bottom panel of
Fig. \ref{Fig2-2} show that the structural transition temperature decreases
systematically down to 722 K as \textit{x} approaches 0.25 from 0.0.
Radaelli \textit{et al} \cite{19} reported an orthorhombic (\textit{Pbnm})
to rhombohedral (\textit{R}$\overline{3}$\textit{c}) transition near 725 K
for La$_{0.7}$Ca$_{0.3}$MnO$_{3}$ from a neutron diffraction study. This
result is consistent with the present transport studies. Therefore, this
high temperature structural transition is attributed to be one from $O$%
-orthorhombic (\textit{Pbnm}) to rhombohedral (\textit{R}$\overline{3}$%
\textit{c}) symmetry. As further shown in Fig. \ref{Fig2-3}, the transition
temperature again increases with \textit{x} until it reaches a broad maximum
near 1000 K for \textit{x}$\approx $0.80, and finally decreases above 
\textit{x}=0.80. The structural transition temperatures for both \textit{x}%
=0.9 and 1.0 are indicated the bottom panel of Fig. \ref{Fig2-3} and its
inset. Systematic doping-dependece of the structural transition temperatures 
is summarized in the phase diagram of Fig. \ref{Fig2-1}.
It is noteworthy that the transition temperatures from $O$-orthorhombic (%
\textit{Pbnm}) to rhombohedral (\textit{R}$\overline{3}$\textit{c})
structure display an interesting electron-hole symmetry as a function of
carrier concentration. In other words, a minimum temperature for the
structural transition occurs near \textit{x}=0.25 (hole doping), while a
maximum exists near \textit{x}=$\sim $0.75 ($\sim $0.25 electron doping).
Those temperatures become lower below \textit{x}=0.50 and higher above 
\textit{x}=0.50, forming an asymmetric phase line centered at \textit{x}%
=0.50. It is an interesting question to be answered how the evolution of
high temperature structural transition is linked to the ground state
properties of the system.

\subsection{Low temperature electronic/thermal transport and commensurate
effects at \textit{x}=N/8}

\subsubsection{Low temperature phase diagram revisited: N/8 anomaly}

\begin{figure}[tbp]
\epsfxsize=60mm
\centerline{\epsffile{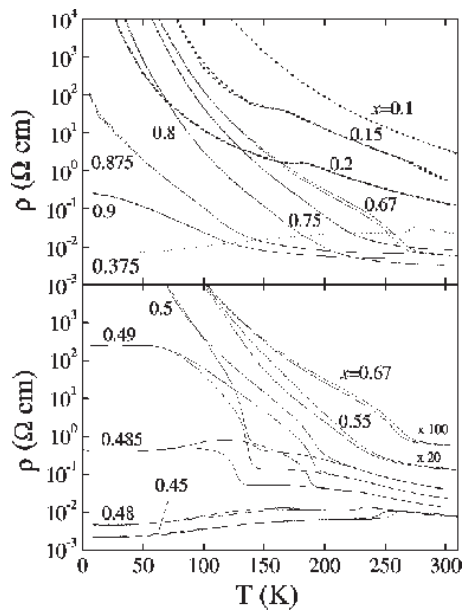}}
\caption{Low temperature-resistivity data of La$_{1-x}$Ca$_{x}$MnO$_{3}$. At
Curie temperature, each compound shows decrease of resistivity while
resistivity increases at charge ordering temperature. Those ordering
temperatures are consistent with the phase diagram shown in Fig. \ref{Fig2-1}%
. See texts for details.}
\label{Fig2-4}
\end{figure}

\begin{figure}[tbp]
\epsfxsize=60mm
\centerline{\epsffile{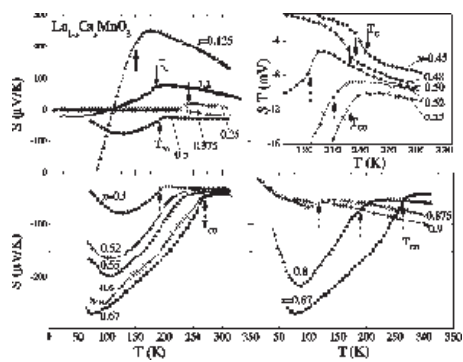}}
\caption{Temperature (\textit{T}) dependent-thermoelectric power \textit{S}
of La$_{1-x}$Ca$_{x}$MnO$_{3}$ at various doping ranges. At low temperature
region of ferromagnetic metallic samples with 0.20\TEXTsymbol{<} \textit{x} $%
\leq $0.48, \textit{S} value is very small, while at low temperature regions
of charge ordered samples with 0.50$\leq $\textit{x}$\leq $0.875, \textit{S}
becomes large negative. At high temperature region above ordering
temperatures, \textit{S} shows a continuous evolution from positive to
negative as doping \textit{x} increases. For \textit{x}$\leq $0.20, \textit{S%
} value is large positive at high temperature and decreases abruptly near 
\textit{T}$_{\text{C}}$, and becomes large negative at lower temperatures.}
\label{Fig2-5}
\end{figure}

\begin{figure}[tb]
\epsfxsize=60mm
\centerline{\epsffile{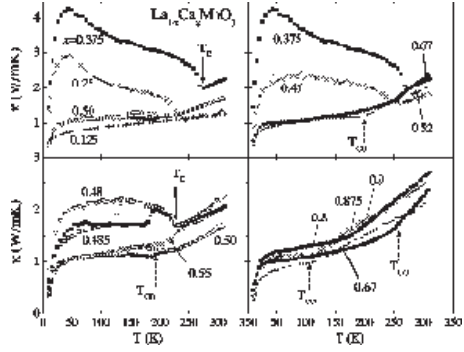}}
\caption{\textit{T-}dependent thermal conductivity $\protect\kappa $ plots
of La$_{1-x}$Ca$_{x}$MnO$_{3}$. $\protect\kappa $ in manganites is mostly
related to the phonon contribution so that increased local anisotropic
lattice distortions can make $\protect\kappa $ values decrease.}
\label{Fig2-6}
\end{figure}

\begin{figure}[tbp]
\epsfxsize=60mm
\centerline{\epsffile{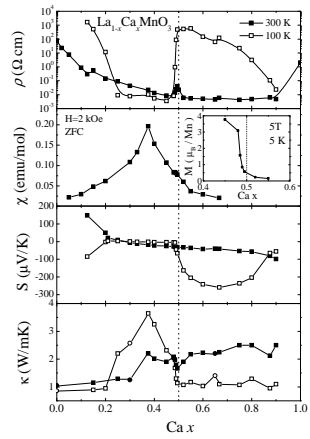}}
\caption{Summaries of resistivity $\protect\rho $, magnetic susceptibility $%
\protect\chi $, thermoelectric power \textit{S}, and thermal conductivity $%
\protect\kappa $ values at \textit{T}=100 (open squares) and 300 K (solid
squares) for La$_{1-x}$Ca$_{x}$MnO$_{3}$. Solid (300 K) and open (100 K)
circles represent the values measured in a previous study by Cohn \textit{et
al}.\protect\cite{22}. The inset of $\protect\chi $ plot shows the saturated
magnetic moment M, estimated at \textit{H}=5 T and \textit{T}=5 K.}
\label{Fig2-7}
\end{figure}

In a previous review by Cheong and Hwang \cite{10}, the phase diagram at low
temperatures was constructed from detailed measurements of magnetic
susceptibility primarily. The resulting phase diagram at low temperatures is
summarized in Fig. \ref{Fig2-1}. To characterize physical properties more
extensively, dc resistivity ($\rho $), thermoelectric power (\textit{S}),
and thermal conductivity ($\kappa $) were measured in the (La,Ca)MnO$_{3}$
system. Temperature-dependent $\rho $, \textit{S}, and $\kappa $ are
displayed in Fig. \ref{Fig2-4}, \ref{Fig2-5}, and \ref{Fig2-6},
respectively, for various Ca concentrations (\textit{x}) in La$_{1-x}$Ca$%
_{x} $MnO$_{3}$. The $\rho $, \textit{S}, $\chi $ ($\equiv $M/H), and $%
\kappa $ values at 100 K and 300 K for various \textit{x} are summarized in
Fig. \ref{Fig2-7}. The electronic/thermal transport data were measured in
a closed-cycle cryostat equipped with a turbo-molecular pump. For \textit{S}
and $\kappa $ measurements, a steady-state method was employed, using a
radiation shield and a precalibrated type E-thermocouple. In particular, for 
$\kappa $ measurements, high-density samples made with similar synthesis
conditions were used to reduce thermal conductivity errors coming from the
grain boundary scattering in polycrystalline specimens. 
Both perovskite LaMnO$_{3}$ and CaMnO$_{3}$ are antiferromagnetic
insulators. The Mn valence in LaMnO$_{3}$ is +3 (\textit{t}$_{\text{2}g}^{3}$%
\textit{e}$_{g}^{1}$) and one in CaMnO$_{3}$ is +4 (\textit{t}$_{\text{2}%
g}^{3}$) so that Mn ions in La$_{1-x}$Ca$_{x}$MnO$_{3}$ becomes
mixed-valent. LaMnO$_{3}$ becomes a good insulator below the Jahn-Teller
distortion temperature 758 K (see Fig. \ref{Fig2-2}), because Mn$^{3+}$ is a
good Jahn-Teller ion, and the doubly degenerate \textit{e}$_{g}$ bands
become split due to the Jahn-Teller distortion. In this special $O$%
%TCIMACRO{\U{b4}}%
%BeginExpansion
\'{}%
%EndExpansion
-orthorhombic structure, LaMnO$_{3}$ is known to become an A-type
antiferromagnet below 140 K, wherein ferromagnetic coupling exists in the
orthorhombic basal plane and antiferromagnetic coupling perpendicular to the
basal plane. Antiferromagnetic superexchange interaction between Mn$^{4+}$
ions via an intervening oxygen makes CaMnO$_{3}$ a normal G-type
antiferromagnet. Therefore, the initial Ca-substitution in LaMnO$_{3}$
produces hole carriers in the Jahn-Teller split \textit{e}$_{g}$ band, and
the low doping of La in CaMnO$_{3}$ induces electrons in the empty \textit{e}%
$_{g}$ band. Consistent with this, thermoelectric power \textit{S} data at
high temperatures (Fig. \ref{Fig2-5}) display large positive values for low
hole doping samples and large negative values for low electron doping ones. 
The real hopping of \textit{e}$_{g}$ electrons between Mn$^{3+}$ and Mn$%
^{4+} $ through an oxygen ion produces a ferromagnetic coupling in the
double exchange (DE) mechanism. Therefore, the increase of Ca-substitution
produces the smaller resistivity and thermoelectric power values. However,
the low Ca doping samples with 0.0\TEXTsymbol{<}\textit{x}\TEXTsymbol{<}$%
\sim $0.2, is still insulating even with ferromagnetic or canted
antiferromagnetic coupling. Similarly, the samples with high Ca doping (%
\textit{x}\TEXTsymbol{>}0.875), i.e., low La doping into CaMnO$_{3}$,
produces insulating behaviour, possibly with a canted-antiferromagnetic
coupling. Other interactions such as Jahn-Teller distortion and phase
segregation tendency in the low-hole doping regime seem to overcome the DE
mechanism, and produce the insulating behaviour even if a slight decreases
of resistivity are observed near \textit{T}$_{\text{C}}$ for both \textit{x}%
=0.15 and 0.2 (Fig. \ref{Fig2-4}). In addition, superexchange coupling
between Mn$^{4+}$ and Mn$^{3+}$ ions can be either antiferromagnetic or
ferromagnetic depending on the relative \textit{e}$_{g}$ orbital orientation %
\cite{18}.

In samples with 0.2$\leq $\textit{x}$\leq $0.5, where the CMR has been
observed, a paramagnetic insulating-to-ferromagnetic metallic transition
occurs upon cooling due to the DE mechanism. At the transition, a sharp
drop of $\mathit{\rho }$ is observed, as shown, for example, 
in the data of \textit{x}=0.375 in Fig. \ref{Fig2-4}. Then, 
\textit{S} values below \textit{T}$_{\text{C}}$ become very small ($\left| \mathit{S}%
\right| $\TEXTsymbol{<}1 $\mu $V/K), consistent with the creation of the
metallic carriers (See Fig. \ref{Fig2-5} and Fig. \ref{Fig2-7}). For the
higher doping range with \textit{x}$\geq $0.5, doped charge carriers
localize and order with stripe modulations at a charge ordering temperature 
\textit{T}$_{\text{CO}}$ \cite{10}, and an antiferromagnetic ordering occurs
at lower temperatures (at the Neel temperature \textit{T}$_{\text{N}}$). The
charge ordering is accompanied by a sharp increase of resistivity for doping
ranges 0.5$\leq $\textit{x}$\leq $0.875. At \textit{T}$_{\text{CO}}$, 
\textit{S} values of these samples decrease sharply upon cooling to become
more negative. The striped ordering of charge carriers reflects a generic
tendency toward microscopic (nano-scale) electronic phase separation in the
system, due to strong electron-lattice and electron-electron interactions.

The systematic behaviors of $\mathit{\rho }$ and \textit{S} observed in Fig. %
\ref{Fig2-2}-\ref{Fig2-5} are well summarized in Fig. \ref{Fig2-7},
displaying the $\mathit{\rho }$ and \textit{S} data at 100 K and 300 K with
varying \textit{x}. The $\rho $\ values at 100 and 300 K for low hole doping
regions are quite larger than those for low electron doping regions. This
indicates that the presence of Jahn-Teller distortion is responsible for the 
$\mathit{\rho }$ increase. For ferromagnetic metallic regions with 0.2$\leq $%
\textit{x}$\leq $0.5, both $\mathit{\rho }$ and \textit{S} become quite
small. The $\mathit{\rho }$ and \textit{S} increases at 100 K for 0.5%
\TEXTsymbol{<}\textit{x}$\leq $0.875 are consistent with localization of
charge carriers with charge ordering. As is evident in the phase diagram,
there are well-defined features at the commensurate carrier concentrations
of \textit{x}=N/8 (N=1,3,4,5, and 7) in La$_{1-x}$Ca$_{x}$MnO$_{3}$. It is
also found that there exist interesting electron-hole symmetry in the phase
diagram; \textit{T}$_{\text{C}}$ becomes maximum at \textit{x}=3/8 and 
\textit{T}$_{\text{CO}}$ peaks at \textit{x}=5/8 with similar \textit{T}$_{%
\text{C}}$ and \textit{T}$_{\text{CO}}$ temperatures. As another interesting
features at commensurate carrier concentrations, Fig. \ref{Fig2-7} reveals
that a slight anomaly of $\mathit{\rho }$ values exists at 300 K near the 
\textit{x}=0.50 compound. In addition, $\chi $ values at \textit{T}=300 K
display a strikingly sharp peak at \textit{x}=3/8 composition. These high
temperature anomalous features observed for \textit{x}=1/2 and 3/8 compounds
will be discussed in detail in Section \ref{sec4}. 
Interestingly, the compound \textit{x}=0.5 is located at the \textit{x}%
-dependent first-order phase-transition line between ferromagnetic (FM)
metals (\textit{x}\TEXTsymbol{<}0.5) and charge-ordered insulators (\textit{%
x}\TEXTsymbol{>}0.5) at low temperature. The \textit{x}=0.5 sample exhibits
both ferromagnetic and charge-ordering temperatures at 220 K and 180 K
(warming), respectively. It turns out that the FM metallic and CO insulating
phases coexist even at the lowest temperature. Therefore, as shown in the
inset of $\chi $ plot in Fig. \ref{Fig2-7}, the saturated magnetic moment M
estimated at \textit{H}=5 T and \textit{T}=5 K displays gradual decrease of
ferromagnetic moment as the system crosses the first-order phase transition
line to become a charge-ordered insulator.

\subsubsection{Thermal conductivity vs local lattice distortions}

To gain further insights on \textit{x}-dependent local lattice distortions,
we measured \textit{T}-dependent thermal conductivity ($\kappa $) of La$%
_{1-x}$Ca$_{x}$MnO$_{3}$ (Fig. \ref{Fig2-6}: heating). The $\kappa $ values
of La$_{1-x}$Ca$_{x}$MnO$_{3}$ are relatively small, showing about 1-3 WK$%
^{-1}$m$^{-1}$ at room temperature. Even the largest value observed in
ferromagnetic samples reaches only about 4 WK$^{-1}$m$^{-1}$ at low
temperatures. At \textit{T} above \textit{T}$_{\text{C}}$ of ferromagnetic
samples with 0.2$\leq $\textit{x}$\leq $0.5, $\kappa $(\textit{T}) increases
linearly with increasing \textit{T}, which is reminiscent of non-crystalline
solids. A previous work on the FM manganites has attributed the anomalous
behavior of $\kappa $, i.e., a linear increase above \textit{T}$_{\text{C}}$
to the phononic $\kappa $, $\kappa _{\text{ph}}$, coupled with large
anharmonic lattice distortions \cite{21}. In addition, it has been known
that an electronic thermal conductivity, $\kappa _{\text{e}}$, of adiabatic
small polarons, which might exist at \textit{T} above \textit{T}$_{\text{C}}$
of the ferromagnetic samples, is also very small \cite{22,23}. For all the
Ca concentrations in La$_{1-x}$Ca$_{x}$MnO$_{3}$, $\kappa _{\text{e}}$ is
negligible in the high temperature regions when it is estimated from the
measured $\mathit{\rho }$ in Fig. \ref{Fig2-4} and the Wiedemann-Franz law.
Even if $\mathit{\rho }$ values for single crystals are used, $\kappa _{%
\text{e}}$ is estimated to be only 20-40 \% of total measured $\kappa $ \cite%
{15}. Cohn \textit{et al}.\cite{22} found out that a magnon contribution to
the observed $\kappa $ is also negligible. Furthermore, they found out that $%
\kappa $ values of either ferromagnetic or charge-ordered samples are
inversely proportional to the degree of anisotropic lattice distortion of
the corresponding compound, which can be estimated from neutron diffraction
data. Thus, for all Ca doping ranges of La$_{1-x}$Ca$_{x}$MnO$_{3}$, it is
concluded that the electronic contribution to $\kappa $ is negligible, and
that the phonon contribution, directly coupled to large anharmonic lattice
distortion, dominates the measured $\kappa $.

From those previous findings and \textit{x}-dependence of $\kappa $ values
at 100 K and 300 K, summarized in Fig. \ref{Fig2-7}, we can get interesting
insights on \textit{x}-dependent local lattice distortions of La$_{1-x}$Ca$%
_{x}$MnO$_{3}$. First of all, $\kappa $ values at both 100 and 300 K are
quite enhanced at a commensurate hole-doping \textit{x}=3/8. Even though
further studies are necessary in samples with more fine doping spacing, Fig. %
\ref{Fig2-7} shows that a sharp increase of $\kappa $ at 300 K resembles
that of $\chi $ in the \textit{x}=3/8. This observation indicates that at
the commensurate doping of \textit{x}=3/8, local anharmonic lattice
distortions are minimized while the ferromagnetic correlation is maximized
at 300 K. It is already well known \cite{10} that \textit{T}$_{\text{C}}$
becomes maximal at \textit{x}=3/8 and, thus, electron hopping does maximal
in the DE mechanism. Therefore, in the context of existing strong
electron-lattice coupling in the manganites, it can be inferred that the
maximum electron hopping at \textit{x}=3/8 is closely associated with the
minimized local lattice distortions and maximized ferromagnetic correlation.

On the other hand, quite low $\kappa $ observed at low temperatures in the
charge-ordered samples, can be also attributed to long-range anisotropic
lattice strains, resulting from the cooperative Jahn-Teller distortion and
orbital ordering. Furthermore, systematic decrease of $\kappa $(100 K)
observed when \textit{x} approaches 0.0 from 0.2, can be linked to the
increased Jahn-Teller distortion and carrier localization. Therefore,
inverses of the $\kappa $ values, summarized in Fig. \ref{Fig2-6} and \ref{Fig2-7}%
, well represent the degree of local lattice distortions and closely related
carrier-localization tendency in both high and low temperature regions of La$%
_{1-x}$Ca$_{x}$MnO$_{3}$.

\section{Phase separation in manganites}

\label{sec3}

\subsection{Percolation model between two electronic phases \ in La$_{1-y}$Pr$_{y}$Ca$_{3/8}$MnO$_{3}$}

Extensive research on CMR in ferromagnetic
manganites has established that magnetoresistance (MR) increases
exponentially when \textit{T}$_{\text{C}}$ is reduced by e.g., varying
chemical pressure \cite{3}. Thus, enormous MR can be realized in low \textit{%
T}$_{\text{C}}$ materials. Since the CMR is realized only in such low 
\textit{T}$_{\text{C}}$ materials, it is important to understand how \textit{%
T}$_{\text{C}}$ is reduced from the optimum value of $\sim $375 K in
(La,Sr)MnO$_{3}$ system with chemical pressure. The chemical pressure can be
varied without changing the valence of Mn ions by substituting trivalent
rare earths with different ionic size into the A site of ABO$_{3}$
perovskites \cite{3}. This chemical pressure modifies the local structural
parameters such as Mn-O bond distance and Mn-O-Mn bond angle, which directly
influence the electron hopping between Mn ions (i.e., electronic band
width). Because ferromagnetic coupling between localized Mn \textit{t}$_{%
\text{2}g}$ spins is mediated by the hopping of \textit{e}$_{g}$ electrons
via avoiding the Hund's rule energy (the DE mechanism) \cite{17}, the band
width (\textit{t}) reduction by chemical pressure can be consistent with the
decrease of \textit{T}$_{\text{C}}$ in the DE mechanism. However, the
detailed structural studies, combined with band structure considerations,
clearly indicated that the \textit{t} change is too small to account for the
large decrease of \textit{T}$_{\text{C}}$ \cite{10,24,25}. Furthermore,
neutron scattering experiments on low-energy magnetic excitations in
manganites with various \textit{T}$_{\text{C}}$ showed that the spin wave
stiffness, proportional to \textit{t} in the double exchange mechanism,
changes little in different \textit{T}$_{\text{C}}$ manganites \cite{26}.

As an alternative scenario, it was suggested that the electron-lattice
coupling ($\lambda $), partially due to the Jahn-Teller effects, as well as 
\textit{t} plays an important role in manganite physics, in particular, in
low \textit{T}$_{\text{C}}$ materials with reduced bandwidth due to the
chemical pressure effect. For instance, it was theoretically predicted that 
\textit{T}$_{\text{C}}$ is a strong function of \textit{t}/$\lambda $ \cite%
{27,28,29}. Therefore, one might speculate that the increase of $\lambda $,
in addition to the \textit{t} decrease, results in the large \textit{T}$_{%
\text{C}}$ reduction by chemical pressure. However, the large change of $%
\lambda $ is not expected by the partial substitution of rare earths for the
chemical pressure variation because $\lambda $ is basically a local physical
parameter. Therefore, it was still puzzling in early 1999 why \textit{T}$_{%
\text{C}}$ is so strongly reduced by chemical pressure, and why CMR is only
realized in the reduced \textit{T}$_{\text{C}}$ materials.

Systematic experimental results in this section will show that the puzzling
question of manganites is a direct consequence of percolative electronic
phase separation in broad compositional ranges. Electronic phase separation,
in general, refers to the coexistence of two or more phases with distinctly
different electronic properties, which does not originate from chemical
inhomogeneity or chemical phase separation. This electronic phase separation
has also attracted considerable attention, particularly in the context of
the high-\textit{T}$_{c}$ superconducting cuprates \cite{30,31}. Systematic
investigations on the prototypical (La,Pr,Ca)MnO$_{3}$ system will
unambiguously indicate that percolative transport through charge-ordered
insulator and ferromagnetic metal mixtures, resulting from electronic phase
separation, plays the principal role in the CMR phenomenon.

\subsubsection{Magnetotransport study and percolative phase separation}

In the phase diagram in Fig. \ref{Fig2-1} and in a previous study \cite{10}, 
\textit{T}$_{\text{C}}$ is known to be optimized for \textit{x}=3/8 in both
La$_{1-x}$Ca$_{x}$MnO$_{3}$ and La$_{1-x}$Sr$_{x}$MnO$_{3}$. In addition, in
Pr$_{\text{1-}x}$Ca$_{x}$MnO$_{3}$ with \textit{x} near 3/8, long-range
(l-r) charge ordering (CO), instead of ferromagnetic (FM) metallicity,
occurs at low \textit{T}. Interestingly, the resistivity ($\rho $) rise at
the CO transition temperature (\textit{T}$_{\text{CO}}$) in Pr$_{\text{1-}x}$%
Ca$_{x}$MnO$_{3}$ was found to be most pronounced at \textit{x}=3/8 even
though \textit{T}$_{\text{CO}}$ changes monotonically near \textit{x}=3/8.
Therefore, the system of La$_{1-x-y}$Pr$_{y}$Ca$_{x}$MnO$_{3}$ with a
commensurate carrier doping \textit{x}=3/8 was chosen to investigate
chemical pressure effects on the FM and CO instability in a series of
samples. 
\begin{figure}[tb]
\epsfxsize=60mm
\centerline{\epsffile{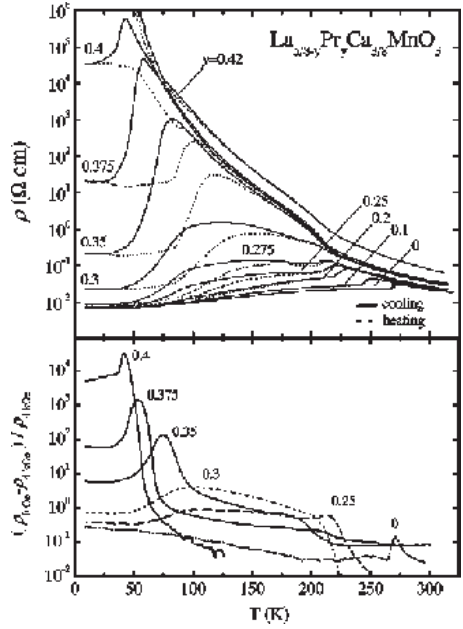}}
\caption{Upper panel: temperature dependent $\protect\rho $ for La$_{\text{%
5/8-}y}$Pr$_{y}$Ca$_{\text{3/8}}$MnO$_{\text{3}}$. Both cooling (solid
lines) and heating (dotted lines) curves are shown. Bottom panel:
Magnetoresistance of La$_{\text{5/8-}y}$Pr$_{y}$Ca$_{\text{3/8}}$MnO$_{\text{%
3}}$ in 4 kOe with field cooling.}
\label{Fig3-1}
\end{figure}

Fig. \ref{Fig3-1} shows \textit{T}-dependence of $\mathit{\rho }$ and MR of
La$_{\text{5/8-}y}$Pr$_{y}$Ca$_{\text{3/8}}$MnO$_{\text{3}}$. First, \textit{%
T}$_{\text{C}}$, defined as \textit{T} for the maximum slope of sudden $%
\mathit{\rho }$ drop, decreases slowly with increasing \textit{y} (0.0$\leq $%
\textit{y}$\leq $0.25), but \textit{T}$_{\text{CO}}$ ($\sim $210 K),
characterized by $\mathit{\rho }$ upturn, does not depend on \textit{y} ($%
\geq $0.3). Surprisingly, \textit{T}$_{\text{C}}$ is strongly suppressed
down to $\sim $80 K after \textit{T}$_{\text{C}}$ coincides with \textit{T}$%
_{\text{CO}}$ of $\sim $210 K (see the \textit{y}=0.3 data), indicating a
competition between FM and CO states. The most striking aspect, however, is
that enormously large residual $\mathit{\rho }$ (e.g., $\rho _{o}\approx $2$%
\times $10$^{4}$ $\Omega $cm for \textit{y}=0.4), much larger than the Mott
metallic limit, develops for large \textit{y} even though \textit{T}%
-dependence of $\mathit{\rho }$ is metallic-like below \textit{T}$_{\text{C}%
} $. Another interesting feature in Fig. \ref{Fig3-1} is that a significant
hysteresis develops when \textit{T}$_{\text{C}}$ becomes lower than \textit{T%
}$_{\text{CO}}\approx $210 K. This hysteresis indicates the first-order-type
transition in the system with two-phase coexistence. Even for 0.2$\leq $%
\textit{y}$\leq $0.275 where no CO transition is indicated in $\rho $(%
\textit{T}), the $\mathit{\rho }$ hysteresis is clearly observable, and this
hysteresis behavior appears to correlate with a broad $\mathit{\rho }$ hump
feature far below \textit{T}$_{\text{C}}$. In the \textit{y}=0 and 0.1 data,
the $\mathit{\rho }$ hysteresis is negligible, but a very broad $\mathit{%
\rho }$ hump far below \textit{T}$_{\text{C}}$ is clearly noticeable. This
finding suggests the persistence of the two-phase coexistence even in La$%
_{5/8}$Ca$_{3/8}$MnO$_{3}$. All these observations naturally suggest the
possibility that FM metallic and CO phases coexist, and percolative
transport through the metallic phase dominates the electronic conductivity.

\begin{figure}[tb]
\epsfxsize=60mm
\centerline{\epsffile{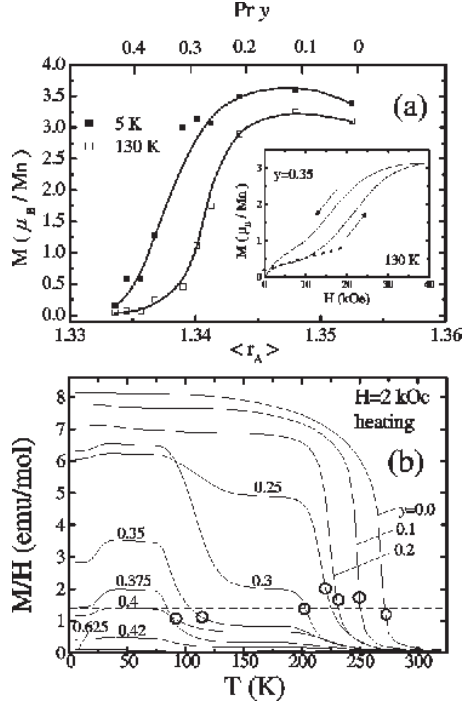}}
\caption{(a) Saturation moments at 5 K (closed squares) and 130 K (open
squares) of La$_{\text{5/8-}y}$Pr$_{y}$Ca$_{\text{3/8}}$MnO$_{\text{3}}$ as
a function of average ionic radius of La$_{\text{5/8-}y}$Pr$_{y}$Ca$_{\text{%
3/8}}$. Solid lines are drawn as guides for the eyes. The inset shows M(H)
for \textit{y}=0.35 at 130 K. The FM component is evident as indicated with
a dashed line, and there exists a field-induced transition in high fields.
(b) \textit{T}-dependent M/H curves of La$_{\text{5/8-}y}$Pr$_{y}$Ca$_{\text{%
3/8}}$MnO$_{\text{3}}$ in \textit{H}=2 kOe after zero field cooling. Open
circles on the M/H curves depict the M-I transition temperatures determined
from the $\protect\rho $ data in Fig. \ref{Fig3-1}. The dotted line
represents an average of the M/H values at the metal-insulator transition
temperatures, which is 17 \% of 8.1 emu/mole, the saturated M/H value of 
\textit{y}=0.0. \textit{T}-dependent volume fraction of the FM domains, 
\textit{f}(\textit{T}), for each \textit{y}, can be determined from \textit{f%
}(\textit{T},\textit{y}) = M$_{y}$(\textit{T})/M$_{0.0}$(\textit{T}).}
\label{Fig3-2}
\end{figure}

To further investigate this possibility, saturation moment and magnetization
of this system was measured as shown in Fig. \ref{Fig3-2}. First, for 
\textit{y}\TEXTsymbol{>}0.3, the saturation moment below \textit{T}$_{\text{C%
}}$ is significantly smaller than the full moment ($\sim $4 $\mu _{\text{B}}$%
) for the optimum \textit{T}$_{\text{C}}$ material of La$_{1-x}$Sr$_{x}$MnO$%
_{3}$ with \textit{x}$\approx $3/8 (see Fig. \ref{Fig3-2} (a)). In addition,
the ratio of the saturation moment ($\sim $0.6 $\mu _{\text{B}}$) for 
\textit{y}=0.4, close to the critical concentration (\textit{y}$_{c}$) for
the metal-insulator transition, and the full moment of 4 $\mu _{\text{B}}$
is within the three-dimensional percolation threshold range of 10-25 \% \cite%
{32}. Furthermore, the M/H curves shown in Fig. \ref{Fig3-2} (b) clearly
suggest that the metal-insulator (M-I) transition occurs at the percolation
threshold value for each sample. The open circles in Fig. \ref{Fig3-2} (b)
represent the M/H values at the metal-insulator (M-I) transition
temperatures estimated from the $\mathit{\rho }$ curves in Fig. \ref{Fig3-1}
(a). Interestingly, the average of those M/H values (dotted line) is about
17 \% of 8.1 emu/mol, the saturated M/H value of \textit{y}=0.0. Thus, with
changing \textit{T}, the M-I transition occurs when M of each sample becomes
about 17 \% of that of \textit{y}=0.0, independent of \textit{y} value. If
we assume that the \textit{T}-dependent volume fraction \textit{f}(\textit{T}%
) of the FM domain is proportional to M(\textit{T}) in \textit{H}=2 kOe, the
M-I transition with changing \textit{T} occurs when \textit{f }reaches $\sim 
$0.17, close to the three-dimensional percolation threshold (\textit{f}$_{c}$%
). 

MR of La$_{\text{5/8-}y}$Pr$_{y}$Ca$_{\text{3/8}}$MnO$_{\text{3}}$ is also
consistent with two-phase coexistence and percolative electronic conduction.
The MR of La$_{\text{5/8-}y}$Pr$_{y}$Ca$_{\text{3/8}}$\\MnO$_{\text{3}}$ was
measured in the relatively low field of 4 kOe, low enough to minimize the
change of the ground state (especially CO phase), but high enough to orient
magnetic domains. As shown Fig. \ref{Fig3-1} (b), the MR peaks near \textit{T%
}$_{\text{C}}$ result from the shift of \textit{T}$_{\text{C}}$ to higher 
\textit{T} by applying field, MR increases with decreasing \textit{T}$_{%
\text{C}}$, and MR becomes ``colossal'' in low \textit{T}$_{\text{C}}$
materials (particularly for 0.25$\leq $\textit{y}$\leq $0.4). One noticeable
feature is the existence of a plateau in the MR for 0.25$\leq $\textit{y}$%
\leq $0.375 at $\sim $80 K$\leq $\textit{T}$\leq \sim $210 K. This MR
plateau is associated with the two-phase coexistence even at \textit{T}$_{%
\text{C}}\leq $\textit{T}$\leq $\textit{T}$_{\text{CO}}$. The FM behavior at
130 K, between \textit{T}$_{\text{CO}}$ and \textit{T}$_{\text{C}}$, for 
\textit{y}=0.35 is clearly visible in the M($H$) curve (inset of Fig. \ref%
{Fig3-2} (a)), and the change of this FM component at 130 K with \textit{y}
is shown in Fig. \ref{Fig3-2} (a). The persistence of the FM component at
130 K for \textit{y} up to 0.41 indicates the two-phase coexistence even for 
\textit{T}$_{\text{C}}\leq $\textit{T}$\leq $\textit{T}$_{\text{CO}}$. The
saturation moment at 5 K in the same figure decreases sharply with \textit{y}
($\geq $0.3) where \textit{T}$_{\text{C}}$ is strongly suppressed below $%
\sim $80 K, corroborating the earlier discussion that two phases coexist for 
\textit{y}$\geq $0.3 below \textit{T}$_{\text{C}}$. It is noteworthy that
the saturation moment at 5 K for 0$\leq $\textit{y}$\leq $0.3, including 
\textit{y}=0, is also slightly suppressed from the optimum value of $\sim $4 
$\mu _{\text{B}}$ in La$_{1-x}$Sr$_{x}$MnO$_{3}$ (\textit{x}$\approx $3/8; 
\textit{T}$_{\text{C}}\approx $375 K). This suggests the presence of a small
amount of CO phase even for \textit{y}=0, consistent with the earlier
speculation from $\mathit{\rho }$ hump.

\subsubsection{Thermal and electronic transport properties vs two-phase model%
}

\begin{figure}[tb]
\epsfxsize=60mm
\centerline{\epsffile{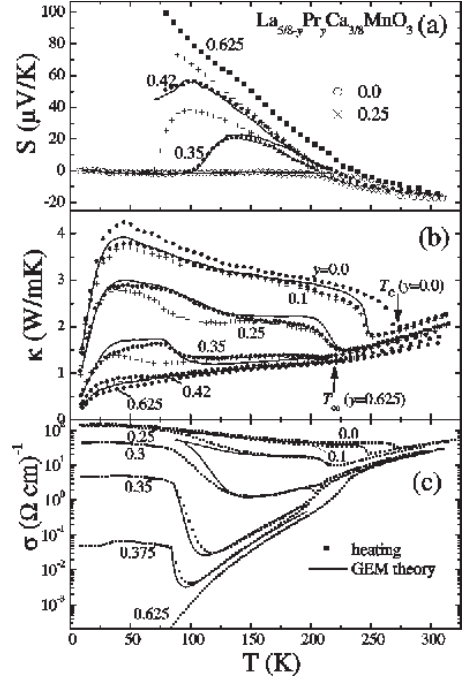}}
\caption{(a) Comparison of the experimental data and theoretical predictions
of \textit{S}(\textit{T}) of La$_{\text{5/8-}y}$Pr$_{y}$Ca$_{\text{3/8}}$MnO$%
_{\text{3}}$. Crosses (+) represent cooling curves for \textit{x}=0.42 and
0.35, and the others represent heating data. The solid lines show the
theoretical predictions, using the Eq. (1), for \textit{x}=0.25, 0.35, and
0.42 with heating. (b) $\protect\kappa $(\textit{T}) of La$_{\text{5/8-}y}$Pr%
$_{y}$Ca$_{\text{3/8}}$MnO$_{\text{3}}$. Solid circles show $\protect\kappa $
data with heating, and crosses (+) represent $\protect\kappa $ data with
cooling. ($\protect\kappa $ of \textit{y}=0.0 and 0.625 with cooling were
nearly the same with that with heating.) Solid lines show the theoretical
predictions by using the generalized effective medium (GEM) equation (Eq.
(2)) with \textit{t}=2 \& \textit{f}$_{\text{\textit{c}}}$=0.17. (c) \textit{%
S}(\textit{T}) by the GEM equation (Eq. (2)). Solid squares are the
experimental (heating) data redrawn from Fig. \ref{Fig3-1}, and solid lines
represent the theoretical predictions with \textit{t}=2 \& \textit{f}$_{%
\text{\textit{c}}}$=0.17 above \symbol{126}80 K.}
\label{Fig3-3}
\end{figure}
\begin{figure}[tbp]
\epsfxsize=60mm
\centerline{\epsffile{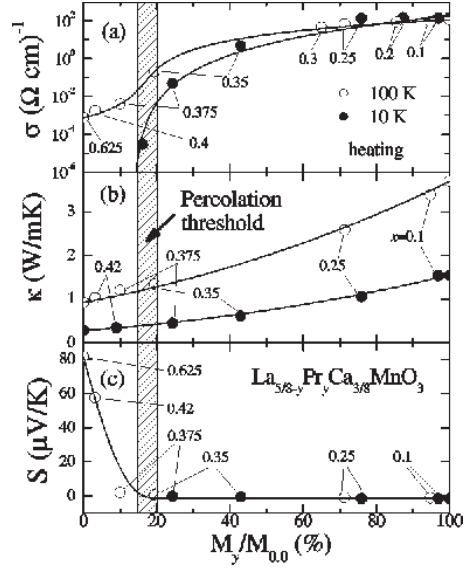}}
\caption{$\protect\sigma $, $\protect\kappa $, and \textit{S} values vs. M$%
_{y}$/M$_{0.0}$ at 10 and 100 K. The lines depict the theoretical
predictions by Eqs. (1) and (2). The solid lines in $\protect\sigma $ show
the theoretical results with \textit{t}=4 \& \textit{f}$_{\text{c}}$=0.15 at
10 K and \textit{t}=4 \& \textit{f}$_{\text{c}}$=0.17 at 100 K. For the
theoretical predictions (solid lines) of $\protect\kappa $ at 10 and 100 K, 
\textit{t}=2 \& \textit{f}$_{\text{c}}$=0.17 were used.}
\label{Fig3-4}
\end{figure}

Various aspects of above electronic transport results are consistent with
the coexistence of FM and CO phases, whose relative volumes change with both 
\textit{T} and \textit{y}, and the percolative M-I transition in FM-CO
mixtures. \textit{T}-dependent $\kappa $ and \textit{S} are also shown in
Figs. \ref{Fig3-3} (a) and (b). In addition, the $\mathit{\rho }$ data in
Fig. \ref{Fig3-1} is redrawn as conductivity vs \textit{T} in Fig. \ref%
{Fig3-3} (c). First, as explained in detail in Section \ref{sec2}, the
measured $\kappa $(\textit{T}) for all \textit{y} has the dominant phonon
contribution. With increasing \textit{y}, $\kappa $(\textit{T}) shows smooth
variation from that of \textit{y}=0.0 to that of \textit{y}=0.625, and the $%
\kappa $ increase at \textit{T}$_{\text{C}}$ becomes smaller. Consistently, $%
\kappa $ vs. M$_{y}$(\textit{T})/M$_{0.0}$(\textit{T}) at 10 and 100 K (Fig. %
\ref{Fig3-4}) shows that $\kappa $ varies monotonically from the maximum ($%
\kappa $ of \textit{y}=0.0) to the minimum ($\kappa $ of \textit{y}=0.625).
In comparison with $\kappa $ (and $\sigma $), \textit{S} exhibits seemingly
different behaviors with \textit{y}. \textit{S} is very close to the
metallic values even near \textit{f}$_{c}$, where $\sigma $ (or $\kappa $)
is still significantly smaller than $\sigma _{\text{M}}$ (or $\kappa _{\text{%
M}}$). \textit{S} vs. M$_{y}$/M$_{0.0}$ at 100 K (Fig. \ref{Fig3-4} (c))
clearly demonstrates this tendency; \textit{S} is close to zero (slightly
negative), and is insensitive to M$_{y}$/M$_{0.0}$ as long as M$_{y}$/M$%
_{0.0}$ $\geq \sim $10 \%. In contrast, $\sigma $ at 100 K of M$_{y}$/M$_{0.0}$ = 10
\% is more than three orders of magnitude smaller than that of M$_{y}$/M$_{0.0}$ = 100
\% (\textit{y}=0). In fact, \textit{T}-dependence of \textit{S} near \textit{T}$_{%
\text{C}}$ is also consistent with the metallic \textit{S} behavior near 
\textit{f}$_{c}$. With decreasing \textit{T} near \textit{T}$_{\text{C}}$, 
\textit{S} starts to decrease, i.e., becomes metallic at \textit{T} higher
than those for $\sigma $ and $\kappa $ changes. For example, in the heating
curves of \textit{y}=0.35, \textit{S} starts to decrease around 130 K,
significantly higher than \textit{T} ($\sim $100 K) for abrupt $\kappa $
increase or \textit{T} ($\sim $110 K) for $\sigma $ minimum. When
thermal/electronic transport properties of the system are viewed as those of
M-I mixtures, the above peculiar \textit{S} behavior is, in fact, consistent
with the theoretical prediction of effective thermoelectric power \textit{S}$%
_{\text{E}}$ by Bergmann and Levy \cite{33}. For an isotropic M-I binary
mixture, they showed that in terms of $\sigma $, $\kappa $, and \textit{S}
of each component, \textit{S}$_{\text{E}}$ is given by%
\begin{equation}
S_{\text{E}}=S_{\text{M}}+(S_{\text{I}}-S_{\text{M}})(\frac{\kappa _{\text{E}%
}/\kappa _{\text{M}}}{\sigma _{\text{E}}/\sigma _{\text{M}}}-1)/(\frac{%
\kappa _{\text{I}}/\kappa _{\text{M}}}{\sigma _{\text{I}}/\sigma _{\text{M}}}%
-1),  \label{Eq1}
\end{equation}%
where the subscripts M and I refer to metallic and insulating components,
respectively. $\kappa _{\text{E}}$ and $S_{\text{E}}$ refer to effective
thermal and electric conductivity, respectively, of the binary mixture. This
equation has been successfully applied to explain \textit{S} behaviors of
binary Al-Ge films \cite{34}. When $\mathit{\sigma }_{\text{I}}$/$\mathit{%
\sigma }_{\text{M}}$\TEXTsymbol{<}\TEXTsymbol{<} $\mathit{\kappa }_{\text{I}%
} $ /$\mathit{\kappa }_{\text{M}}$ \TEXTsymbol{<}1, which applies to our
system, and for \textit{f }= \textit{f}$_{c}$, the above equation leads to 
\textit{S}$_{\text{E}}$ $\approx $\textit{S}$_{\text{M}}$, which explains
our experimental results noted above.

To quantitatively compare experimental results in Fig. \ref{Fig3-3} with Eq. %
\ref{Eq1} , we calculated $\mathit{\sigma }_{\text{E}}$ at every \textit{T}.
In this comparison, experimental $\mathit{\sigma }$ and $\kappa $, shown in
Fig. \ref{Fig3-3}, were used for $\sigma _{\text{E}}$ and $\kappa _{\text{E}%
} $ and ($\mathit{\sigma }_{\text{I}}$, $\mathit{\kappa }_{\text{I}}$, and $%
\mathit{S}_{\text{I}}$) and ($\mathit{\sigma }_{\text{M}}$, $\mathit{\kappa }%
_{\text{M}}$, and \textit{S}$_{\text{M}}$) are assumed to be identical with
those of \textit{y}=0.0 and 0.625, respectively. The solid lines in the
bottom panel of Fig. \ref{Fig3-3} (a) depict the calculated \textit{S}$_{%
\text{E}}$, using Eq. \ref{Eq1}, for heating curves of \textit{y}=0.25,
0.35, and 0.42. The calculated curves match with our experimental \textit{S}
surprisingly well at all \textit{T} below \textit{T}$_{\text{C}}$ or \textit{%
T}$_{\text{CO}}$.

For $\sigma _{\text{E}}$ (or $\kappa _{\text{E}}$) of a binary M-I mixture,
Mclachlan \cite{35} proposed the generalized effective medium (GEM) equation,%
\begin{equation}
(1-f)(\frac{\sigma _{\text{I}}^{1/t}-\sigma _{\text{E}}^{1/t}}{\sigma _{%
\text{I}}^{1/t}+A\sigma _{\text{E}}^{1/t}})+f(\frac{\sigma _{\text{M}%
}^{1/t}-\sigma _{\text{E}}^{1/t}}{\sigma _{\text{M}}^{1/t}+A\sigma _{\text{E}%
}^{1/t}})=0,  \label{Eq2}
\end{equation}%
where \textit{A} $\equiv $(1-\textit{f}$_{c}$)/\textit{f}$_{c}$. The same
equation also works for $\kappa $. The critical exponent \textit{t} is close
to 2 in three dimension. This equation has been successfully applied to
isotropic inhomogeneous media in wide \textit{f} regions including
percolation regime \cite{34,35,36}.

By using above GEM equation, the $\kappa $(\textit{T}) for various \textit{y}
can be calculated with the assumption that $\mathit{\kappa }_{\text{I}}$(%
\textit{T})=$\kappa $(\textit{T}) of \textit{y}=0.625, $\mathit{\kappa }_{%
\text{M}}$(\textit{T})=$\kappa $(\textit{T}) of \textit{y}=0.0, \textit{t}%
=2, and \textit{f}$_{\text{\textit{c}}}$=0.17. The solid lines in Fig. \ref%
{Fig3-3} (b) represent the estimated $\kappa _{\text{E}}$ for \textit{y}%
=0.1, 0.25, 0.35, and 0.42. In addition, the calculated $\kappa _{\text{E}}$
as a function of M$_{y}$/M$_{0.0}$ at 10 and 100 K is depicted as solid
lines in Fig. \ref{Fig3-4} (b). Estimated $\kappa _{\text{E}}$ lines in
Figs. \ref{Fig3-3} and \ref{Fig3-4} coincide with the experimental data
well. In order to confirm self-consistency, \textit{S}$_{\text{E}}$ at 10
and 100 K is evaluated by using the calculated $\sigma _{\text{E}}$ (see
below) and $\kappa _{\text{E}}$ (solid lines of Figs. \ref{Fig3-3} (b) and
(c)), and the Eq. \ref{Eq1}. The calculated \textit{S}$_{\text{E}}$ (solid
lines of Fig. \ref{Fig3-4} (c)) with the variation of \textit{y} is in good
agreement with the experimental values. 
To apply the GEM equation to $\sigma $(\textit{T}), we assumed that \textit{f%
}(\textit{T},\textit{y}) = M$_{y}$(\textit{T})/M$_{0.0}$(\textit{T}), $%
\mathit{\sigma }_{\text{M}}$(\textit{T}) = $\sigma $(\textit{T}) of \textit{y%
}=0.0, and $\mathit{\sigma }_{\text{I}}$(\textit{T}) = $\sigma $(\textit{T})
of \textit{y}=0.625. With the parameters \textit{t}=2 \& \textit{f}$_{\text{%
\textit{c}}}$=0.17, the calculated $\sigma _{\text{E}}$ for various \textit{y%
} are shown as solid lines in Fig. \ref{Fig3-3} (c). At \textit{T} $\sim $80
K, $\sigma _{\text{E}}$(\textit{T}) nicely matches the experimental $\sigma $%
(\textit{T}) even if $\mathit{\sigma }$ changes by 6 orders of magnitude
with \textit{T} and \textit{y}. However, this agreement does not hold at
very low \textit{T}. The calculated $\sigma _{\text{E}}$(\textit{T}) at 
\textit{T} \TEXTsymbol{<} 80 K with the same parameters \textit{t}=2 \& 
\textit{f}$_{\text{\textit{c}}}$=0.17 significantly deviated from the
experimental $\sigma $(\textit{T}). We found that at \textit{T} \TEXTsymbol{<%
} 80 K, the calculated $\sigma _{\text{E}}$(\textit{T}) matches the
experimental $\sigma $(\textit{T}) better when \textit{t} is increased to $%
\sim $4. 
\begin{figure}[tbp]
\epsfxsize=60mm
\centerline{\epsffile{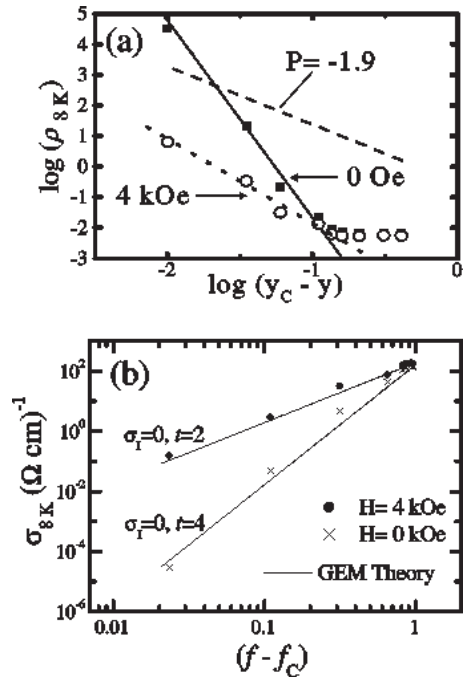}}
\caption{(a) log $\protect\rho _{\text{o}}$ vs. log (\textit{y}$_{c}$-%
\textit{y}) plot for zero field (closed circle) and 4 kOe (open circle) with 
\textit{y}$_{c}$=0.41. For 0.275$\leq $\textit{x}$\leq $0.4, there exists a
linear relationship with the slope of -6.9 for zero field (-2.5 for 4 kOe).
The dotted line with the slope of -1.9 represents the prediction of a
percolation theory $\protect\rho _{\text{o}}$\symbol{126}(\textit{y}$_{c}$-%
\textit{y})$^{P}$. (b) $\protect\sigma _{o}$ vs. (\textit{f}-\textit{f}$_{%
\text{\textit{c}}}$) log scale plot redrawn from the data of (a), for zero
field (closed circle) and 4 kOe (open circle) with \textit{f}$_{\text{%
\textit{c}}}$=0.17. Temperature-dependent volume fraction of the FM domains, 
\textit{f}(\textit{T}), for each \textit{y}, was determined from the
relation of \textit{f}(\textit{T},\textit{y}) = M$_{y}$(\textit{T})/M$_{0.0}$%
(\textit{T}) and the data in Fig. \ref{Fig3-2} (b). The two solid lines
represent the prediction of GEM theory with $\protect\sigma _{\text{I}}$=0, $%
\protect\sigma _{\text{M}}$\symbol{126}(\textit{f}-\textit{f}$_{\text{%
\textit{c}}}$)$^{t}$, with the critical exponents \textit{t}, 2 and 4,
respectively. Note the GEM equation reduces to the percolation equation when 
$\protect\sigma _{\text{I}}$=0.}
\label{Fig3-5}
\end{figure}
This anomalous critical exponent at zero-field becomes more evident in the 
\textit{y} dependence of $\mathit{\rho }$ at 8 K. In Fig. \ref{Fig3-5} (a),
log $\rho _{\text{o}}$ is plotted as a function of log (\textit{y}$_{c}$-%
\textit{y)} with \textit{y}$_{c}\equiv $0.41. Surprisingly, there exists a
linear region (0.4$\leq $\textit{y}$\leq $0.275) in the log $\rho _{\text{o}%
} $ vs. log (\textit{y}$_{c}$-\textit{y}) plot where $\rho _{\text{o}}$
changes about 7 orders of magnitude. The slope ($\sim $-6.9) of the linear
region turns out to be much larger than the three-dimensional percolation
prediction ($\sim $-1.9). In other words, with the decreasing volume of FM
phase, $\rho _{\text{o}}$ increases much faster than the simple percolation
consideration. Furthermore, as seen in Fig. \ref{Fig3-5} (b), the slope
still remains about --4 even when it is calculated from the log $\sigma _{%
\text{o}}$ vs. log (\textit{f}-\textit{f}$_{c}$) plot with \textit{f}$%
_{c}\equiv $0.17, where \textit{f} and \textit{f}$_{c}$ was determined from
the FM volume fraction in the magnetization data of Fig. \ref{Fig3-2}. [Note
that the GEM theory reduces to a percolation equation $\mathit{\sigma }_{%
\text{M}}$ $\sim $(\textit{f}-\textit{f}$_{c}$)$^{t}$ when $\mathit{\sigma }%
_{\text{I}}$ =0.] These observations demonstrate that \textit{t}, normally
close to a three-dimensional exponent of $\sim $2, becomes $\sim $4 at very
low \textit{T}. A similar, drastic increase of \textit{t} has been noted in
the case of tunneling transport for M-I mixtures \cite{34,37}, suggesting
that a tunneling process between FM domains is important for $\mathit{\sigma 
}$ of our system at very low \textit{T} \cite{38}. In other words, as will
be also evident in a next section, the reduced electron conduction through
FM regions can be responsible for the anomalous \textit{t} value at zero
field when the magnetizations of neighboring FM regions are not aligned.

In summary, the unambiguous agreement between the measured
thermal/electronic transport properties and the calculated values based on
Eqs. \ref{Eq1} and \ref{Eq2} strongly indicates that: (1) transport
properties are dominated by thermal/electrical conduction in M-I mixtures,
(2) the relative volume of the (FM) metallic phase is proportional to the
measured M(\textit{T},\textit{y}), and (3) the \textit{T}-dependent
transport and magnetic properties of metallic and insulating phases are
always that of \textit{y}=0 and 5/8, respectively.

\subsubsection{Electron microscopy study and colossal magnetoresistance}

\begin{figure}[tbp]
\epsfxsize=60mm
\centerline{\epsffile{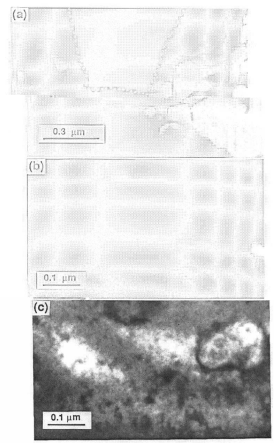}}
\caption{Dark-field images of electron microscopy, obtained by using one
superlattice peak caused by CO, for La$_{\text{5/8-}y}$Pr$_{y}$Ca$_{\text{3/8%
}}$MnO$_{\text{3}}$ with \textit{y}=0.375 at 20 K (top), \textit{y}=0.4 at
17 K (middle) and \textit{y}=0.4 at 120 K (bottom). Bright regions are due
to CO, and dark areas show charge-disordered regions. The river-like dark
lines in bright regions are due to discommensurations.}
\label{Fig3-6}
\end{figure}

Direct evidence of the two-phase coexistence is provided by the change of
the CO correlation length with \textit{y} and \textit{T} investigated by
electron diffraction experiments. It is well established that superlattice
peaks due to CO can be readily detected by electron diffraction \cite{20,39}%
. The dark-field image, obtained from a superlattice peak due to CO, at 20 K
for \textit{y}=0.375 in Fig. \ref{Fig3-6} (a) clearly exhibits the existence
of CO regions even below \textit{T}$_{\text{C}}$. The white domains are CO
regions, and the dark domain (upper and middle part of the picture) is a
charge-disordered, presumably, FM region. It was also unexpectedly found
that the typical size of CO and FM regions at low \textit{T} is on the order
of several thousand angstroms. This micro-scale mixture is commonly observed
in other compositions at low \textit{T}, and is in contrast with the
nano-scale mixture of CO and FM phases in La$_{0.5}$Ca$_{0.5}$MnO$_{3}$ at $%
\sim $150 K\TEXTsymbol{<}\textit{T}\TEXTsymbol{<}220 K. 

In order to observe the \textit{T}-evolution of CO, dark-field images at 17
K (Fig. \ref{Fig3-6} (b)) and 120 K (Fig. \ref{Fig3-6} (c)) are taken for 
\textit{y}=0.4. It is found that the CO regions are micro-scale (a few
thousand angstroms in size) domains at 17 K, similarly to those of \textit{y}%
=0.375, and become short-ranged, nano-scale domains at 120 K, as can be seen
from the white spotted areas in Fig. \ref{Fig3-6} (c). From the non-zero
saturation moment at 130 K even for \textit{y}\TEXTsymbol{>}0.25 and the
presence of nano-scale CO regions in the dark field image at 120 K for 
\textit{y}=0.4, it is inferred that the nano-scale coexistence of CO and FM
observed in La$_{0.5}$Ca$_{0.5}$MnO$_{3}$ is realized for 0.25\TEXTsymbol{<}%
\textit{y}$\leq $0.4 at $\sim $120-130 K. One more surprise in the electron
diffraction results was the observation of the \textit{x}=1/2-type
superlattice peaks (not shown) in all our specimens with the carrier
concentration of 3/8 (i.e., the modulation wave length in orthorhombic basal
planes is about 2$\times $\textit{a} with \textit{a}$\approx $5.5 \AA ,
indicating the strong stability of the \textit{x}=1/2-type CO state.) In
fact, the \textit{x}=1/2-type CO for \textit{x} less than 1/2 has been
observed in earlier reports in, e.g., Pr$_{0.7}$Ca$_{0.3}$MnO$_{3}$ \cite{40}%
. In spite of the presence of the \textit{x}=1/2-type CO state, the carrier
concentration of CO regions is presumably close to 3/8, because it must be
difficult to break the charge neutrality on large length scale as large as
thousands of angstroms. This difference in carrier concentration could be
accommodated by charge defects in the \textit{x}=1/2-type CO state.

All of above experimental studies unequivocally suggest a clear physical
picture to understand the large $\rho _{o}$ and the large MR in reduced 
\textit{T}$_{\text{C}}$ manganites. Since micro-scale FM regions exist in
these samples, electron conduction between neighboring FM regions should be
reduced if their magnetizations are not aligned. This conduction reduction
must be particularly large in manganites where hopping electrons are
strongly aligned with the local orientation of magnetization (i.e.
manganites are the so-called half-metallic ferromagnet) \cite{38}. The
electron conduction between neighboring FM regions could be dominated by
tunneling if adjacent FM regions are separated by this insulating region.
The degree of this magnetization misalignment will increases when the system
approaches the percolation threshold due to the increasingly poor connection
between the FM regions. Therefore, with decreasing volume of FM regions, $%
\rho _{o}$ increases much faster than the prediction of simple percolation.
It is then expected that when the FM regions are aligned in applied fields, $%
\rho _{o}$ should once again follow the percolation prediction. This appears
to be consistent with our MR results. The variation of $\rho _{o}$ (open
circles) in 4 kOe with \textit{y} is shown in Fig. \ref{Fig3-5} (a). The
same plot is also drawn as $\sigma _{o}$ vs \textit{f} (FM volume fraction)
in Fig. \ref{Fig3-5} (b). The slope of the linear region for $\rho _{o}$ and 
$\sigma _{o}$ in 4 kOe is about $\sim $-2.6 and 2 in Fig. \ref{Fig3-5} (a)
and (b), respectively. Thus, both plots are showing the exponents closer to
the percolation prediction of $\sim $-1.9.

\begin{figure}[tbp]
\epsfxsize=60mm
\centerline{\epsffile{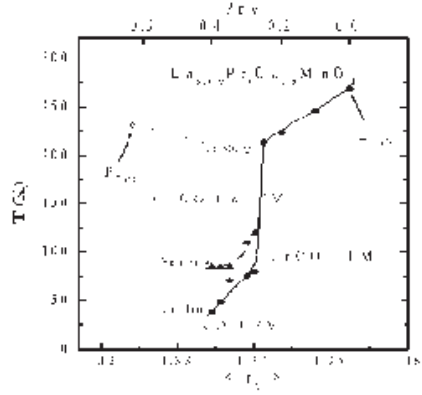}}
\caption{The phase diagram of La$_{\text{5/8-}y}$Pr$_{y}$Ca$_{\text{3/8}}$MnO%
$_{\text{3}}$ as a function of average ionic radius of La$_{5/8-y}$Pr$_{y}$Ca%
$_{3/8}$. \textit{T}$_{\text{C}}$ and \textit{T}$_{\text{CO}}$ are shown
with closed circles (or triangles) and open circles, respectively. There
exists a large thermal hysteresis at \textit{T}$_{\text{C}}$ for \textit{y}%
*=0.275. The coexistence of long range (l-r) FM/short range (s-r) CO, s-r
CO/s-r FM, or l-r CO/l-r FM is schematically shown in the phase diagram.}
\label{Fig3-7}
\end{figure}

All of our findings can be summarized within the two-phase picture. (1) When 
\textit{T}$_{\text{C}}$ is above $\sim $210 K, but less than the optimum
value of $\sim $375 K, there exists a long-range (l-r) FM state with a
short-range (s-r) CO phase. (2) When the $\mathit{\rho }$ upturn at \textit{T%
}$_{\text{CO}}$ is clearly visible (for 0.3$\leq $\textit{y}$\leq $0.4), 
\textit{T}$_{\text{C}}$ is strongly suppressed below $\sim $80 K. In this
case, s-r CO and s-r FM coexist at \textit{T}$_{\text{C}}\leq $\textit{T}$%
\leq $\textit{T}$_{\text{CO}}$, and l-r FM with l-r CO develops below 
\textit{T}$_{\text{C}}$. This view, as well as the experimental values of 
\textit{T}$_{\text{C}}$ and \textit{T}$_{\text{CO}}$, is summarized in the
phase diagram of Fig. \ref{Fig3-7}. Note that the experimental values of 
\textit{T}$_{\text{C}}$ and \textit{T}$_{\text{CO}}$ are determined from our 
$\mathit{\rho }$ data, and they coincide with those from our M(\textit{T})
data in Fig. \ref{Fig3-2} (b). Furthermore, our M(\textit{T}) data show a
spin glass-like transition below $\sim $25 K for \textit{y}$\geq $0.25;
there exists a slight difference between zero-field-cooled and field-cooled
M(\textit{T}) data below $\sim $25 K. Note that the average ionic radius, 
\TEXTsymbol{<}\textit{r}$_{A}$\TEXTsymbol{>}, was estimated from the
tabulated values for twelve coordination \cite{41}. The initial slow \textit{%
T}$_{\text{C}}$ decrease for small \textit{y} (0.0$\leq $\textit{y}$\leq $%
0.25) may originate from the decrease of \textit{t}/$\lambda $ by increasing
Mn-O bond distance and decreasing Mn-O-Mn bond angle, induced by the
substitution of the small Pr ions.

All of above results clearly indicate that the two-phase coexistence is
directly associated with (1) CMR for low \textit{T}$_{\text{C}}$ manganites,
(2) the first-order-like \textit{T}$_{\text{C}}$ for low \textit{T}$_{\text{C%
}}$ manganites, (3) the sharp drop of \textit{T}$_{\text{C}}$ with chemical
pressure, (4) the $\mathit{\rho }$ hump far below \textit{T}$_{\text{C}}$,
and (5) the small saturation moment at low \textit{T} for reduced \textit{T}$%
_{\text{C}}$ materials. Since the nature of the two-phase coexistence
changes drastically with \textit{T}, this coexistence is not the direct
consequence of chemical inhomogeneity, but results from electronic phase
separation. The possibility of electronic phase separation has been
discussed theoretically between antiferromagnetic insulating and FM metallic
states \cite{16}. However, our results clearly indicate the electronic phase
separation relevant to CMR at low temperature is the one between FM metallic
and the \textit{x}=1/2-type CO states.

In this section, one succinct picture was provided to explain the vast
amount of earlier experimental results and ambiguities, as well as our own
data. Basically, FM (metallic) and the \textit{x}=1/2-type CO states coexist
in a broad range of phase space through electronic phase separation in real
space, and this two-phase coexistence is responsible for the anomalous
behavior of low \textit{T}$_{\text{C}}$ manganites. The subtle balance
between these two states with distinctly different electronic properties can
be readily influenced by varying physical parameters such as applied field,
chemical (hydrostatic) pressure, and oxygen isotope, producing various
``colossal'' effects \cite{1,2,3,40,42,43}.

\subsection{Martensitic transformation and multiphase coexistence}

\subsubsection{Relaxation between charge order and ferromagnetism}

A simple percolation model between two-phases of FM metallic and CO
insulating states explains the experimental results above quite well in a
broad sense. However, there are emerging experimental results that cannot be
explained with a standard percolation model between the two electronic
phases. The most surprising point is the existence of different structural
phases as well as the distinct two electronic phases. The first can be
intimately associated with the lattice degree of freedom, while the latter
can be related to the spin and charge degree of freedom in the system. In
view of the strong coupling among lattice, spin, and charge degrees of
freedom in manganites, it is expected that these structural and electronic
phase separations can be intimately linked to each other. 
\begin{figure}[tb]
\epsfxsize=60mm
\centerline{\epsffile{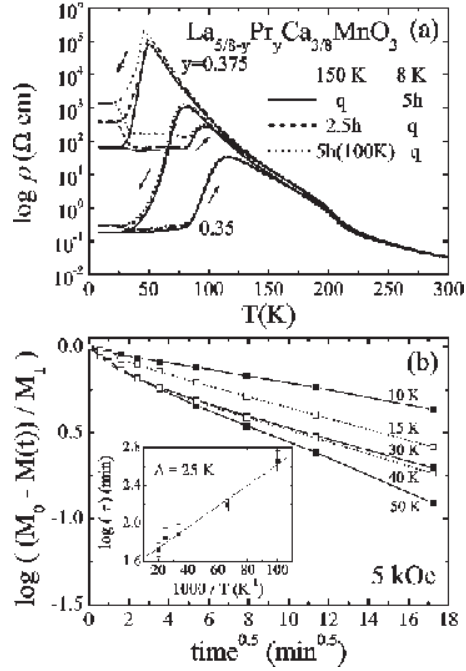}}
\caption{(a) Temperature-dependent resistivity $\protect\rho $ for La$_{%
\text{5/8-}y}$Pr$_{y}$Ca$_{\text{3/8}}$MnO$_{\text{3}}$, \textit{y}=0.35 and 
\textit{y}=0.375 with various cooling rates. $\protect\rho $ at low \textit{T%
} becomes smaller when the specimen was cooled slower (faster) through 
\textit{T}$_{\text{C}}$ (\textit{T}$_{\text{CO}}$). (b) Time dependence of
magnetization M(\textit{t}) at various, fixed temperatures after quick
cooling for 5 minutes under 5 kOe for La$_{\text{5/8-}y}$Pr$_{y}$Ca$_{\text{%
3/8}}$MnO$_{\text{3}}$, with \textit{y}=0.375. M(\textit{t}) follows the
stretched exponential form, M$_{0}$- M$_{1}\cdot $exp(-($\protect\tau $/%
\textit{t})$^{\protect\beta }$). In this panel, log of the exponential part,
log ((M$_{0}$-M$_{1}$)/M$_{1}$) $\propto $-($\protect\tau $/\textit{t})$^{%
\protect\beta }$, is plotted as a function of square root of time. The inset
shows the temperature dependence of relaxation time \textit{t}, obtained by
fitting data to the stretched exponential form. The fitting results are as
follows; M$_{0}$=0.71,1.01,1.80,1.61 and 1.15 emu/mol, M$_{1}$=0.015, 0.267,
0.432, 0.322 and 0.192 emu/mol, \textit{t}=451, 155, 76, 70 and 52 min, and $%
\protect\beta $=0.46, 0.44, 0.34, 0.35 and 0.41 at 10,15, 30, 40 and 50 K,
respectively.}
\label{Fig3-8}
\end{figure}
P. B. Littlewood first suggested an idea that strain induced phase
separation may exist in low temperature regions of (La,Pr,Ca)MnO$_{3}$ \cite%
{44}. He pointed out that the FM and the CO phases have a large strain
mismatch so that if part of a single FM crystallite nucleates into the CO
phase, that domain is under a huge stress from the surrounding crystal that
discourages further growth. Thus, the coexistence of CO and FM domains can
produce random strain fields populated in each domain. To tackle the issue
more systematically how structural (lattice) responses can be related to the
electronic phase separation, we carefully studied the physical properties of
La$_{1-x}$Ca$_{x}$MnO$_{3}$ (LCM) with \textit{x} near 1/2 and La$_{5/8-y}$Pr%
$_{y}$Ca$_{3/8}$MnO$_{3}$ (LPCM) with \textit{y} near 0.35, representing two
important regions of manganite phase space. Magnetization study (not shown)
clearly indicated that the ground state of LPCM is FM, and CO is a high-%
\textit{T} state \cite{45}. On the other hand, LCM shows the opposite way.
Resistivity ($\rho $) data in Figs. \ref{Fig3-8} (a) and \ref{Fig3-9} (a)
were consistent with the M(\textit{T}) results. In LPCM, CO transition,
which is characterized by a sudden $\mathit{\rho }$ upturn, appears at a
high \textit{T}, and the abrupt $\mathit{\rho }$ downturn, indicating FM
transition, occurs at a low-\textit{T} whereas the order of transitions is
reversed in LCM (Fig. \ref{Fig3-9} (a)). There exists a large $\mathit{\rho }
$ thermal hysteresis at the low \textit{T} transitions, indicating a
first-order nature of the transition.

\begin{figure}[tb]
\epsfxsize=60mm
\centerline{\epsffile{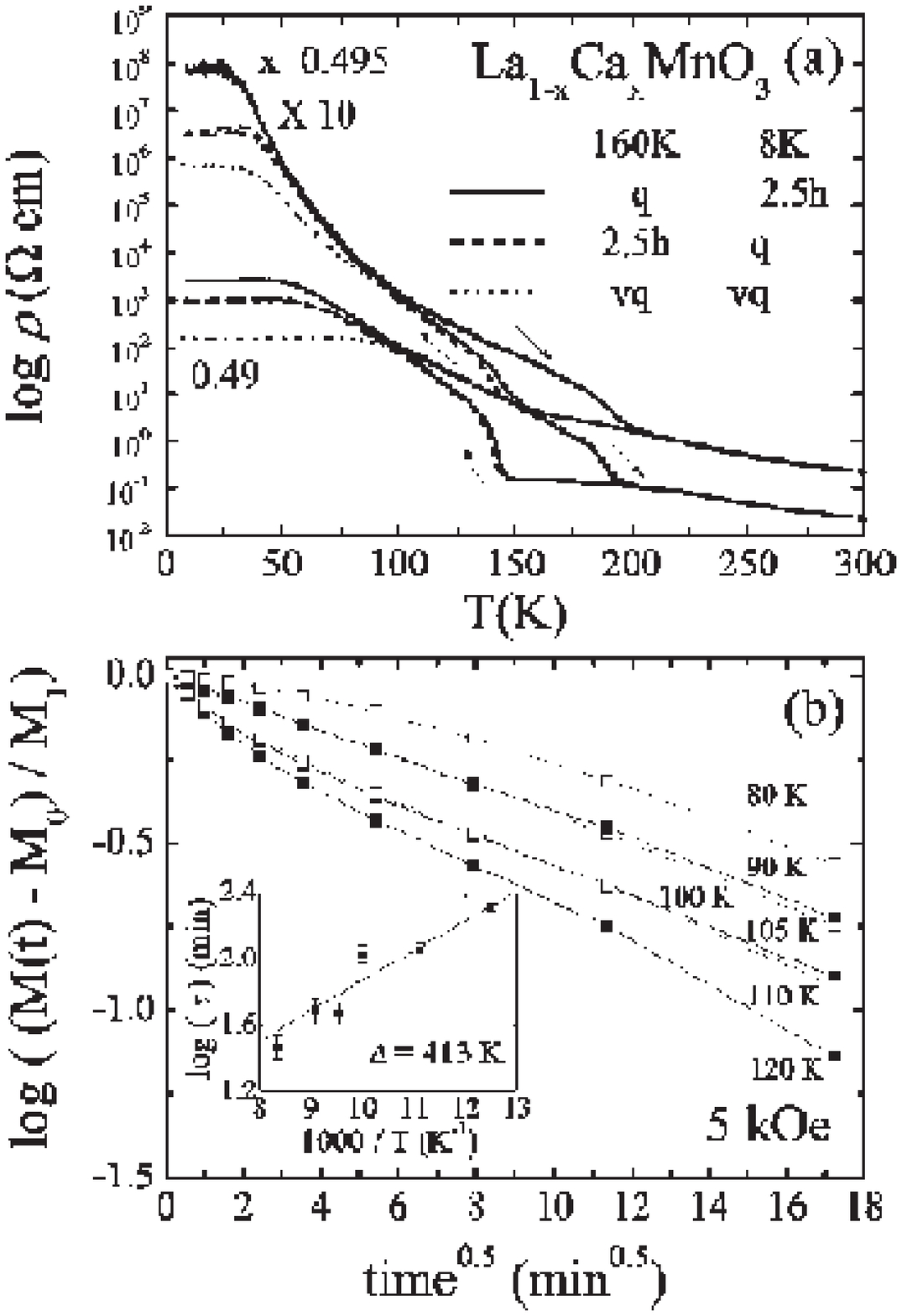}}
\caption{(a) Temperature-dependent resistivity $\protect\rho $ for La$_{1-x}$Ca$%
_{x}$MnO$_{3}$ with \textit{x}=0.49 and \textit{x}=0.495 with various
cooling rates. ``q'' and ``vq'' represents quick cooling within \symbol{126}%
20 and \symbol{126}1.5 minutes, respectively. (b) Time dependence of
magnetization M(\textit{t}) at various, fixed temperatures after quick
cooling for 5 minutes under 5 kOe for La$_{1-x}$Ca$_{x}$MnO$_{3}$ with 
\textit{x}=0.495. M(\textit{t}) follows the stretched exponential form, M$%
_{0}$- M$_{1}\cdot $exp(-($\protect\tau $/\textit{t})$^{\protect\beta }$).
In this figure, log of the exponential part, log ((M$_{0}$-M$_{1}$)/M$_{1}$) 
$\propto $-($\protect\tau $/\textit{t})$^{\protect\beta }$, is plotted as a
function of square root of time. The inset shows the temperature dependence
of relaxation time \textit{t}, determined by fitting data to the stretched
exponential form. The fitting results are as follows; M$_{0}$=0.364, 0.359,
0.359, 0.357, 0.358 and 0.378 emu/mol, M$_{1}$=0.014, 0.027, 0.031, 0.047,
0.049 and 0.058 emu/mol, \textit{t}=206, 112, 107, 48, 49 and 29 min and $%
\protect\beta $=0.71, 0.51, 0.52, 0.38, 0.41 and 0.39 at 80, 90, 105, 110
and 120 K, respectively.}
\label{Fig3-9}
\end{figure}

\begin{figure}[tb]
\epsfxsize=60mm
\centerline{\epsffile{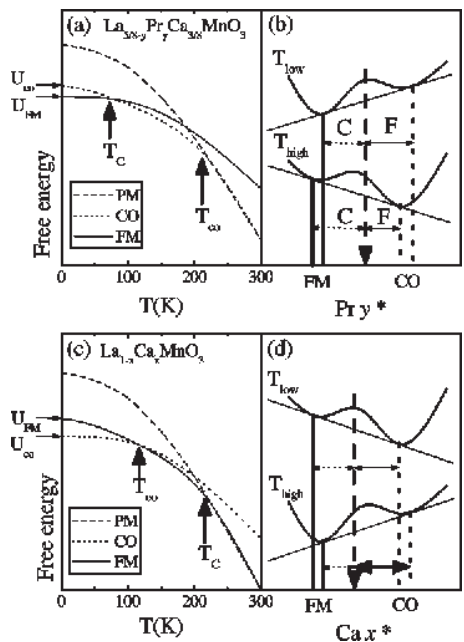}}
\caption{(a) Temperature- and (b) effective Pr doping \textit{y}*-dependence
of the free energy for La$_{\text{5/8-}y}$Pr$_{y}$Ca$_{\text{3/8}}$MnO$_{%
\text{3}}$ and (c) temperature and (d) effective Ca doping \textit{x}%
*-dependence of free energy for La$_{1-x}$Ca$_{x}$MnO$_{3}$.}
\label{Fig3-10}
\end{figure}

Thermodynamic view of these two systems in terms of free energy (F=U-\textit{%
T}S; U: internal energy, S: entropy) is depicted in Fig. \ref{Fig3-10} (a)
and (c). U of FM (U$_{\text{FM}}$) is lower than the CO U (U$_{\text{CO}}$)
in LPCM with \textit{x}=3/8, but U$_{\text{CO}}$ is lower than U$_{\text{FM}%
} $ in LCM with \textit{x} near 1/2. Now, the CE-type (\textit{n}$_{e}$=1-%
\textit{x}=1/2-type) CO state is commonly observed in the system with 
\textit{n}$_{e}$ near 5/8, and excess electrons may be accommodated as
charge defects or discomensurations in CO domains \cite{39}. Thus, there
ought to exist large configurational entropy associated with CO for \textit{n%
}$_{e}$ near 5/8. In LPCM with \textit{n}$_{e}$=5/8, CO state can be, thus,
``softer'' than FM state, i.e. the F line for CO decreases faster than that
for FM with increasing temperature. U of paramagnetic (PM) state is higher
than U$_{\text{FM}}$ and U$_{\text{CO}}$, but PM state is softer than FM and
CO states because of large entropy. In the case of LCM with \textit{x} near
1/2, the system is near the phase boundary between FM and AFM states, and
thus FM state can be susceptible to various thermally induced magnetic
excitations. On the other hand, little configurational entropy exists for
the CO state of LCM with \textit{n}$_{e}$ $\sim $1/2. Therefore, FM state is
naturally softer than CO state in LCM. The free energy vs. \textit{T} plots
constructed from these general observations are consistent with our various
results shown in Figs. \ref{Fig3-7}-\ref{Fig3-9}. For example, in LPCM, 
\textit{T}$_{\text{CO}}$ and \textit{T}$_{\text{C}}$ are the crossing points
of the F lines for PM-CO and CO-FM, respectively. In LCM, \textit{T}$_{\text{%
C}}$ and \textit{T}$_{\text{CO}}$ correspond to the crossing points of the F
lines between PM-FM and FM-CO, respectively

Even though the transitions are evident, FM state in LPCM (or CO state in
LCM) does not develop fully at low \textit{T}. In LPCM, enormously large $%
\mathit{\rho }$ and reduced M (full saturation M$_{s}$= 4 $\mu _{\text{B}}$)
at low \textit{T} indicate that CO phase remains partially at low \textit{T}%
. Similarly, non-diverging $\rho $(\textit{T}) and significant magnitude of
M in LCM at low \textit{T} indicate that FM phase does not disappear
completely at low \textit{T}. The electron-diffraction study on La$_{\text{5/8-}%
y} $Pr$_{y}$Ca$_{\text{3/8}}$MnO$_{\text{3}}$ in the above section
corroborates the two-phase coexistence with the length-scale associated $%
\sim $1/2 $\mu $m.

Intimately associated with the origin of this large-scale coexistence of FM
and CO phases, it was found that the two-phase situation can be influenced
by cooling rate across \textit{T}$_{\text{C}}$ and \textit{T}$_{\text{CO}}$.
As shown in Figs. \ref{Fig3-8} (a) and \ref{Fig3-9} (a), $\rho $(\textit{T}%
), especially at low \textit{T}, is sensitive on the cooling rates across
the transition temperatures. For example, solid line data in Fig. \ref%
{Fig3-8} (a) represent when the sample was cooled down to 150 K quickly for $%
\sim $20 minutes, and then cooled down to 8 K slowly for 5 hours. The
residual resistivity ($\rho _{\text{0}}$) changes more than a factor $\sim $%
100 in \textit{x}=0.495 and $\sim $10 in \textit{y}=0.375 by varying the
cooling rate. In both LPCM and LCM, $\rho _{\text{0}}$ becomes smaller when
the specimen was cooled slower through \textit{T}$_{\text{C}}$ or faster
through \textit{T}$_{\text{CO}}$. Correspondingly, saturation M at 5 K
becomes larger for slower (faster) cooling through \textit{T}$_{\text{C}}$ (%
\textit{T}$_{\text{CO}}$) (not shown). These findings indicate that the
relative volume of FM with respect to the CO volume increases when the
system was cooled slower (faster) though \textit{T}$_{\text{C}}$, and the
associated time scale is on the order of hours. This slow time-scale
associated with the conversion between FM and CO suggests the involvement of
the large-scale structure or lattice in the two-phase coexistence.

This slow dynamics is also reflected in the results of aging effect. The
evolution of M was measured with time after quick cooling (for $\sim $5
min.) from room \textit{T} to various, fixed temperatures. In LPCM (see Fig. %
\ref{Fig3-8} (b)), M increases with time, and obeys the stretched
exponential form M$_{0}$+M$_{1}$exp(-($\tau $/\textit{t})$^{\beta }$) with
the exponent $\beta $ near 0.5, indicating a wide distribution of relaxation
process \cite{46}. It is found that $\mathit{\tau }$ decreases from about $%
\sim $7.5 (at 10 K) to $\sim $1 (at 50 K) hour with increasing \textit{T},
and the \textit{T} dependence of $\mathit{\tau }$ is thermal-activation-type
with the energy gap ($\Delta $) of 25$\pm $2K (see the inset of Fig. \ref%
{Fig3-8} (b)). In the LCM case, M, in contrast to M in LPCM, decreases with
time, due to further development of CO phase. However, M even in LCM follows
the stretched exponential form, and $\mathit{\tau }$(\textit{T}) is
thermal-activation-type with $\Delta $ of 413$\pm $5 K. Interestingly, this
large difference of $\Delta $'s for LPCM and LCM is consistent with the fact
that the strength of magnetic field to melt CO phase for LCM is much lager
than that for LPCM ($\sim $15 T for LCM \cite{47}, $\sim $3 T for LPCM
(inset of Fig. \ref{Fig3-2})). 

Above experimental results lead us to construct a scenario for the two-phase
coexistence with slow dynamics as well as large-length scales. In La$_{0.5}$%
Ca$_{0.5}$MnO$_{3}$, the system is quasi-cubic (slightly orthorhombic) at
high temperatures, but the difference between \textit{a}$\approx $\textit{b}
and \textit{c}/$\sqrt{2}$ of orthorhombic structure becomes very large
(about 3 \%) below \textit{T}$_{\text{CO}}$ \cite{19}. This large structural
anisotropy originates from the cooperative Jahn-Teller distortions
associated with CO \cite{48}. During the nucleation of CO phase, this large
structural anisotropy associated with CO cannot be readily accommodated in
bulk materials, and results in large-scale anisotropic strain. Now, in the
bulk specimen of LPCM, the concentration of small Pr ions, which can
accommodate lattice distortions more easily than larger La ions, can fix or
``clamp'' the total anisotropic strain that the bulk system can sustain. In
other words, Pr concentration fixes the relative volume of CO phase, and the
non-CO regions may become FM, resulting in the two-phase coexistence. This
large-scale strain can, naturally, show slow dynamics or relaxation, for
example through slowly moving the two-phase interfaces \cite{49}. When the
specimen was cooled slowly through \textit{T}$_{\text{CO}}$ (\textit{T}$_{%
\text{C}}$), the strain is ``annealed'' so that CO (FM) develops better,
leading to larger (smaller) $\rho _{0}$ and smaller (larger) M at low 
\textit{T}. This view in terms of strain for LPCM is schematically
represented in Fig. \ref{Fig3-10} (b). Consistent with Fig. \ref{Fig3-10}
(a), there are local free energy minima for FM and CO states, F of FM (F$_{%
\text{FM}}$) is lower than F of CO (F$_{\text{CO}}$) at low \textit{T} (%
\TEXTsymbol{<}\textit{T}$_{\text{C}}$), and F$_{\text{CO}}$ is lower than F$%
_{\text{FM}}$ at high \textit{T} (\textit{T}$_{\text{C}}$\TEXTsymbol{<}%
\textit{T}\TEXTsymbol{<}\textit{T}$_{\text{CO}}$). The \textit{x}-axis
represents the degree of the total anisotropic strain that the bulk system
can accommodate, which is denoted as ``effective'' Pr concentration \textit{y%
}*. FM and CO phases correspond to small and large \textit{y}*,
respectively. When Pr concentration is fixed in a specimen, the averaged 
\textit{y}* is fixed for the bulk material. Therefore, the Maxwell
construction provides the lowest total free energy state in bulk materials,
leading to the coexistence of FM and CO \cite{50}. As shown in Fig. \ref%
{Fig3-10} (b), the amount of FM and CO phases is fixed by the length of ``C
(dotted line)'' and ``F (solid line)'', respectively. (Note that the broad
distribution of relaxation process, indicated from the stretched exponential
form of M(\textit{t}), can be ascribed to a wide distribution of the energy
barrier height between F$_{\text{CO}}$ and F$_{\text{FM}}$.) Similar free
energy argument holds for LCM with \textit{x}$\sim $1/2 as shown in Fig. \ref%
{Fig3-10} (d). CO and FM are the lowest free energy states at low \textit{T}
(\TEXTsymbol{<}\textit{T}$_{\text{CO}}$) and high \textit{T} (\textit{T}$_{%
\text{CO}}$\TEXTsymbol{<}\textit{T}\TEXTsymbol{<}\textit{T}$_{\text{C}}$),
respectively. The \textit{x}-axis is the ``effective'' Ca concentration 
\textit{x}*, similar with \textit{y}* for LPCM. Here again, the Maxwell
construction leads to the two-phase coexistence. This scenario for the
two-phase coexistence resembles the origin of the so-called tweed structure
formation in Martensitic systems \cite{51,52}. However, what is unique about
manganite is that the two phases with the different structures, induced by
strain, show distinct electronic properties (i.e., one FM metal and the
other CO insulator). This can be referred as \textit{structural} phase
separation, in comparison with \textit{chemical} or \textit{electronic}
phase separation.

In this section, a thermodynamic view combined with quasi-long-range strain
consideration was proposed to explain the intriguing structural phase
coexistence in manganites. This proposition naturally explains why the low-%
\textit{T} transition, independent from whether it is CO or FM transition,
is always strongly first-order. Furthermore, the large-length-scale
associated with the FM-CO-phase coexistence, and also the slow relaxation
between FM and CO phases can be attributed to the quasi-l-r, anisotropic
strain, resulting from the cooperative Jahn-Teller distortions associated
with CO.

\subsubsection{Accommodation of strain and the metal-insulator transition}

Martensitic transformations, i.e. cooperative (diffusionless) motion of
atoms resulting in a formation of different crystal structure within a
parent crystal, have been known for more than a century \cite{51,52,53}. In
metals and alloys, important physical and metallurgical properties are
determined by long-range strains associated with the structural distortion
of martensitic phases. In transition metal oxides, where strong
electron-electron and electron-lattice interactions govern such phenomena as
magnetic ordering, metal-insulator transition (MIT) and superconductivity,
phase transitions are often accompanied by structural deformation \cite%
{7,8,9}. However, the structural transformations are often considered as a
secondary or even a cumbersome effect. Here another experimental evidence is
summarized to support that the synergy of the martensitic accommodation
strain and strongly correlated electrons can lead to the extraordinary
electronic properties of manganites, and that martensitic \textit{%
accommodation strain}, produced by the CO phase, plays an important role in
the MIT \cite{54}.

The martensitic transformation is a structural phase transition of part of a
parent crystal involving large structural distortions. This phase,
structurally different from the parent structure, is called martensitic
phase or martensite. Physical properties of martensitic alloys are governed
by the long-range elastic or plastic deformation of the parent crystal
lattice surrounding the martensitic particles, so called accommodation
strain \cite{51,52,53}. Growth of the accommodation strain with lowering 
\textit{T} below the temperature of the martensitic transformation dominates
an establishment of the thermo-elastic equilibrium between the parent phase
and the martensite. This type of strain is inherent to martensitic
transformations as well as to the transformation twins and is produced at
the martensite/parent boundary or at the twin faults \cite{51}. In this
respect, both the internal twinning and martensitic transformations are
accompanied by the accommodation strain.

Applying the terminology of martensitic transformations to manganites, we
refer to the orthorhombic CO phase as a martensite and to the high-\textit{T}
cubic paramagnetic phase as a parent phase. Thus, for example, the CO
transition in (Nd,Sm)$_{1-x}$Sr$_{x}$MnO$_{3}$ (\textit{x}$\approx $1/2) %
\cite{55} and in La$_{0.5}$Ca$_{0.5}$MnO$_{3}$ \cite{56} could be described
by martensitic phenomenology. Since martensitic transformation, such as
charge ordering, takes place by cooperative motion of atoms, the growth of
martensite crystals across the grain boundary is prohibited and the
accommodation strain is expected to be very sensitive to the grain size of
polycrystalline samples. In order to investigate the influence of grain
boundaries on the accommodation strain and, therefore, on physical
properties of manganites, systematic transport studies were conducted in
single crystalline and polycrystalline samples with different average grain
size \TEXTsymbol{<}\textit{d}\TEXTsymbol{>} = 3, 6, 9, 12 and 17 $\mu $m.
All the polycrystalline samples were prepared from one high-quality La$%
_{0.275}$Pr$_{0.35}$Ca$_{0.375}$MnO$_{3}$ pellet, which was carefully
grounded to become fine powder and then was sintered additionally at 1380 $%
^{\circ }$C (and 1300 $^{\circ }$C) for different time periods $\Delta t$.

\begin{figure}[tb]
\epsfxsize=60mm
\centerline{\epsffile{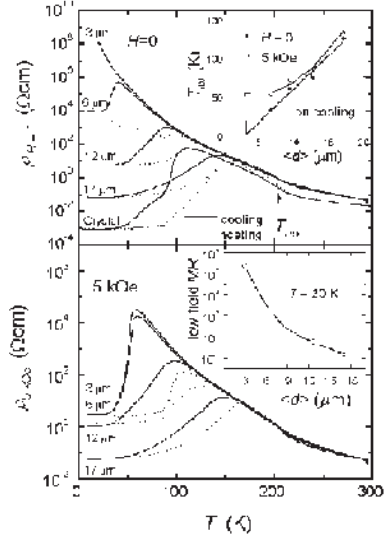}}
\caption{Temperature-dependent resistivity in zero magnetic field $H$ = 0
(upper panel) and in 5 kOe (lower panel) of the polycrystalline La$_{0.275}$%
Pr$_{0.35}$Ca$_{0.375}$MnO$_{3}$ samples with different grain size 
\TEXTsymbol{<}\textit{d}\TEXTsymbol{>} = 3 - 17 mm and a single crystal of
the same composition on cooling and heating. For simplicity, not all of the
measured samples are presented in the figure. The upper inset shows the
dependence of the insulator-metal transition temperature, \textit{T}$_{\text{%
MI}}$, on \TEXTsymbol{<}$d$\TEXTsymbol{>} for zero field cooling (closed
circles) and for field cooling in 5 kOe (open circles). The lower inset
shows the grain-size dependence of the magnetoresistance MR =($\protect\rho %
_{\text{\textit{H}=0}}$- $\protect\rho _{\text{5kOe}}$)/$\protect\rho _{%
\text{5kOe}}$ in 5 kOe at \textit{T} = 20 K.}
\label{Fig3-11}
\end{figure}

Remarkable dependence of the transport properties on grain size is revealed
by the resistivity ($\rho $) and magnetoresistance (MR$\equiv $($\rho _{H=0}$%
-$\rho _{\text{5kOe}}$)/\newline
$\rho _{\text{5kOe}}$) measurements of polycrystalline La$_{0.275}$Pr$%
_{0.35} $Ca$_{0.375}$MnO$_{3}$ (Fig. \ref{Fig3-11}). Data for a single
crystal of the same composition are shown for comparison. Although \textit{T}%
$_{\text{CO}}\approx $210 K is similar for all samples, \textit{T}$_{\text{MI%
}}$, defined as the temperature of the maximum of d(log$\rho $)/d\textit{T}
taken on cooling, systematically decreases from 125 K to 30 K when 
\TEXTsymbol{<}\textit{d}\TEXTsymbol{>} is reduced from 17 to 6 $\mu $m, (the
inset in Fig. \ref{Fig3-11}). Finally, the specimen with the smallest grain
size - 3 $\mu $m, does not exhibit the MIT down to 20 K, below which $%
\mathit{\rho }$ becomes too large to be measured reliably. Surprisingly, a
change of the grain size by a factor of two is sufficient to switch the low-%
\textit{T} ground state of the same compound from metallic to insulating.
The low-\textit{T} resistivity, $\mathit{\rho }_{0}\equiv \mathit{\rho }$(20
K), appears to be the most grain-size-sensitive characteristic of the
samples. For \TEXTsymbol{<}\textit{d}\TEXTsymbol{>} = 3 - 17 $\mu $m, $%
\mathit{\rho }_{0}$ varies systematically over the range 0.1-10$^{8}$ $%
\Omega $cm.

Sensitivity of $\rho _{0}$ and \textit{T}$_{\text{MI}}$ to \TEXTsymbol{<}%
\textit{d}\TEXTsymbol{>} becomes much less pronounced when a small magnetic
field \textit{H} = 5 kOe is applied (Fig. \ref{Fig3-11}, the lower panel).
First, this field induces the MIT in the sample with \TEXTsymbol{<}\textit{d}%
\TEXTsymbol{>} = 3 $\mu $m. In fact, $\rho $(\textit{T}) for all \TEXTsymbol{%
<}\textit{d}\TEXTsymbol{>} becomes much more similar in \textit{H} = 5 kOe,
indicating that the sharp increase of $\mathit{\rho }_{0}$ observed in 
\textit{H} = 0 cannot be attributed to the increasing contribution of
topological defects and scattering at the grain boundaries to $\mathit{\rho }
$ in the samples with small \TEXTsymbol{<}\textit{d}\TEXTsymbol{>}. Second,
low-field MR$_{\text{5kOe}}$ at \textit{T} = 20 K varies systematically with 
\TEXTsymbol{<}\textit{d}\TEXTsymbol{>} = 3 - 17 $\mu $m over the range 1 - 10%
$^{8}$ (see the inset). Extremely large low-field MR, up to 10$^{10}$ \%, in
samples with small \TEXTsymbol{<}\textit{d}\TEXTsymbol{>} cannot be
explained by the spin-polarized transport across magnetic domain boundaries,
which has been shown to result in the 20-30 \% change of $\mathit{\rho }$ in
polycrystalline films and epitaxial films grown on a bicrystal substrate %
\cite{57,58,59}. The drastic suppression of the grain-size dependence of $%
\mathit{\rho }$ by a small magnetic field, observed here, supports the
scenario of the MIT where the insulating phase responsible for the high $%
\rho _{0}$\ is not the charge-ordered phase \cite{60}. The magnetic field of
5 kOe is not strong enough to ``melt'' the CO phase \cite{6}.

Based on this transport measurement, a following scenario can be proposed
for the MIT in manganites. The accommodation strain, introduced by the CO
domains into the surrounding lattice at \textit{T}$_{\text{CO}}$ $\approx $%
210 K, strongly affects the properties of the latter. When the parent phase
is loaded with the strain, the FM transition becomes suppressed, e.g. 
\textit{T}$_{\text{MI}}$ is shifted to a lower temperature. Therefore, it
tends to retain properties of the high-\textit{T} paramagnetic phase,
remaining charge-disordered and insulating (CDI) even at low \textit{T}. It
is known that with a decreasing grain size of a sample, it is more difficult
to accommodate the martensitic strain \cite{51,52}. As a result, in the
samples with a smaller grain size the amount of the strain-loaded phase,
e.g. CDI, increases. The presence of a considerable amount of this
insulating phase results in the unusually high $\mathit{\rho }_{0}$ and
leads to the low temperature shift of the MIT with decreasing \TEXTsymbol{<}%
\textit{d}\TEXTsymbol{>}. An experimental evidence for the existence of CDI
phase at low temperatures will be provided in a next section, where optical
conductivity probes at least three phases in a single crystal sample.
Furthermore, recent \textit{x}-ray and neutron scattering experiments \cite%
{60} revealed that the volume fraction of the CO phase remains constant when
the system is driven through the MIT, e.g. the CO phase is not involved in
the MIT directly. Thus, the transition occurs within the parent phase, which
is separated into the FM metallic and strain-stabilized CDI phases at 
\textit{T}$\leq $\textit{T}$_{\text{MI}}$. It's worth mentioning, that the
ability of strain to stabilize phases, which do not exist at all without the
strain, is well known in martensites \cite{53}.

The results of this section show that sensitivity of the martensitic phase
to the grain boundaries leads to the observed striking dependence of the
transport properties of polycrystalline La$_{0.275}$Pr$_{0.35}$Ca$_{0.375}$%
MnO$_{3}$ samples on grain size. In contrast to the conventional
insulator-to-metal transition, wherein the FM phase grows at the expense of
the CO phase, e.g., La$_{0.5}$Ca$_{0.5}$MnO$_{3}$ under high magnetic field,
the insulator-metal transition in La$_{0.275}$Pr$_{0.35}$Ca$_{0.375}$MnO$%
_{3} $ is suggested not to be a transition from the charge ordered into the
ferromagnetic state. Instead, it is a transformation of the
charge-disordered insulating phase into the ferromagnetic one. The former
phase, charge-disordered insulating, is a result of the stabilization of the
parent paramagnetic phase at low temperature by the martensitic strain.
Typical signatures of martensitic transformations manifested by other
manganites indicate the general applicability of the martensitic approach
and phenomenology to the structural phase transitions in oxides with
strongly correlated electrons.

\subsection{Optical evidences of multiphase coexistence}

\subsubsection{Mid-infrared absorption peaks in La$_{\text{5/8-}y}$Pr$_{y}$Ca%
$_{\text{3/8}}$MnO$_{\text{3}}$}

To provide further insights on the nature of the phase coexistence,
temperature (\textit{T})- and magnetic field ($H$)-dependent optical
conductivity $\sigma $($\omega $) of a La$_{\text{5/8-}y}$Pr$_{y}$Ca$_{\text{%
3/8}}$MnO$_{\text{3}}$ (\textit{y}$\approx $0.35) (LPCMO) single crystal was
investigated \cite{61}. \textit{T}-dependent $\mathit{\rho }$ data for the
LPCMO, which is very similar to the single crystal data in Fig. \ref{Fig3-11}%
, indicated that the sample undergoes a charge-ordering transition at 
\textit{T}$_{\text{CO}}\approx $220 K and then a relatively sharp
insulator-metal transition around \textit{T}$_{\text{C}}\approx $120 K. The
reflectivity spectra $R$($\omega $) was measured with increasing \textit{T} at \textit{%
H}=0 T and with increasing \textit{H} at 4.2 K and the Kramers-Kronig transformation
produced $\sigma $($\omega $) from the measured $R$($\omega $). Detailed
techniques for the $R$($\omega $) measurements were described in our
previous report \cite{62}. 
\begin{figure}[tb]
\epsfxsize=60mm
\centerline{\epsffile{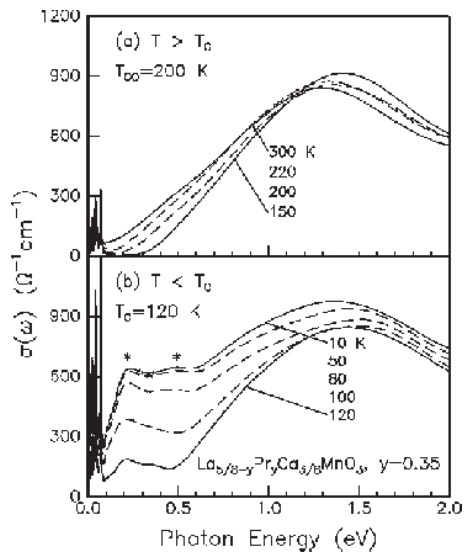}}
\caption{Temperature-dependent $\protect\sigma $($\protect\omega $) of a La$%
_{\text{5/8-}y}$Pr$_{y}$Ca$_{\text{3/8}}$MnO$_{\text{3}}$ (\textit{y}=0.35)
crystal (a) above and (b) below \textit{T}$_{\text{C}}$. At \textit{T}$_{%
\text{C}}$\TEXTsymbol{<}\textit{T}, the optical gap energy due to the CO
phase is determined by drawing a linearly extrapolated line (dotted) at the
inflection point of $\protect\sigma $($\protect\omega $). At \textit{T} $%
\leq $\textit{T}$_{\text{C}}$, at least two absorption bands appear in the
mid-infrared region. The peak positions of the bands are indicated as
asterisks.}
\label{Fig3-12}
\end{figure}

\begin{figure}[tb]
\epsfxsize=60mm
\centerline{\epsffile{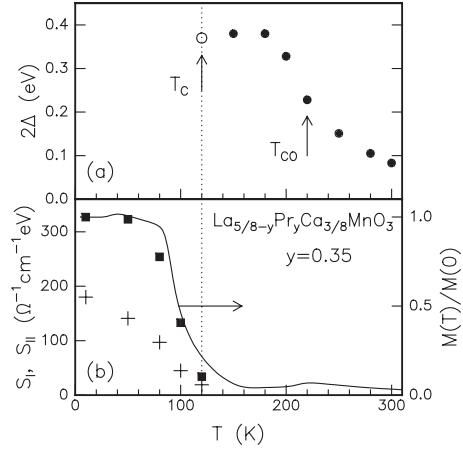}}
\caption{(a) Temperature dependence of the optical gap energy 2$\Delta $
(solid circles). An open circle represents a 2$\Delta $ at \textit{T}$_{%
\text{C}}$. (b) Temperature-dependence of spectral weight \textit{S}$_{\text{%
I}}$ (solid squares) and \textit{S}$_{\text{II}}$ (crosses) at \textit{T}%
\TEXTsymbol{<}\textit{T}$_{\text{C}}$. See, Fig. \ref{Fig3-14} and texts for
definition. A solid line represents a normalized magnetization curve.}
\label{Fig3-13}
\end{figure}

Fig. \ref{Fig3-12} (a) shows \textit{T}-dependent $\sigma $($\omega $) above 
\textit{T}$_{\text{C}}$. As \textit{T} decreases from 300 K, the low
frequency $\sigma $($\omega $) below 0.5 eV are systematically suppressed
and an optical gap is clearly developed. Therefore, it is likely that the
charge ordering (CO) phase is dominant at \textit{T}$_{\text{C}}$\TEXTsymbol{%
<}\textit{T}\TEXTsymbol{<}\textit{T}$_{\text{CO}}$ and that the broad band
around 1.4 eV can be attributed to the characteristic optical response of
the CO domains. The optical gap energy 2$\Delta $ at \textit{T}$\geq $%
\textit{T}$_{\text{CO}}$, obtained from a crossing energy between a linear
tangential line at the inflection point of $\sigma $($\omega $) and the 
\textit{x}-abscissa, is shown as a dotted line in Fig. \ref{Fig3-12} (a).
Figure \ref{Fig3-13} (a) shows the 2$\Delta $ vs. \textit{T} plot. The 2$%
\Delta $ just above \textit{T}$_{\text{C}}$ is found to be as large as 0.38
eV. This value remains nearly the same between \textit{T}$_{\text{C}}\leq $%
\textit{T}$\leq $180 K and slightly decreases near \textit{T}$_{\text{CO}}$.
It should be noted that the 2$\Delta \approx $0.38 eV at \textit{T}$\approx $%
150 K is comparable to that of La$_{0.5}$Ca$_{0.5}$MnO$_{3}$ (i.e., 2$\Delta 
$(0)$\approx $0.45 eV at the ground state). Since the CO phase in LPCMO is the La$%
_{0.5}$Ca$_{0.5}$MnO$_{3}$ -type, i.e., the so-called CE-type, the large
value of 2$\Delta $ of LPCMO can be ascribed to a characteristic of the
CE-type CO phase. The slightly smaller gap value of LPCMO might be related
to the presence of small FM phase or carrier defects due to \textit{x}=3/8
doping in the CE-type CO pattern. It is also noted in Fig. \ref{Fig3-13} (a)
that 2$\Delta \approx $0.22 eV at \textit{T}$_{\text{CO}}$ and 2$\Delta
\approx $0.1 eV at 300 K. Thus, 2$\Delta $ does not become zero at \textit{T}
far above \textit{T}$_{\text{CO}}$. This is an anomalous behaviour because
many CO materials show nearly zero value of 2$\Delta $ at \textit{T}$_{\text{%
CO}}$. This nonzero 2$\Delta $ behaviour indicates that there exists
enhanced spatial and/or temporal fluctuation of CO correlation far above 
\textit{T}$_{\text{CO}}$ in LPCMO. This enhanced CO fluctuation is regarded
as a generic feature of LPCMO that has mixed-phases near the phase boundary
where a phase separation occurs even when \textit{T}$\rightarrow $0 K \cite%
{63}. This subject will be discussed in Section \ref{sec4} in detail. 

Figure \ref{Fig3-12} (b) shows that as \textit{T} decreases below \textit{T}$%
_{\text{C}}$, new absorption bands appear below $\sim $0.4 eV, and their
strengths grow. As indicated by asterisks, $\sigma $($\omega $) at 10 K show
at least two mid-infared absorption bands, centered around 0.22 and 0.49 eV,
respectively. As \textit{T} is lowered below \textit{T}$_{\text{C}}$, the
former is located nearly at the same frequency, while the latter one shifts
to a higher frequency from 0.35 eV at 120 K to 0.49 eV at 10 K, indicating
that the origin of the lower frequency peak might be different with that of
the higher frequency one. Note that the strength of a broad absorption band
around 1.4 eV does not decrease even below \textit{T}$_{\text{C}}$. This is
in contrast with the $\sigma $($\omega $) behaviours of homogeneous FM
metallic samples that show a significant spectral weight transfer from above
1.0 to below 1.0 eV \cite{64,65}. This observation suggests that the volume
fraction of the CO phase does not change significantly below \textit{T}$_{%
\text{C}}$. 
\begin{figure}[tb]
\epsfxsize=60mm
\centerline{\epsffile{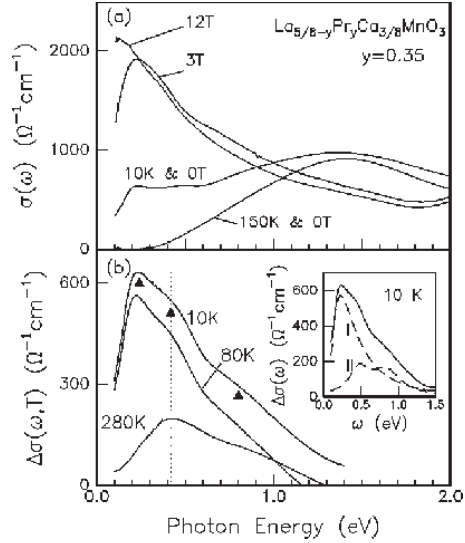}}
\caption{(a) Temperature- and magnetic field-dependent $\protect\sigma $($%
\protect\omega $) of a La$_{\text{5/8-}y}$Pr$_{y}$Ca$_{\text{3/8}}$MnO$_{%
\text{3}}$ (\textit{y}=0.35) crystal. (b) $\Delta \protect\sigma $($\protect%
\omega ,T$) =$\protect\sigma $($\protect\omega ,T$)-$\Delta \protect\sigma $(%
$\protect\omega ,$150 K) for various temperatures. The $\Delta \protect%
\sigma $($\protect\omega ,T$) curves at \textit{T}\TEXTsymbol{<}\textit{T}$_{%
\text{C}}$ are composed of an asymmetric absorption band and small
additional bands (filled triangles). The spectral shape of the additional
bands are very similar to the shape of $\Delta \protect\sigma $($\protect%
\omega ,T$) at \textit{T}\TEXTsymbol{>}\textit{\textit{T}}$_{\text{CO}}$.
The inset shows a $\Delta \protect\sigma $($\protect\omega $,10 K) curve and
its fitting results using an asymmetric line shape (for Band I) and two
Lorentzians (for Band II).}
\label{Fig3-14}
\end{figure}

Figure \ref{Fig3-14} (a) shows \textit{H}-dependent $\sigma $($\omega $) at
fixed \textit{T}=4.2 K. It is found that, with increasing $H$, the spectral
weight above 1.0 eV becomes strongly suppressed and transferred to a lower
frequency region. Although the sample shows a metallic resistivity at 3 T,
the corresponding optical spectrum has a very asymmetric mid-infrared band.
At \textit{H}=12 T, a clear Drude-like peak is observed with a saturated
suppression of the absorption peak around 1.4 eV. This suggests that the CO
insulating domains are melted to become the FM metallic ones and that the
volume fraction of the FM metallic domains is increasing. The optical
spectrum at 12 T is characterized as a single asymmetric absorption band
below 1.5 eV, which is very similar to the shape of $\sigma $($\omega $) in
some homogeneous FM metallic manganites at low temperature \cite{64}. In
addition, under the high $H$, the feature of two absorption bands in Fig. %
\ref{Fig3-12} (b) is not observed. Therefore, the $\sigma $($\omega $) at 12
T seems to represent the optical response of the dominant FM metallic phase.

To elucidate the origin of the two peak structure, the optical response of
the CO phase needs to be subtracted from the measured $\sigma $($\omega $)
at each temperature, $\sigma $($\omega $,$T$). It was assumed that $\sigma $(%
$\omega $,150 K) could represent the $\sigma $($\omega $) of CO domains.
Figure \ref{Fig3-14} (b) shows the results of $\Delta \sigma $($\omega ,T$)$%
\equiv \sigma $($\omega $,\textit{T})-$\sigma $($\omega $,150 K) at various
temperatures. The $\Delta \sigma $($\omega $,10 K) curve is composed of an
asymmetric absorption band peaked around 0.2 eV and a broad band with peaks
around 0.4 and 0.8 eV. While the absorption band around 0.2 eV appears below 
\textit{T}$_{\text{C}}$, the broad band with peaks around 0.4 and 0.8 eV
already exists above \textit{T}$_{\text{CO}}$.

To quantitatively estimate the \textit{T}-dependent spectral weight of each
absorption band, we fitted the $\Delta \sigma $($\omega ,T$) below \textit{T}%
$_{\text{C}}$ as a sum of an asymmetric band (Band I) around 0.2 eV and two
Lorentzians (Band II) around 0.4 and 0.8 eV. The inset of Fig. \ref{Fig3-14}
(b) shows the $\Delta \sigma $($\omega $,10 K) curve and its fitting
results. It is found that Band I is very similar to $\sigma $($\omega $) at 
\textit{H}=3 T in Fig. \ref{Fig3-14} (a). Especially, Band II is very
similar to $\Delta \sigma $($\omega $,280 K) in Fig. \ref{Fig3-14} (b).
These analyses strongly suggest that Band I at low \textit{T} should be due
to the FM metallic phase, while Band II can be attributed to another phase,
of which physical properties are very similar to those of the high \textit{T}
charge-disordered insulating (CDI) phase.

Based on these analyses, we estimated integrated spectral weights of Band I, 
$\mathit{S}_{\text{I}}$, and of Band II, $\mathit{S}_{\text{II}}$, below 
\textit{T}$_{\text{C}}$. It is interesting to compare \textit{T}-dependence
of $\mathit{S}_{\text{I}}$ and $\mathit{S}_{\text{II}}$ with a normalized
magnetization value, M(\textit{T})/M(0). Figure \ref{Fig3-13} (a) shows 
\textit{T}-dependences of $\mathit{S}_{\text{I}}$ (solid squares), $\mathit{S%
}_{\text{II}}$ (solid triangles), and M(\textit{T})/M(0). It is found that $%
\mathit{S}_{\text{I}}$ is roughly proportional to M(\textit{T})/M(0),
indicating that Band I is linked to the FM spin ordering. However, rather
gradual increase of M(\textit{T})/M(0) is not consistent with the FM
transition in a homogeneous system. Thus, $\mathit{S}_{\text{I}}$ can be
attributed to the spectral weight of FM metallic domains in an inhomogeneous
system. On the other hand, $\mathit{S}_{\text{II}}$ starts to increase near 
\textit{T}$_{\text{C}}$, and continuously does below \textit{T}$_{\text{C}}$
even when M(\textit{T})/M(0) is saturated. This observation again supports
that Band II is the absorption band of a CDI phase, of which volume fraction
continuously increases with the development of the FM phase below \textit{T}$%
_{\text{C}}$. How the CDI phase can be developed below \textit{T}$_{\text{C}%
} $ will be discussed further below. In partial summary, all of above
experimental findings unequivocally suggest that there exist at least three
phases, i.e. the FM metallic, the CO insulating, and the CDI phases in LPCMO.

\subsubsection{Implication of structural phase separation in optical
conductivity spectra}

In LPCMO with the charge ordering domains inside, anisotropic lattice
strains can be developed due to the Jahn-Teller (JT) distortion and the
concomitant orbital ordering. In particular, the anisotropic strain in the
CE-type charge ordering is known to be quite large, due to a cooperative JT
distortion and a $d_{z}^{2}$ orbital ordering. In a nearly homogeneous FM
metallic state, the JT distortion becomes small. If FM metallic domains
appear below \textit{T}$_{\text{C}}$ in the backbone of CO domains, the
strain of this region will be released. Due to a large strain mismatch
between the two phases, the interfacial region will have inhomogeneous
strains larger than that of the FM domains, but smaller than that of the CO
domains. This physical situation is quite similar to the case of
ferroelastic materials with the martensitic transformation. This postulation
suggests an appealing idea that at low \textit{T} region of LPCMO, each
domain has different structural distortions so that electronic phase
separation can be accompanied by structural phase separation.

As a result of a strong electron-phonon coupling, presence of a polaronic
absorption band is another important spectral feature in the $\sigma $($%
\omega $) of manganites. However, in case of the manganites, the polaron
absorption band is also strongly coupled to spin and orbital degrees of
freedom \cite{66}; a single polaron theory should be clearly extended
further to include multi-polaron nature and many body effects in this
compound. To our knowledge, theoretical understandings on the multi-polaron
effects are still lacking. Thus, it should be noted that the discussion
below is yet qualitative and based on a single polaron picture. With this in
mind, the structural multiphase coexistence can still be well described by
the existing single polaron theory. According to the polaron absorption
theory \cite{67,68}, the peak energy of the incoherent band is roughly
proportional to the binding energy of a polaron that increases as the local
lattice distortion increases. In the FM metallic manganites well below 
\textit{T}$_{\text{C}}$, the mid-infrared polaron band becomes quite
asymmetric and it is centered below $\sim $0.3 eV, which is close to the
line shape of incoherent absorption of a large polaron \cite{64,67}. Above 
\textit{T}$_{\text{C}}$, as local JT lattice distortions increase, the band
becomes rather symmetric and the peak frequency shifts to a higher
frequency. This polaron band can be close to the small polaron absorption
band.

The asymmetric line shape of Band I indicates that the lattice distortion of
the FM domains is not large. On the other hand, the $\sigma $($\omega $) of
the CO domains with large strains showed a band centered $\sim $1.4 eV with
a large 2$\Delta \approx $0.4 eV. Most of the spectral weights of Band II
appear in an energy region of 0.3 -1.0 eV, indicating that the lattice
strain of the CDI region can be larger than that of the FM metallic domains,
but smaller than that of the CO domains. According to Fig. \ref{Fig3-12} (b)
and Fig. \ref{Fig3-13} (b), both a lower peak frequency of Band II and $%
\mathit{S}_{\text{II}}$ increased as \textit{T} decreased below \textit{T}$_{%
\text{C}}$. This seems to indicate that the volume fraction as well as the
strain of the CDI domains, possibly located at the interface of FM and CO
domains, increases with deceasing \textit{T}. Therefore, the $\sigma $($%
\omega $) data suggest that the lattice strains and their interplay with 
\textit{T} inside the three main phases play a crucial role on the
electronic and the magnetic properties of the LPCMO.

In this section, \textit{T}-and $H$-dependent optical conductivity spectra
of a La$_{\text{5/8-}y}$Pr$_{y}$Ca$_{\text{3/8}}$MnO$_{\text{3}}$ (\textit{y}%
$\approx $0.35) single crystal revealed that at least two absorption bands
newly appeared below 0.4 eV at \textit{T}\TEXTsymbol{<}\textit{T}$_{\text{C}%
}\approx $120 K. The new absorption bands can be attributed to a FM metallic
and a charge-disordered phase, coexisting with a charge-ordered phase. Quite
different peak frequency of each absorption band suggests that the
coexisting multi-phases can have different lattice strains. In addition,
LPCMO had a rather large charge gap due to fluctuating charge-ordering
correlation above \textit{T}$_{\text{C}}$. This $\sigma $($\omega $) study
supports that the structural as well as the electronic phase separation
occurs in the LPCMO below \textit{T}$_{\text{C}}$.

\section{High temperature charge-ordering fluctuation and nano-scale phase
coexistence}

\label{sec4} In mixed-valent manganites, orbital degree of freedom
associated with Mn$^{3+}$ ions, in addition to charge and spin degrees of
freedom, plays an important role. The static charge/orbital ordering with
stripe patterns is now well established, especially in La$_{1-x}$Ca$_{x}$MnO$%
_{3}$ with \textit{x}$\geq $0.5 at low \textit{T} region. CO
in manganites occurs as periodic arrays of the sheet-like arrangement of Mn$%
^{3+}$ ions \cite{39}. In this scheme, the CO state of La$_{0.5}$Ca$_{0.5}$%
MnO$_{3}$ is special in the sense that the density of Mn$^{3+}$-Mn$^{4+}$
pairs is the highest. In La$_{0.5}$Ca$_{0.5}$MnO$_{3}$, all of the charge,
orbital and spin degrees of freedom freeze into the so-called CE-type stable
configuration below 180 K (for heating) \cite{11,69}.

One of the best examples of electronic phase separation, known from earlier
research stage of manganites, was the nano-scale coexistence of an
insulating phase with striped charge-order (CO) and a metallic phase with
ferromagnetism (FM) in a narrow temperature ($\sim $180\TEXTsymbol{<}\textit{%
T}\TEXTsymbol{<}220K) range of La$_{0.5}$Ca$_{0.5}$MnO$_{3}$ \cite{20,39}.
An early study of synchrotron \textit{x}-ray scattering for La$_{0.5}$Ca$%
_{0.5}$MnO$_{3}$ showed a drastic broadening of all the Bragg peaks in the
FM region between \textit{T}$_{\text{C}}$ and \textit{T}$_{\text{N}}$ ($%
\approx $\textit{T}$_{\text{CO}}$) \cite{19}. After this discovery, an
electron diffraction study showed that the fine scale ($\sim $100 \AA )
coexistence of CO and FM phases is responsible for the drastic broadening of 
\textit{x}-ray Bragg peaks \cite{20}. This coexistence is not due to
chemical inhomogeneity because it disappears below $\sim $150 K. This
nano-scale two-phase coexistence at \textit{x}=0.5 at finite temperatures
was thought to originate from the fact that \textit{x}=0.5 is the phase
boundary composition between FM (\textit{x}\TEXTsymbol{<}0.5) and CO (%
\textit{x}\TEXTsymbol{>}0.5) states. It was also presumed that the ground
state of La$_{1-x}$Ca$_{x}$MnO$_{3}$ with \textit{x}=0.5 becomes the CO
state below \textit{T}$_{\text{CO}}\approx $180 K.

This section will cover electronic/thermal transport, \textit{x}-ray
diffraction, and optical studies to investigate short-ranged, and/or dynamic
CO correlation in the paramagnetic state of the half-doped manganite. The
enhancement of $\mathit{\rho }$ up to 900 K in La$_{0.5}$Ca$_{0.5}$MnO$_{3}$
indicates that the spatial or dynamic CO fluctuation probably persists even
at \textit{T}-ranges much higher than long-range ordering temperature \cite%
{70}. Furthermore, accumulated experimental and theoretical investigations
now suggest that this robust charge fluctuation up to very high temperatures
be intimately associated with the fact that the ground state of La$_{0.5}$Ca$%
_{0.5}$MnO$_{3}$ has, in fact, intrinsically two competing order parameters,
i.e., FM metallic and CO insulating phases. This new view on the ground
state of La$_{0.5}$Ca$_{0.5}$MnO$_{3}$ is quite consistent with the idea of
electronic phase separation at zero temperature.

\subsection{Transport and structural studies}

\label{subsec4-1} 
\begin{figure}[tb]
\epsfxsize=60mm
\centerline{\epsffile{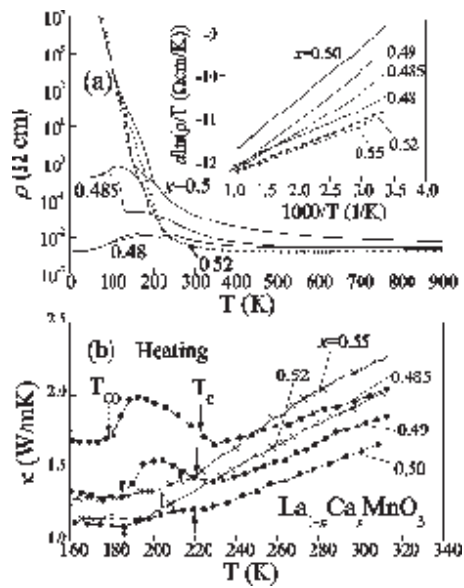}}
\caption{(a) Temperature-dependent resistivity $\protect\rho $ curves (for
heating and cooling) of La$_{1-x}$Ca$_{x}$MnO$_{3}$ near \textit{x}=0.5 from
4 to 900 K. The inset shows ln($\protect\rho $/\textit{T}) vs. 1/\textit{T}
curves. The activation energy, \textit{E}$_{a}$/\textit{k}$_{\text{B}}$, and
prefactor $\protect\rho _{0}$, of adiabatic small polarons vs. \textit{x}
plot, obtained from the curve. (b) $\protect\kappa $(\textit{T}) (for
heating) for La$_{1-x}$Ca$_{x}$MnO$_{3}$, near \textit{x}=0.5. The solid and
dotted arrows show \textit{T}$_{\text{C}}$ and \textit{T}$_{\text{CO}}$
determined from M(\textit{T}) and $\protect\rho $(\textit{T}).}
\label{Fig4-1}
\end{figure}

\begin{figure}[tb]
\epsfxsize=60mm
\centerline{\epsffile{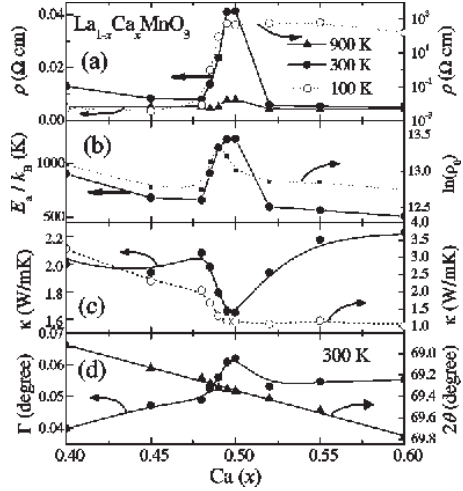}}
\caption{(a) and (c) $\protect\rho $ and $\protect\kappa $ values at 100 and
300 K for La$_{1-x}$Ca$_{x}$MnO$_{3}$ near \textit{x}=0.5, respectively. (b)
The activation energy \textit{E}$_{a}$/\textit{k}$_{\text{B}}$ and prefactor 
$\protect\rho _{0}$ vs. \textit{x} plot of adiabatic small polarons,
obtained from the inset of Fig. \ref{Fig4-1}. (d) The \textit{x}-dependence
of the peak width, $\Gamma $, and the center position of the (242) \textit{x}%
-ray Bragg peak at 300 K, estimated from Gaussian fitting of the data in
Fig. \ref{Fig4-3}. The dotted and solid lines are guides for the eyes.}
\label{Fig4-2}
\end{figure}

Figure \ref{Fig4-1} shows the $\rho $(\textit{T}) curves
for La$_{1-x}$Ca$_{x}$MnO$_{3}$ with \textit{x} near 0.5 from 4 to 900 K.
For \textit{x}=0.48, the FM metallic phase is dominant below 220 K even
though a short-range CO phase probably coexists as indicated by a broad hump
and small hysteresis at 100-180 K \cite{71,72}. However, with \textit{x}
approaching 0.5 from below, the CO state stabilizes at low temperature.
Thus, low temperature $\mathit{\rho }$ near \textit{x}=0.5 increases
systematically with \textit{x}, and shows the insulating \textit{T}%
-dependence when \textit{x} $\approx $0.485. As shown in Fig. \ref{Fig4-2}
(a), $\rho $(100 K) (open circles) shows such a systematic increase with 
\textit{x}, consistent with a crossover from the FM metallic to the CO
insulating states. If CO is stabilized at low temperature region for \textit{%
x} $\approx $0.5, $\rho $(100 K) becomes insensitive on \textit{x}. 

It is interesting that in Fig. \ref{Fig4-2} (a), $\rho $ of La$_{0.5}$Ca$%
_{0.5}$MnO$_{3}$ at \textit{T}\TEXTsymbol{>}\textit{T}$_{\text{CO}}$ is
considerably larger than that of any neighboring compositions, and this
behavior persists up to 900 K. This unexpected behavior was confirmed
systematically in samples with fine spacing of \textit{x} near 0.5. The $%
\mathit{\rho }$(300 K) vs. \textit{x} plot in Fig. \ref{Fig4-2} (a)
summarizes the results, showing a clear maximum at \textit{x}=0.5. Moreover,
the $\mathit{\rho }$(900 K) vs. \textit{x} plot confirms that the maximum
behavior at \textit{x}=0.5 persists up to, at least, 900 K. A previous study
revealed that an adiabatic small polaron model, with $\mathit{\rho }$\textit{%
=}$\mathit{\rho }_{0}$\textit{T}exp(\textit{E}$_{a}$/\textit{k}$_{\text{B}}$%
\textit{T}), describes high temperature $\mathit{\rho }$ of La$_{1-x}$Ca$%
_{x} $MnO$_{3}$ in broad doping and temperature ranges (0$\leq $\textit{x}$%
\leq $1 and $\sim $300K $\leq $\textit{T} $\leq $1100K) \cite{73}. Here, 
\textit{E}$_{a}$ represents the activation energy of small polarons, i.e.
the potential barrier that polarons must overcome to hop to the next site.
The inset of Fig. \ref{Fig4-1} shows that ln($\rho $\textit{/T}) vs. 1/%
\textit{T} plot of our data at high \textit{T} region is almost linear,
corroborating with the adiabatic small polaron model. Interestingly, \textit{%
E}$_{a}$ is systematically enhanced at \textit{x}=0.5 even if $\mathit{\rho }%
_{0}$ becomes maximum at \textit{x}=0.49, slightly lower than 0.5, as shown
in Fig. \ref{Fig4-2} (b). Therefore, the strong charge localization tendency
at \textit{x}=0.5 up to very high \textit{T}, far above \textit{T}$_{\text{CO%
}}$, is closely associated with the enhancement of polaron activation
energy. 

To gain further insights into understanding this surprising result, we
measured \textit{T}-dependent $\kappa $ (Fig. \ref{Fig4-1}(b): heating). As
pointed out in Section \ref{sec2}, the electronic $\kappa $ estimated from $%
\mathit{\rho }$ by using the Wiedemann-Franz law is negligible, and the
anomalous behavior of $\kappa $, i.e., a linear increase above \textit{T}$_{%
\text{C}}$, is related to the phononic $\kappa $ coupled with large
anharmonic lattice distortions \cite{21}. Thus, it is evident that at high 
\textit{T} region of La$_{1-x}$Ca$_{x}$MnO$_{3}$ (\textit{x}$\approx $0.5),
phononic contribution dominates the measured $\kappa $. The abrupt $\kappa $
increase near \textit{T}$_{\text{C}}$ in $\kappa $(\textit{T}) of \textit{x}%
=0.485, 0.49, and 0.50 is due to the reduced lattice distortions in the
FM-metallic state, and this $\kappa $ increase at \textit{T}$_{\text{C}}$
becomes smaller with \textit{x} approaching 0.5. For \textit{x}$\leq $0.5, $%
\kappa $ tends to decrease at \textit{T}$_{\text{CO}}$, which can be
attributed to the large (JT-type) lattice distortion associated with CO. For 
\textit{x}\TEXTsymbol{>} 0.5, $\kappa $ shows only slight slope changes near 
\textit{T}$_{\text{CO}}$, as seen in the data of \textit{x}=0.52 and 0.55.
As summarized in Fig. \ref{Fig4-2} (c), $\kappa $ at 100 K decreases
systematically with \textit{x} approaching 0.5 (due to the stabilization of
the CO state), and remains small when \textit{x}\TEXTsymbol{>}0.5. It is
noted that even if the $\kappa $ values of \textit{x}=0.52 and 0.55 are
similar to that of \textit{x}=0.5 at low \textit{T}, they become
considerably larger than that of \textit{x}=0.5 for \textit{T}\TEXTsymbol{>}%
\textit{T}$_{\text{CO}}$. This behavior is well illustrated in the $\kappa $%
(300 K) vs. \textit{x} plot (solid circles in Fig. \ref{Fig4-2} (c)),
demonstrating a clear minimum at \textit{x}=0.5. This suppression of $\kappa 
$(300 K) at \textit{x}=0.5 correlates well with the $\mathit{\rho }$ peaking
near 0.5 at 300 K. Therefore, the results in Figs. \ref{Fig4-1} and \ref%
{Fig4-2} show that the lattice thermal conductivity as well as electrical
transport is suppressed in the high temperature region of the half-doped
manganite.

\begin{figure}[tb]
\epsfxsize=60mm
\centerline{\epsffile{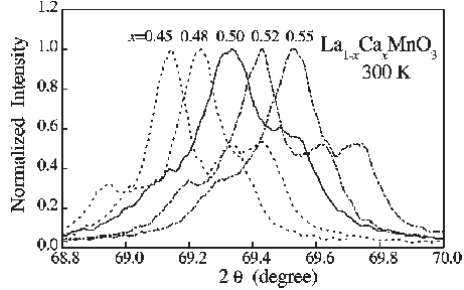}}
\caption{\textit{x}-ray intensity profiles of the (242) Bragg peak (center)
at 300 K for La$_{1-x}$Ca$_{x}$MnO$_{3}$ near \textit{x}=0.5.}
\label{Fig4-3}
\end{figure}

Directly related to the suppression of phononic $\kappa $, there exists a
slight, but noticeable broadening of the \textit{x}-ray Bragg peaks for 
\textit{x}=0.5 at room temperature. One example of the broadened \textit{x}%
-ray peaks is shown in Fig. \ref{Fig4-3}, displaying the compositional
change of the (242) Bragg peak (in the orthorhombic \textit{Pbnm} notation)
of \textit{x}-ray powder diffraction at 300 K. The (242) Bragg peak,
centered at 2$\theta \approx $69.3$^{\circ }$ for \textit{x}=0.5, changes
its position to higher angles as \textit{x} increases. The left and right
sides of the (242) Bragg peak are due to the (004)-(400) peaks and K$%
_{\alpha 2}$ of the (242) and (004)-(400) peaks, respectively. As evident in
Fig. \ref{Fig4-3}, the peak width, $\Gamma $, of the central (242) peak is
considerably broad at \textit{x}=0.5. To extract \textit{x}-dependence of $%
\Gamma $, the intensity profiles were fitted as a sum of three Gaussian
peaks (by neglecting the weak K$_{\alpha 2}$ peaks of (004) and (400)). The
solid squares and triangles in Fig. \ref{Fig4-2}(d) represent fitting
results for $\Gamma $ and the center position of the (242) peak,
respectively. The center position increases almost linearly with \textit{x},
indicating the linear lattice contraction with increasing \textit{x}. On the
other hand, $\Gamma $ shows a clear maximum at \textit{x}=0.5. This
broadening of $\Gamma $ indicates a slight distribution of lattice constants
in La$_{0.5}$Ca$_{0.5}$MnO$_{3}$ at room temperature \cite{19}. 

Based on the results from the synchrotron \textit{x}-ray scattering study of
La$_{0.5}$Ca$_{0.5}$MnO$_{3}$ that showed drastic broadening of all the
Bragg peaks between \textit{T}$_{\text{C}}$ and \textit{T}$_{\text{N}}$ ($%
\approx $\textit{T}$_{\text{CO}}$) \cite{19}, this new data indicate that
the Bragg peak broadening for \textit{x}$\approx $0.5 persists even at room 
\textit{T}, far above \textit{T}$_{\text{C}}$ and \textit{T}$_{\text{CO}}$.
This observation strongly suggests that short-range CO exists in the
paramagnetic state of the half-doped manganite. We cannot rule out the
possibility of dynamic correlation of CO at room \textit{T}. Furthermore,
the enhancement of $\mathit{\rho }$ up to 900 K in La$_{0.5}$Ca$_{0.5}$MnO$%
_{3}$ indicates that the spatial or dynamic CO fluctuation probably persists
even at \textit{T} ranges much higher than long-range ordering \textit{T} %
\cite{70}. Naturally, such a spatial variation of lattice constants,
indicated by the Bragg peak broadening, will shorten phonon lifetime, and
thus suppresses the phononic $\kappa $. It is emphasized that the observed
Bragg peak broadening indicates various anomalous behaviors of \textit{x}$%
\approx $0.5 as \textit{bulk effects}. In other words, the $\mathit{\rho }$
enhancement and the $\kappa $ suppression at \textit{x}$\approx $0.5 are not
due to, for example, grain boundaries in the polycrystalline specimens. It
is also noted that the findings are not consistent with La/Ca ionic ordering
because the La/Ca ordering should reduce $\rho $.

\begin{figure}[tb]
\epsfxsize=60mm
\centerline{\epsffile{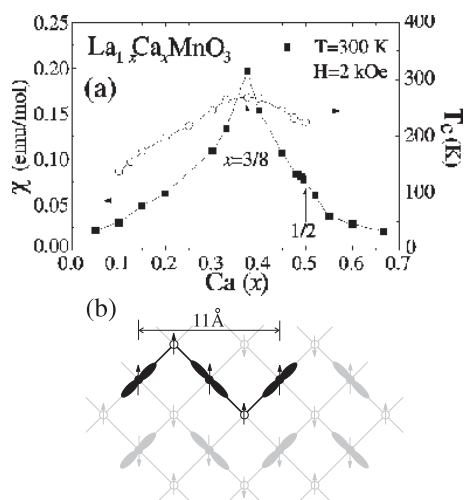}}
\caption{(a) The \textit{x}-dependence of magnetic susceptibility $\protect%
\chi $(closed squares) at 300 K (left axis). The open circles show the 
\textit{T}$_{\text{C}}$ variation determined from $\protect\chi $(\textit{T}%
) (right axis). (b) A schematic of FM zigzag chains, coupled
antiferromagnetically each other. Open circles are Mn$^{4+}$ and the lobes
show the \textit{e}$_{g}$ orbitals of Mn$^{3+}$. \symbol{126}11 \AA\ FM
zigzag is shown with dark hue.}
\label{Fig4-4}
\end{figure}

Because CO correlation can influence magnetic correlation, the evolution of
magnetic susceptibility ($\chi \equiv $M/H) was also investigated as a
function of \textit{x} at room \textit{T} (above \textit{T}$_{\text{C}}$). A
few surprising results can be found in Fig. \ref{Fig4-4} (a). First, $\chi $(%
\textit{T}) above \textit{T}$_{\text{C}}$ roughly follows the Curie-Weiss
law, and the Curie-Weiss \textit{T} is FM for all \textit{x} studied.
Consistent with earlier results, the \textit{T}$_{\text{C}}$ vs. \textit{x}
plot shows a broad bump near \textit{x}=3/8 \cite{10}. On the other hand, $%
\chi $(\textit{x}) at 300 K sharply peaks at \textit{x}=3/8. This
observation might be related to the fact that \textit{T}$_{\text{C}}$ is
maximized smoothly at \textit{x}=3/8, and $\chi $(\textit{T}) is
proportional to 1/\textit{T} at high temperature. However, this pronounced
peaking behavior of $\chi $(\textit{x}) above \textit{T}$_{\text{C}}$ again
corroborates that the commensurate carrier concentration of \textit{x}=3/8
has a specially enhanced FM correlation. This observation naturally suggests
an extraordinary possibility; the presence of short-range or dynamic
correlation of charge/orbital ordering in such a way as to produce special
FM coupling in addition to the double exchange-type FM coupling.

How can such a short-range charge correlation promotes FM coupling? In
manganites, it has been well established that the CE-type CO is very stable
in broad \textit{x} ranges, at least at low temperature. For example, the
CE-type CO has been commonly observed in La$_{1-x}$Ca$_{x}$MnO$_{3}$ and Nd$%
_{1-x}$Sr$_{x}$MnO$_{3}$ for \textit{x}$\approx $0.5. In addition, the
CE-type CO has been reported even when \textit{x} deviates significantly
from 0.5. For example, the CE-type CO occurs in Pr$_{\text{1-}x}$Ca$_{x}$MnO$%
_{3}$ with 0.3\TEXTsymbol{<}\textit{x}$\leq $0.5, and also in (La,Pr)$_{5/8}$%
Ca$_{3/8}$MnO$_{3}$ at low \textit{T} \cite{71}. Thus, it is appealing to
assume that short-range or dynamic CO at high \textit{T} of La$_{1-x}$Ca$%
_{x} $MnO$_{3}$ (0.2\TEXTsymbol{<}\textit{x}$\leq $0.5) is also the CE-type.
In the CE-type CO, there exist FM zigzag chains (Fig. \ref{Fig4-4} (b)),
which couple to each other antiferromagnetically \cite{69,74}. It is
conceivable that at high \textit{T}, the CO correlation is so short that the
short-range CO state may contain only one short FM zigzag (Mn$^{3+}$- Mn$%
^{4+}$- Mn$^{3+} $- Mn$^{4+}$- Mn$^{3+}$: shown with dark hue in Fig. \ref%
{Fig4-4} (b)) with $\sim $11 \AA\ in length or one short FM ``zig or zag''
(Mn$^{3+}$-Mn$^{4+}$-Mn$^{3+}$) with $\sim $5.5 \AA\ in length. Then, these
extended objects can enhance FM correlation overall. The short FM zigzag can
be considered as correlated polaronsor a ferromagnetic polaron cluster, and
may exhibit dynamic nature. Note that the carrier concentration of the short
FM zigzag (zig or zag) corresponds to \textit{x}=0.4 (1/3), which is close
to \textit{x}$\approx $3/8 for the enhanced FM correlation. However, if the
range of CO becomes slightly longer, then the zigzag may couple with the
neighboring zigzags antiferromagnetically so that FM correlation can be
reduced. This effect can be significant at \textit{x}$\approx $0.5 where CO
tendency is strong because carrier concentration matches the CE-type
ordering. We also found that, in general, $\rho $(300 K) for \textit{x}%
\TEXTsymbol{<}0.5 (including \textit{x}$\approx $3/8 where \textit{T}$_{%
\text{C}}$ is maximized) is larger than that for \textit{x}\TEXTsymbol{>}0.5
(including \textit{x}$\approx $5/8 where \textit{T}$_{\text{CO}}$ is
optimized) even though the ground state is metallic (insulating) for \textit{%
x}\TEXTsymbol{<}(\TEXTsymbol{>})0.5, which corroborates with short-range CO
at high \textit{T} for \textit{x} near or smaller than 0.5. This remarkable
scenario remains to be confirmed by local probe measurements such as \textit{%
x}-ray or neutron scattering experiments.

This section showed that the suppression of electronic conductivity as well
as phononic thermal conductivity, and the broadening of Bragg peaks exist in
a narrow composition range near \textit{x}=0.5, but in a very broad
temperature range up to 900 K. All these findings suggest the presence of
spatial or temporal fluctuation of CO at high \textit{T}. On the other hand,
FM correlation is strongly enhanced for \textit{x} near 3/8, which can be
related to the presence of FM zigzags that can be coupled or decoupled,
depending on \textit{x}. The ``decoupled'' short FM zigzag can enhance the
overall FM correlation at \textit{x} near 3/8, and the AFM coupling of FM
zigzags can progressively increase with \textit{x}, and be maximized at 
\textit{x}$\approx $0.5 where charge localization tendency is strong.

\subsection{Optical pseudogap and charge ordering fluctuation}

Optical conductivity study has been one of unique tools to probe short-range
fluctuating order parameters of solids. It has successfully revealed a
phase-correlation time of superconducting order parameter in the normal
state of the Bi$_{2}$Sr$_{2}$CaCu$_{2}$O$_{8+\delta }$ family and
fluctuating charge-density-wave (CDW) order parameters of K$_{0.3}$MoO$_{3}$
and (TaSe$_{4}$)$_{2}$I at high temperature \cite{75,76}. However, up to
this point, there are few optical conductivity studies to probe the
short-range charge correlation above long-range charge ordering temperature, 
\textit{T}$_{\text{CO}}$.

In parallel with the findings in Section \ref{subsec4-1}, many recent
experiments, such as Raman \cite{77}, neutron \cite{78,79}, and \textit{x}%
-ray \cite{70} scattering studies, showed the existence of the spatial
and/or temporal fluctuations of the CE-type CO in the FM metallic compounds
above the Curie temperature \textit{T}$_{\text{C}}$. In particular, the
neutron scattering studies found dynamic ($\geq \sim $1 ps) or short-ranged (%
$\sim $10 \AA ) CE-type CO correlation above \textit{T}$_{\text{C}}$.
Recently, evidences for the existence of incipient charge ordering were
observed in optical conductivity spectra of a FM bilayer manganite La$_{1.2}$%
Sr$_{1.8}$Mn$_{2}$O$_{7}$, just above \textit{T}$_{\text{C}}$ \cite{80}.
Based on the experimental results in the Section \ref{subsec4-1}, the
Ca-doped manganites including La$_{0.5}$Ca$_{0.5}$MnO$_{3}$ can be an ideal
system to investigate the fluctuating CO through optical conductivity
studies. 
Thus, optical conductivity spectra $\sigma $($\omega $) of the La$%
_{1-x}$Ca$_{x}$MnO$_{3}$ system (0.48$\leq $\textit{x}$\leq $0.67) were
systematically investigated to probe the fluctuating CO correlation \cite{81}%
. For this purpose, high-density polycrystalline specimens of La$_{1-x}$Ca$%
_{x}$MnO$_{3}$ (\textit{x}=0.48, 0.50, 0.52, 0.60, and 0.67) were
investigated, of which characteristics are well described in Section \ref%
{subsec4-1}. For \textit{x}=0.48, a FM metallic state is dominant below 
\textit{T}$_{\text{C}}\approx $220 K with a possible short-range CO phase
below 180 K. However, for \textit{x}$\geq $0.52, the antiferromagnetic
long-range CO becomes stabilized at low \textit{T}; \textit{T}$_{\text{CO}}$
values of \textit{x}=0.52, 0.60, and 0.67 were 190, 250, and 257 K,
respectively. The \textit{x}=0.50 sample has \textit{T}$_{\text{CO}}\approx $%
180 K and \textit{T}$_{\text{C}}\approx $220 K. Note that the \textit{x}%
=0.50 sample can have coexistence of the FM metallic and the CO insulating
states \textit{at lower} \textit{T} as well as between \textit{T}$_{\text{CO}%
}$ (\textit{T}$_{\text{N}}$) and \textit{T}$_{\text{C}}$. 
\begin{figure}[tb]
\epsfxsize=60mm
\centerline{\epsffile{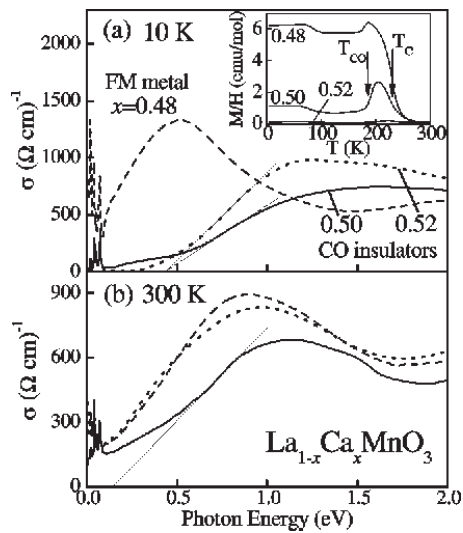}}
\caption{Optical conductivity spectra $\protect\sigma $($\protect\omega $)
of \textit{x}=0.48, 0.50, and 0.52 at (a) 10 K and (b) 300 K. The crossing
points between \textit{x}-axis and linear extrapolation lines give an
estimate of the charge gap 2$\Delta $. The inset shows that M/H values
measured at 2 kOe after zero field cooling.}
\label{Fig4-5}
\end{figure}

Figure \ref{Fig4-5} (a) shows $\sigma $($\omega $) of the \textit{x}=0.48,
0.50, and 0.52 samples at 10 K. $\sigma $($\omega $) of \textit{x}=0.48 have
a large absorption band centered at $\sim $0.5 eV, which can be related to
incoherent hopping motion of polarons from Mn$^{3+}$ to Mn$^{4+}$ sites \cite%
{64,82,83}. On the other hand, an optical gap due to the long-range CO,
namely, a charge gap is clearly observed in $\sigma $($\omega $) of \textit{x%
}=0.50 and 0.52 at 10 K. A charge gap, 2$\Delta $, an onset energy of the
steeply rising part of $\sigma $($\omega $), can be determined from a
crossing point between \textit{x}-abscissa and a linear extrapolation line
drawn at the inflection point of $\sigma $($\omega $). This procedure has
been a common practice to evaluate 2$\Delta $ of various CO materials \cite%
{84}, resulting in 2$\Delta $(10 K)$\approx $0.45 eV for both \textit{x}%
=0.50 and 0.52 samples. The charge gap energy at the ground state, 2$\Delta $%
(0)$\approx $0.45 eV is the largest among numerous CDW systems \cite{76} and
charge-ordered oxides \cite{84}. This large 2$\Delta $(0) can be a peculiar
characteristic of the CE-type CO, indicating unusual stability of the
special CO pattern.

Interestingly, $\sigma $($\omega $) of \textit{x}=0.50 at 10 K show
significant in-gap absorption below 0.5 eV, while in-gap absorption of 
\textit{x}=0.52 is negligible at 10 K. One key finding from our experiments
is that the spectral weight of the in-gap absorption is proportional to the
amount of FM phase inside the samples. For example, decrease of M/H values
at 10 K from \textit{x}=0.48 to 0.52 (the inset of Fig. \ref{Fig4-5} (a)) is
well correlated with decrease of spectral weight of $\sigma $($\omega $)
below 0.5 eV at 10 K. In addition, the larger M/H values of \textit{x}=0.50
than \textit{x}=0.52 are consistent with the increased FM regions in \textit{%
x}=0.50, located near the CO/FM phase boundary. Therefore, the in-gap
absorption of \textit{x}=0.50 is attributed to a FM phase coexisting with a
CO phase at low temperature.

In Fig. \ref{Fig4-5} (b), $\sigma $($\omega $) of \textit{x}=0.50 at 300 K
reveal anomalous spectral features; $\sigma $($\omega $) below 1.0 eV are
smaller than those of neighbouring compounds. Furthermore, $\sigma $($\omega 
$) increase steeply with $\omega $, showing a positive curvature at low
photon energy, which is very similar to the gap-feature observed at 10 K. It
is noted that this spectral response is not compatible with a single polaron
absorption model \cite{64}. Instead, the gap feature at 300 K suggests the
presence of short-range CO or correlated multi-polarons even far above 
\textit{T}$_{\text{CO}}$. It is noteworthy that it is not yet known if a
theory for correlated multi-polaron absorption could account for the
peculiar $\sigma $($\omega $) of \textit{x}=0.50 above \textit{T}$_{\text{CO}%
}$. 
\begin{figure}[tb]
\epsfxsize=60mm
\centerline{\epsffile{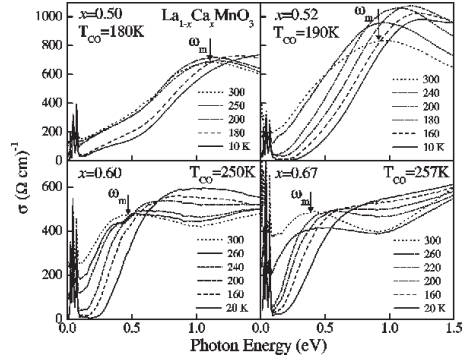}}
\caption{$\protect\sigma $($\protect\omega $) of La$_{1-x}$Ca$_{x}$MnO$_{3}$
(0.50$\leq $\textit{x}$\leq $0.67) at various temperature. The arrow for
each \textit{x} represents an energy of maximum $\protect\sigma $($\protect%
\omega $), $\protect\omega _{\text{m}}$, at 300 K. With decreasing \textit{x}
from 0.67 to 0.50, 2$\Delta $ at the lowest temperature increases (while 
\textit{T}$_{\text{CO}}$ decreases), and the suppression of spectral weight
below wm, i.e., the optical pseudogap feature, is systematically enhanced.}
\label{Fig4-6}
\end{figure}

To understand further the anomalous $\sigma $($\omega $) of \textit{x}=0.50,
we systematically investigated $\sigma $($\omega $) of La$_{1-x}$Ca$_{x}$MnO$%
_{3}$ with \textit{x}=0.50, 0.52, 0.60, and 0.67 (Fig. \ref{Fig4-6}). It is
found that \textit{T}-dependence of the charge gap for \textit{x}=0.50 is
also peculiar; at 10 K$\leq $\textit{T}$\leq $180 K, large 2$\Delta $ values
are observed in $\sigma $($\omega $) of \textit{x}=0.50. In particular, 2$%
\Delta $ values at \textit{T}=250 and 300 K still remain finite, remarkably
having almost the same magnitude with 2$\Delta $ at 200 K. The $\sigma $($%
\omega $) of \textit{x}=0.52 with \textit{T}$_{\text{CO}}\approx $190 K also
show that 2$\Delta $ remains nonzero up to \textit{T}$\approx $240 K. These
observations strongly suggest that the strong fluctuations of short-range CO
(or correlated multi-polaron) can be responsible for the finite charge gap
far above \textit{T}$_{\text{CO}}$ in the \textit{x}=0.50 and 0.52 samples.

Even at high \textit{T}\TEXTsymbol{>}240 K, where 2$\Delta $ is no longer
finite, there exists significant suppression of spectral weight below a
photon energy of maximum $\sigma $($\omega $), $\omega _{\text{m}}$. (arrows
in Fig. \ref{Fig4-6}). This suppression of spectral weight is accompanied by
a pseudogap in $\sigma $($\omega $), i. e., decreasing $\sigma $($\omega $)
below $\omega _{\text{m}}$ at each \textit{T}. This pseudogap is also observed in the $\sigma 
$($\omega $) of \textit{x}=0.60 and 0.67 at temperatures up to at least 300
K, even if 2$\Delta $ of \textit{x}=0.60 and 0.67 becomes zero just above 
\textit{T}$_{\text{CO}}$. Surprisingly, it is found that the pseudogap
feature shows systematic doping dependence. First, $\omega _{\text{m}}$ at
300 K systematically increases as \textit{x} approaches 0.50 from above (See
Fig. \ref{Fig4-6}). Second, the suppression of spectral weight below $\omega
_{\text{m}}$ at 300 K becomes more evident as \textit{x} approaches 0.50,
which finally produces a nonzero 2$\Delta $ even at 300 K. This systematic
enhancement of the pseudogap feature and its proximity to the finite 2$%
\Delta $ far above \textit{T}$_{\text{CO}}$ near \textit{x}=0.50
consistently suggest that the optical pseudogap can be attributed to the
spatially fluctuating CO correlation of La$_{1-x}$Ca$_{x}$MnO$_{3}$ (\textit{%
x}$\geq $0.50) at high \textit{T} region. 
\begin{figure}[tb]
\epsfxsize=60mm
\centerline{\epsffile{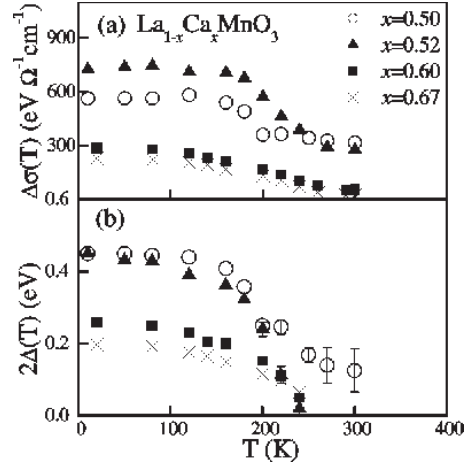}}
\caption{(a) Spectral weight difference, $\Delta \protect\sigma $(\textit{T})%
$\equiv \protect\sigma $($\protect\omega _{\text{m}}$)$\protect\omega _{%
\text{m}}$- $\protect\int_{0}^{\protect\omega _{\text{m}}}\protect\sigma $($%
\protect\omega $)$d\protect\omega $, and (b) optical gap, 2$\Delta $(\textit{%
T}) plots of La$_{1-x}$Ca$_{x}$MnO$_{3}$ (0.50$\leq $\textit{x}$\leq $67).}
\label{Fig4-7}
\end{figure}

To investigate the charge gap and its pseudogap developments quantitatively,
we determined \textit{T}-dependent suppressed spectral weight, $\Delta
\sigma $(\textit{T})=$\sigma $($\omega _{\text{m}}$)$\omega _{\text{m}}$-$%
\int_{0}^{\omega _{\text{m}}}\sigma $($\omega $)$d\omega $ and 2$\Delta $(%
\textit{T}) for each \textit{x}, as shown in Fig. \ref{Fig4-7}. Below 
\textit{T}$_{\text{CO}}$, $\Delta \sigma $(\textit{T}) values of \textit{x}%
=0.50 are smaller than those of \textit{x}=0.52. The proximity to FM phase
boundary at \textit{x}=0.50 may be responsible for the reduction of $\Delta
\sigma $(\textit{T}) of \textit{x}=0.50 at low \textit{T}. It is noted that
high-\textit{T} $\Delta \sigma $(\textit{T}) values of all the samples are
clearly nonzero up to at least 300 K. In particular, $\Delta \sigma $(%
\textit{T}) values at 300 K increase systematically as \textit{x} approaches
0.50. At the same time, 2$\Delta $(0) values in Fig. \ref{Fig4-7} (b)
increase as \textit{x} approaches 0.50; 2$\Delta $(0)$\approx $0.2, 0.26,
0.45, and 0.45 eV for \textit{x}=0.67, 0.60, 0.52 and 0.50, respectively.
This increase of the 2$\Delta $(0) is well correlated with the increase of $%
\Delta \sigma $(\textit{T}) at 300 K as \textit{x} approaches 0.50. These
findings indicate that the strength of CO stability is clearly maximized for 
\textit{x}$\approx $0.50, being responsible for the enhanced CO fluctuation
at high temperature. Because \textit{T}$_{\text{CO}}$ of these compounds
decreases as \textit{x} approaches 0.50, 2$\Delta $(0)/\textit{k}$_{\text{B}%
} $\textit{T}$_{\text{CO}}$ values systematically increase: 2$\Delta $(0)/%
\textit{k}$_{\text{B}}$\textit{T}$_{\text{CO}}$ $\approx $9, 12, 28, and 30
for \textit{x}=0.67, 0.60, 0.52 and 0.50, respectively. The 2$\Delta $(0)/%
\textit{k}$_{\text{B}}$\textit{T}$_{\text{CO}}$ up to 30, an unusually large
value among many charge-ordered oxides, indicates strongly enhanced electron
correlation near \textit{x}=0.50 \cite{84,85}. 

\begin{figure}[tb]
\epsfxsize=60mm
\centerline{\epsffile{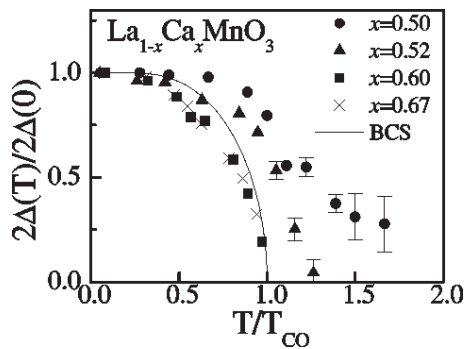}}
\caption{The 2$\Delta $(\textit{T})/2$\Delta $(0) vs \textit{T}/\textit{T}$_{%
\text{CO}}$ plot. The solid line represents the BCS functional form.}
\label{Fig4-8}
\end{figure}

Related with the enhanced pseudogap feature and large 2$\Delta $(0)/\textit{k}$_{\text{B}}$\textit{T}$%
_{\text{CO}}$ value for \textit{x}$\approx $0.50, 2$\Delta $(\textit{T})/2$%
\Delta $(0) vs $T$/\textit{T}$_{\text{CO}}$ curves of \textit{x}=0.50 and
0.52 clearly deviate from the BCS functional form (Fig. \ref{Fig4-8}). For
example, 2$\Delta $(\textit{T})/2$\Delta $(0) values of \textit{x}=0.50 are
still about 0.25 at $T$/\textit{T}$_{\text{CO}}\approx $ 1.7 (\textit{T}=300
K) and those values of \textit{x}=0.52 are nonzero up to at least $T$/%
\textit{T}$_{\text{CO}}$ $\approx $ 1.2 ($T\approx $240 K). These unique 2$%
\Delta $(\textit{T})/2$\Delta $(0) curves of non-BCS-type are quite
consistent with the greatly enhanced spatial and/or temporal CO fluctuation
near \textit{x}=0.50 at high temperature regions. However, as \textit{x} is
increased, 2$\Delta $(\textit{T})/2$\Delta $(0) curves recover the BCS form
for \textit{x}=0.60 and 0.67, as observed in most of CO materials \cite{85}. 

\begin{figure}[tb]
\epsfxsize=60mm
\centerline{\epsffile{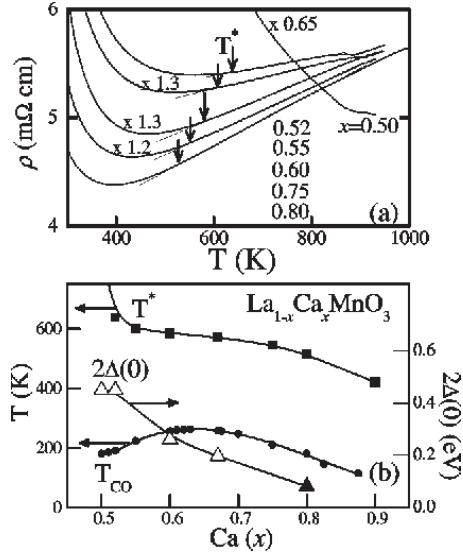}}
\caption{(a) High temperature-$\protect\rho $ data for La$_{1-x}$Ca$_{x}$MnO$%
_{3}$ for \textit{x}$\geq $0.50. \textit{T}*(arrow) for each \textit{x} was
determined as the \textit{T} where the linear metallic resistivity starts to
deviate. (b) Phase diagram of La$_{1-x}$Ca$_{x}$MnO$_{3}$ with \textit{x} $%
\geq $0.50 showing \textit{T}*, \textit{T}$_{\text{CO}}$, and 2$\Delta $(0).
The solid lines are guides to eye. A solid triangle represents a 2$\Delta $%
(0) value of \textit{x}=0.80 from Ref. \protect\cite{86}.}
\label{Fig4-9}
\end{figure}

To check how high temperatures this CO fluctuation survives, we studied high-%
$T$ resistivity of La$_{1-x}$Ca$_{x}$MnO$_{3}$ with \textit{x}$\geq $0.50 in
Fig. \ref{Fig4-9} (a). The \textit{x}=0.50 sample shows insulating behaviour
up to above $\sim $850 K, where orthorhombic (low \textit{T}) to
rhombohedral (high \textit{T}) structural transition occurs. Surprisingly,
for \textit{x}$\geq $0.52, there exists a crossover from metallic to
insulating states as \textit{T} decreases. Moreover, the crossover
temperature defined as, \textit{T}*, decreases systematically as \textit{x}
increases from 0.50. This \textit{T}* evolution with \textit{x} is well
correlated to 2$\Delta $(0) and $\Delta \sigma $(300 K) behaviours for 0.50$%
\leq $\textit{x}$\leq $0.67. Furthermore, in Fig. \ref{Fig4-9} (b), \textit{T%
}* decreases as \textit{T}$_{\text{CO}}$ does above \textit{x}=0.67. Our
previous study of \textit{x}=0.80 showed 2$\Delta $(0)$\approx $0.08 eV,
indicating that both \textit{T}$_{\text{CO}}$ and 2$\Delta $(0) decrease
together above \textit{x}=0.67 \cite{86}. Therefore, \textit{T}* of
charge-ordered La$_{1-x}$Ca$_{x}$MnO$_{3}$ is clearly linked to 2$\Delta $%
(0). This observation supports that \textit{T}* can be the temperature where
high \textit{T}-CO correlation starts and thus optical pseudogap appears.

Why is 2$\Delta $(0)/\textit{k}$_{\text{B}}$\textit{T}$_{\text{CO}}$ is so
large near \textit{x}=0.50? In a quasi-1-dimensional (D) CDW system, the CDW
fluctuation, induced by low dimensionality, results in gap-like features in $%
\sigma $($\omega $) above the 3D long-range ordering temperature, \textit{T}$%
_{3\text{D}}$ \cite{76}. The mean field transition temperature of \textit{T}$%
_{\text{MF}}$ (equiv 2$\Delta $(0)/3.5) is usually quite larger than \textit{%
T}$_{\text{3D}}$. Between \textit{T}$_{\text{MF}}$ and \textit{T}$_{\text{3D}%
}$, a crossover from 1-D to 3-D correlation occurs at \textit{T}*. These
phenomena have some similarities with what observed in La$_{1-x}$Ca$_{x}$MnO$%
_{3}$ with \textit{x}$\geq $0.50. However, it is not clear whether the
striped charge and orbital ordering, and resultant effective low
dimensionality can induce such unusually large charge fluctuation in 3-D
materials. In addition, if the low dimensionality induced by striped
charge/orbital ordering is the main source for the fluctuation, the samples
near \textit{x}=0.67 would have similar amount of charge fluctuations due to
well-defined charge/orbital ordering at the commensurate case. 

The more appealing origin for the enhanced CO fluctuation for \textit{x}$%
\geq $0.50, can be a competition of order parameters in La$_{1-x}$Ca$_{x}$MnO%
$_{3}$ (\textit{x}=0.50). Obviously, the commensurate 1:1 ratio of Mn$^{3+}$
and Mn$^{4+}$ ions is compatible with the strong CO tendency at \textit{x}%
=0.50. At the same time, the double exchange mechanism predicts that the
strength of the FM correlation in La$_{1-x}$Ca$_{x}$MnO$_{3}$ should be
varied as \textit{x}(1-$x$) and optimized at \textit{x}=0.50 \cite{29}.
Indeed, La$_{1-x}$Ca$_{x}$MnO$_{3}$ (\textit{x}=0.50) has a thermodynamic
bicritical point, where the FM metallic and the CO insulting states meet
with the paramagnetic insulating state. Therefore, near the critical point,
competing order parameters can lead to the suppression of ordering
temperatures as well as increased spatial/temporal fluctuation among those
phases. This scenario is further supported by a recent computational study,
predicting that large charge fluctuations could be a generic feature of the
mixed-phase systems at or near the regimes where a phase separation occurs
as \textit{T}$\rightarrow $0 K \cite{63}. 

Finally, it will be interesting to
check the time and the length scales of the fluctuating short-range order
that can be observed at a $\sigma $($\omega $) study. The recent neutron
scattering study indicated that CO fluctuations above \textit{T}$_{\text{C}}$
occur with a time scale slower than 1 ps \cite{87,88,89}. Thus,
frequency-dependent optical spectroscopy can probe the presence of
fluctuations, because 0.5-1 eV corresponds to time scales of 5-10 fs.
Besides, an incoming light takes an average over a length scale of $\lambda $%
/$n$, where $\lambda $ is the wavelength of the light and $n$ is the
refractive index of the medium. For the 0.5-1 eV energy ranges, $n\approx $3
and $\lambda $/$n\approx $300-700 nm. Therefore, when the volume fraction of
the CO domains is large, 10-100 nm scale CO correlations can be observed by $%
\sigma $($\omega $) study \cite{84}. If the correlation length is too small,
for example, $\sim $10 \ \AA\ for the La$_{1-x}$Ca$_{x}$MnO$_{3}$ (\textit{x}%
=0.30) sample, the effects of the CO correlation on the $\sigma $($\omega $)
might not be easily distinguished with a strong small polaron absorption
band. 

This section presented doping-dependent evolutions of charge-ordering
gap and its pseudogap in La$_{1-x}$Ca$_{x}$MnO$_{3}$ (0.48$\leq $\textit{x}$%
\leq $0.67) from systematic optical conductivity and transport studies. With
decreasing \textit{x} from 0.67 to 0.50, the low temperature charge gap
systematically increased while charge ordering temperature decreased.
Simultaneously, the optical pseudogap, indicating charge-ordering
fluctuation at high temperatures, is greatly enhanced as \textit{x}
approaches 0.50. This $\sigma $($\omega $) study proves that short-range
charge ordering fluctuation is anomalously strong in manganites.

\section{X-ray scattering studies of high-temperature charge/orbital
correlations}

\label{sec5} Because of the strong coupling between the electronic,
magnetic, and structural degrees of freedom in manganites, charge/orbital
fluctuations in these materials are always accompanied by local structural
distortions. These distortions can be measured directly by \textit{x}-ray
and neutron diffraction techniques. In recent years, these techniques have
been extensively utilized for the investigation of the local structural
distortions in the high-temperature paramagnetic insulating (PI) phase of
manganites \cite{87,88,89,90,91,92,93,94,95,96,97}. In this section, we
briefly outline some recent results of these studies, concentrating on 
\textit{x}-ray diffraction and on three-dimensional perovskite manganites.
We note that this is not a comprehensive review. In particular, layered
manganites are not discussed here; the interested reader is referred to
Refs. \cite{87,95,97}. 

The importance of local structural distortions was
first pointed out in connection to the anomalously large resistivity of the
PI phase \cite{27}. It was proposed that small lattice polarons are present
in this state. A lattice polaron forms when an \textit{e}$_{g}$ electron
localizes on a Mn$^{3+}$ ion, and the surrounding oxygen octahedron distorts
due to Jahn-Teller effect. Formation of the lattice polarons leads to the
increase of the electrical resistivity. The polarons are strongly suppressed
in the ferromagnetic metallic (FM) state, and therefore it was suggested
that they play a key role in the CMR effect. Numerous experiments \cite%
{83,98,99,100,101} have confirmed the presence of lattice polarons in the PI
state including, in particular, diffuse \textit{x}-ray scattering
measurements \cite{87,90} and scattering measurements of the pair
distribution function (PDF) \cite{101}. 

An important step in understanding
the nature of the PI phase was made in 1999 when it was realized that the
local lattice distortion in this phase is not necessarily confined to the
single oxygen octahedron surrounding a Mn ion. Neutron and \textit{x}-ray
diffraction measurements have shown that nanoscale structural correlations
are present in the PI phase \cite{88,89,90,91,92,93,94,96}. These
correlations were initially described as correlated polarons. It was also
pointed out that in the \textit{x}=0.3 compounds, these correlations occur
at the same scattering vector as the charge/orbital peaks in the samples
exhibiting long-range charge/orbital order of the CE type. Therefore, it was
proposed that these correlations reflect the presence of nanoscale regions
possessing charge and orbital order. Importantly, it was shown that the
electrical resistivity and the intensity of the peaks due to the nanoscale
correlations follow very similar temperature dependences \cite{88,89,91}.
The following picture therefore emerged: the PI state contains nanoscale
insulating regions possessing charge/orbital order, and it is these regions
that are responsible for the enhanced resistivity of the PI state. The
nanoscale correlations, therefore, were concluded to play an important role
in the CMR effect. 

Since the original observation of the nanoscale
correlations in \newline
La$_{1-x}$Ca$_{x}$MnO$_{3}$, they were found in a number of different
manganite compounds \cite{91,92,93,94,96}. It appears that these
correlations are present in the PI state in a broad range of carrier
concentration, which we tentatively define as \textit{x}=0.2-0.5, provided
that the lattice symmetry is of the orthorhombic$\ O$ type. Interestingly,
the correlations are not significantly affected by the nature of the
low-temperature ordered state. In particular, very similar nanoscale
correlations were found in Pr$_{0.7}$Ca$_{0.3}$MnO$_{3}$ which exhibits the
CE-type order at low temperatures, and in La$_{0.7}$Ca$_{0.3}$MnO$_{3}$,
which is a ferromagnetic metal \cite{91}. Nanoscale structural correlations,
therefore, appear to be a generic feature of the orthorhombic PI state in
hole-doped manganites. 

Despite significant amount of research activity
devoted to the local inhomogeneities in the PI state of the manganites, the
exact structure of the correlated regions is still largely unknown.
Moreover, recent studies indicate that the scenario described above, in
which nanoscale charge/orbital ordered insulating regions are present in the
PI state, is in all likelihood oversimplified. Below, we describe some
recent \textit{x}-ray scattering studies of the structural correlations in
the PI state, discuss the implications of these experiments for the local
structure of the PI state, and pose questions for future research. 
\begin{figure}[tb]
\epsfxsize=60mm
\centerline{\epsffile{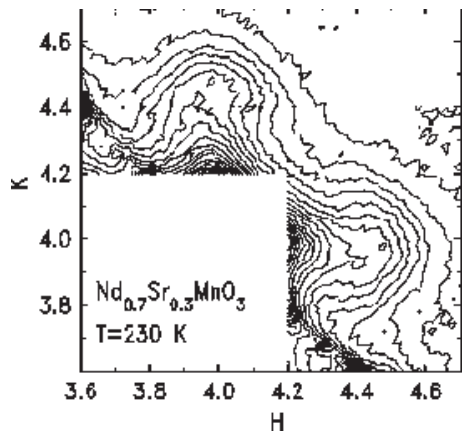}}
\caption{Contour plot of the \textit{x}-ray intensity around the (4, 4, 0)
Bragg peak at \textit{T}=230 K in Nd$_{0.7}$Sr$_{0.3}$MnO$_{3}$ (reprinted
from {\protect\cite{96}}).}
\label{Fig5-1}
\end{figure}

\begin{figure}[tb]
\epsfxsize=60mm
\centerline{\epsffile{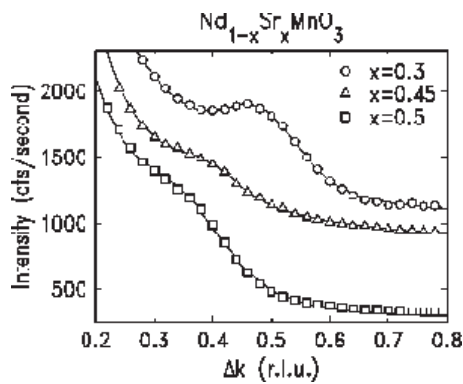}}
\caption{\textit{x}-ray scans along the (4, 6+$\Delta $\textit{k}, 0)
direction (\textit{x}=0.45, 0.5), and the (4, 4+$\Delta $\textit{k}, 0)
direction (\textit{x}=0.3). The temperatures are 210 K, 275 K, and 260 K for
the \textit{x}=0.3, 0.45, and 0.5 samples, respectively (reprinted from {%
\protect\cite{96}}).}
\label{Fig5-2}
\end{figure}

We will focus our discussion on the properties of the Nd$_{1-x}$Sr$_{x}$MnO$%
_{3}$, \textit{x}=0.3-0.5 samples. These properties are representative of
the manganites with the FM ground state; materials with the charge-ordered
ground state will be discussed later. The nanoscale correlations give rise
to broad diffraction peaks at the (0, $k$, 0), and ($h$, 0, 0) reduced
scattering vectors, with $h$,$k$=0.35-0.55 in different manganites \cite%
{88,89,90,91,92,93,94,96}. (We are using the orthorhombic \textit{Pbnm}
notation, in which $a\sim \sqrt{2}a_{c}$, $b\sim \sqrt{2}a_{c}$, and $c\sim $%
2$a_{c}$, $a_{c}$ is the cubic perovskite lattice constant.) A
representative overall scattering pattern is illustrated in Fig. \ref{Fig5-1}%
. The broad peaks at (4.5, 4, 0) and (4, 4.5, 0) reflect the presence of the
nanoscale correlations. The intensity of these peaks is believed to reflect
the concentration of the correlated regions, and their width is inversely
proportional to the region size \cite{96}. The position of the peaks defines
the period of the lattice modulation in the correlated regions. The peaks
are observed on top of the ``butterfly-like'' shaped background which is
attributed to scattering from uncorrelated polarons, also known as Huang
scattering, and to thermal-diffuse scattering \cite{87,90}. To separate this
background from the correlated peaks, scans along the $a$ or $b$ directions
are taken, see Fig. \ref{Fig5-2}. These scans are fitted to a Gaussian or
Lorentzian line shape and a monotonically sloping background. 

\begin{figure}[tb]
\epsfxsize=60mm
\centerline{\epsffile{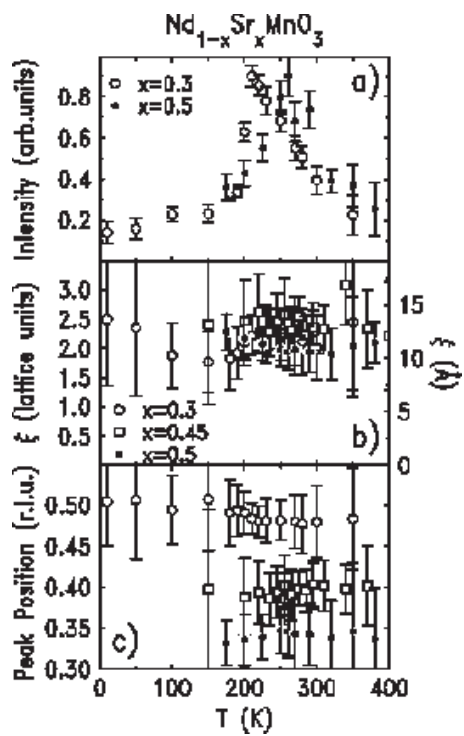}}
\caption{(a) Temperature dependence of the intensity of the peak due to the
structural correlations in Nd$_{1-x}$Sr$_{x}$MnO$_{3}$. The single-polaron
background is subtracted. (b) The correlation length of the ordered regions.
(c) The position of the peak relative to the nearest Bragg peak (the lattice
modulation wave vector). Reprinted from \protect\cite{96}.}
\label{Fig5-3}
\end{figure}

The parameters of the correlated peaks in the Nd$_{1-x}$Sr$_{x}$MnO$_{3}$
samples are shown in Fig. \ref{Fig5-3}. The peak intensity is strongly
reduced in the FM state and traces the behavior of the electrical
resistivity over the entire temperature range in the PI and FM states.
Interestingly, in many samples the structural correlations do not disappear
completely below the Curie temperature even at the lowest temperatures.
These results are consistent with the earlier PDF measurements \cite{101}
indicating the presence of local Jahn-Teller distortions in the FM state.
The data of Fig. \ref{Fig5-3} show that the correlations survive deep in the
FM region of the phase diagram, indicating that the FM state is
inhomogeneous. 

\begin{figure}[tb]
\epsfxsize=60mm
\centerline{\epsffile{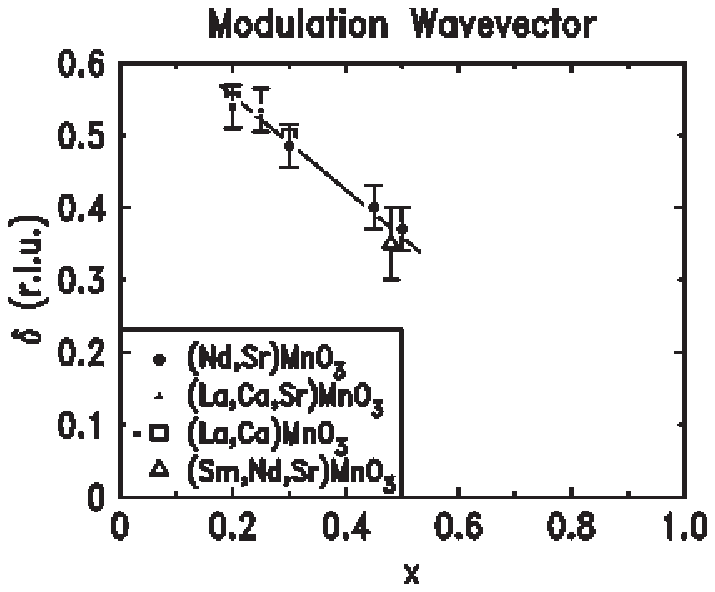}}
\caption{The wave vector $\protect\delta $ of the lattice modulation in the
correlated regions as a function of doping \textit{x} (filled symbols). Open
symbols show the wave vector of the structures with long range charge and
orbital order observed in manganites with \textit{x}\TEXTsymbol{>}0.5
(reprinted from {\protect\cite{96}}).}
\label{Fig5-4}
\end{figure}

Fig. \ref{Fig5-3}\ also illustrates two other important observations. First,
the correlation length of the nanoscale regions is the same in all samples
and is independent on temperature. The same correlation length is also found
in other manganites, including La$_{1-x}$Ca$_{x}$MnO$_{3}$ (\textit{x}=0.2,
0.3) \cite{88,89}, and La$_{0.75}$(Ca$_{0.45}$Sr$_{0.55}$)$_{0.25}$MnO$_{3}$ %
\cite{96}. Second, the period of the lattice modulation is also independent
on temperature. It does depend on the sample composition, but in a very
regular manner. In Fig. \ref{Fig5-4} we plot the period of the lattice
modulation as a function of doping, \textit{x}. Remarkably, all the results
shown in Fig. \ref{Fig5-4} appear to fall on the same line, which begins at 
\textit{x}=0 and $\delta $=2/3, and ends at \textit{x}=1 and $\delta $=0. 

\begin{figure}[tb]
\epsfxsize=60mm
\centerline{\epsffile{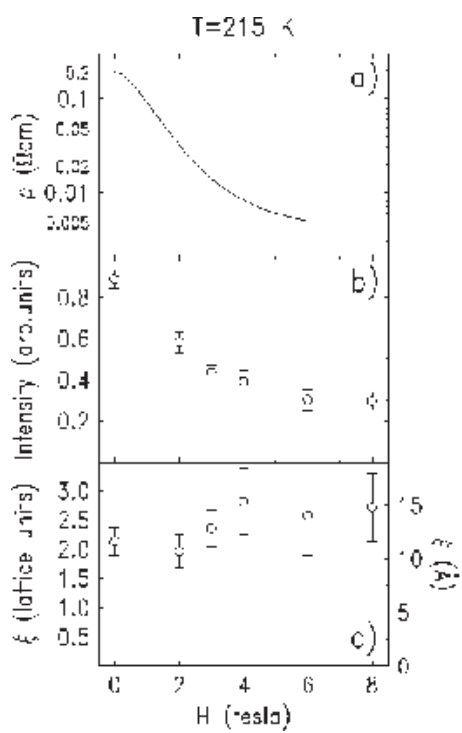}}
\caption{Magnetic field-dependence of (a) the electrical resistivity, (b)
the intensity of the (4, 4.5, 0) peak due to the correlated distortions, and
(c) the correlation length of the ordered regions in Nd$_{0.7}$Sr$_{0.3}$MnO$%
_{3}$. All the data were taken at \textit{T}=215 K (reprinted from {%
\protect\cite{94}}).}
\label{Fig5-5}
\end{figure}

In its mostly widely investigated form, the CMR effect is a field-induced
transition from the PI to the FM phase. Since the nanoscale correlations are
suppressed in the FM state, application of a magnetic field should also
result in their suppression. The data of Fig. \ref{Fig5-5} show that this is
indeed the case: as the magnetic field is applied, the electrical
resistivity is reduced, and the intensity of the correlated peak is
diminished, tracing the behavior of the resistivity \cite{94}.
Interestingly, a substantial correlated peak is still present at the highest
magnetic field obtainable in this experiment. The field-induced transition
is gradual. Assuming that the intensity of the correlated peak reflects the
concentration of the correlated regions, one concludes that the
field-induced state is very inhomogeneous. Finally, we note that the
magnetic field affects neither the correlation length, nor the period of the
lattice modulation in the ordered regions. 

\begin{figure}[tb]
\epsfxsize=60mm
\centerline{\epsffile{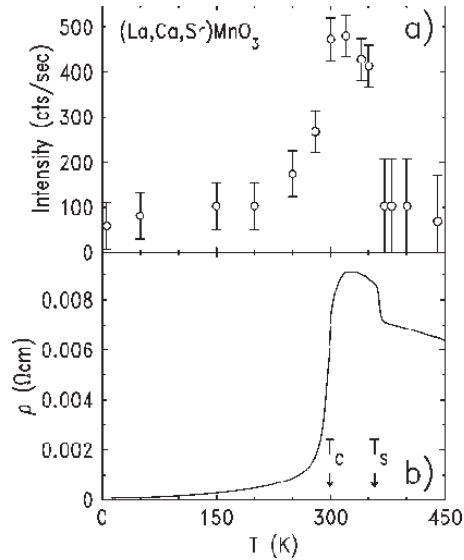}}
\caption{(a) Temperature dependence of the intensity of the peak due to the
structural correlations in La$_{0.75}$(Ca$_{0.45}$Sr$_{0.55}$)$_{0.25}$MnO$%
_{3}$. The single-polaron bakcground is subtracted. (b) Temperature
dependence of the electrical resistivity. The data were taken on heating.}
\label{Fig5-6}
\end{figure}

Pseudo-cubic perovskite
manganites exhibit a number of different structural phases, among them
orthorhombic, rhombohedral, and tetragonal. The nanoscale correlations
described above have thus far only been found in the orthorhombic$\ O$
phase. It remains to be seen whether they exist in other structural phases.
Preliminary measurements suggest the possibility that the rhombohedral phase
does not support the nanoscale fluctuations \cite{102}. An example of the
data supporting this claim is shown in Fig. \ref{Fig5-6} which shows
temperature dependence of the correlated peak intensity and the electrical
resistivity in La$_{0.75}$(Ca$_{0.45}$Sr$_{0.55}$)$_{0.25}$MnO$_{3}$. This
compound exhibits a PI state for \textit{T}\TEXTsymbol{>}300 K. With
increasing temperature, the paramagnetic insulating state undergoes a
structural transition from the orthorhombic \textit{O} state to a
rhombohedral state. The data of Fig. \ref{Fig5-6} show that the nanoscale
correlations abruptly disappear at the orthorhombic-to-rhombohedral
transition. At the same time, the electrical resistivity is significantly
reduced. These observations are in general agreement with the picture in
which the correlations reflect the presence of insulating regions in the
sample. 

The above experimental results could be summarized as follows: (i)
The nanoscale structural correlations are present in the orthorhombic PI
state. They are suppressed (but not completely destroyed) on the transition
to the FM state, independent on whether the transition is induced by
changing temperature or magnetic field. (ii) The correlation length
characteristic to the nanoscale regions is the same in all samples and does
not depend on temperature or magnetic field. (iii) The period of the lattice
modulation is the same linear function of the concentration of the divalent
ion (doping level) \textit{x} in all the samples. Thus, it appears that the
structure of the correlated regions in the manganites with the FM
low-temperature state is defined by a single parameter, \textit{x}.
 
We now turn to the interpretation of these data. We note that the precise
determination of the structure of the correlated regions, similar to that
carried out for the layered manganites \cite{95}, is yet to be done. At this
stage, therefore, we can only present some current ideas on this subject.
From the very beginning, the correlated regions were described as small
regions possessing charge and orbital order \cite{88,89,103}. This
suggestion was made based on the observed scattering vectors of the maxima
of the broad correlated peaks. In the case of \textit{x}=0.3, these peaks
are observed at the same positions as the so-called orbital-ordering peaks
in the CE-type charge/orbital ordered state. Thus, it was proposed that
small regions with the CE-type order are present in the PI state. As \textit{%
x} deviates from 0.3, different charge/orbital ordered structures, possibly
containing discommensurations, were speculated to be realized \cite{96}. In
particular, as \textit{x} approaches 0.5, the lattice modulation period
approaches 3, and one of the striped structures observed in highly doped
manganites could be realized. It was also pointed out that if the ordered
regions are well defined, their actual size could be significantly larger
than the correlation length shown in Figs. \ref{Fig5-3} and \ref{Fig5-5},
which was defined as inverse half-width at half-maximum of the peaks \cite%
{96}. Thus, the charge and orbital ordered structures described above are
compatible with the experimental results. 

In the above scenario, the charge-ordered structures are observed at doping
levels different from those of the corresponding long-range-ordered
counterparts. The CE-type order, for example, is most stable at \textit{x}%
=0.5, and the striped structure with a period 3 is observed at \textit{x}%
=2/3 (see open symbols in Fig. \ref{Fig5-4}). As a possible reason for this
discrepancy, it was proposed that charge-poor and charge-rich regions are
formed in the sample \cite{96}. The charge concentration in the
charge-depleted regions could then be close to the ideal concentration
needed for the formation of the charged-ordered structures with the observed
periodicity. Several possible reasons for such variations of the charge
density were proposed. First, it can result from chemical inhomogeneities.
It is, however, difficult to see how chemical inhomogeneities would explain
the observed common features in a variety of samples with quite different
chemical compositions. 

Second, even in the absence of chemical
inhomogeneities, the formation of charge-rich and charge-poor regions is
still possible. A number of theoretical calculations, in fact, predict such
a phase separation \cite{16}. Moreover, electronic phase separation appears
to be a generic feature of almost any current theory describing
mixed-valence manganites. Note, the usual argument that Coulomb interaction
would prevent significant spatial charge segregation should not necessarily
work in the case of nanoscale domains. This interaction could, of course, be
one of the factors limiting the ordered region size. 

It is interesting to
note that manganites with the long-range CE-type and striped order exhibit
the $O$-orthorhombic structure with the lattice constants $c$/$\sqrt{2}$ 
\TEXTsymbol{<}$b\sim a$ \cite{104,105}. The $c$-axis contraction in these
compounds reflects the Jahn-Teller distortion of the Mn$^{3+}$O$_{6}$
octahedra with the long axis lying in the $ab$ plane. Thus, on average, the 
long axes of the MnO$_6$ octahedra lie in the $ab$ plane in both the long-range charge ordered
phase, and in the $O$ paramagnetic insulating state.  In
contrast, the MnO$_{6}$ octahedra are undistorted in the rhombohedral (%
\textit{R}$\overline{3}$\textit{c}) state \cite{106}. It appears, therefore,
that the average lattice distortion characteristic to the $O$ state promotes
(or, possibly, reflects) formation of the nanoscale correlations. Since
there is no {\it a priori} requirement that the local symmetry of the nanoscale
ordered regions should match the average symmetry of the crystal lattice,
this is quite an interesting observation. 

Finally, we note that the case of
the transition from the PI or FM phase directly to the charge-ordered phase,
such as in Pr$_{\text{1-}x}$Ca$_{x}$MnO$_{3}$ or La$_{0.5}$Ca$_{0.5}$MnO$%
_{3} $, is more complex. In such transitions, the periodicity of the lattice
modulation and the correlation length vary on warming as the
low-temperature, long-range charge order disappears and the high-temperature
correlations arise \cite{91,92,93}. The existing experimental data are not
completely consistent, and it is unclear as to what extent they reflect the
intrinsic properties of the PI state. Systematic measurements at
temperatures much larger than the charge-ordering transition temperature are
needed to answer this question. 

Summing up, it appears that the important
role of nanoscale structural correlations in manganites is currently quite
well established. In particular, the presence of these correlations is one
of the factors leading to the large magnetoresistance observed in these
compounds. The correlations are found in a number of different manganite
compounds, they show a systematic behavior, and, therefore, these
correlations are not likely to be the artifacts resulting from unclean
samples or secondary phases. 

However, both the experimental investigation of
the local structure of the phases exhibiting nanoscale correlations, and the
theoretical work on the mechanism of their formation are still in their
beginning stage. While the simple interpretation of the experimental data
given above appears to be reasonable, there are several important problems
with it. First, the integrated intensity of the correlated peaks is
surprisingly small. Calculations based on the assumption that the magnitude
of the lattice distortion in the correlated regions is the same as that in
the corresponding long-range ordered phases produce extremely small
estimates for the volume fraction of the charge-ordered regions in the PI
state. Different authors give estimates for this fraction ranging from
several percent \cite{107} to a small fraction of 1 percent \cite{102}. It
is hard to see how such a small volume fraction could have a profound effect
on the transport properties. Second, the ``uncorrelated polarons'' and the
``nanoscale correlations'' are now often treated as two separate objects. In
our opinion, this separation is not altogether justified, and both the 
\textit{q}=0 and the \textit{q}$\sim $0.5 signals might originate from the
same object. Further studies of diffuse scattering in these materials,
including energy-resolved studies, are clearly needed to address these
problems. 

The example of manganites clearly shows that local nanoscale
inhomogeneities could have a great influence on the bulk properties.
Numerous recent studies show that similar inhomogeneities are likely to play
an important role in other correlated materials, cuprates and nickelates
among them. Understanding the intriguing properties of the correlated
materials, therefore, will require understanding details of their local
structure. Experimental and theoretical studies addressing this issue will,
therefore, play an important role in future work on correlated materials.

\section{Conclusions}

\label{sec6} In this review chapter, we presented extensive experimental
evidence to support the unambiguous existence of phase modulations in
mixed-valent manganites. It is revealed that two important non-trivial
inhomogeneities (or phase modulations) play an important role in producing
colossal magnetoresistance in mixed-valent manganites, i.e.,
sub-micrometer-scale (macroscopic) and nano-scale (microscopic) phase
coexistence/fluctuation. The first (macroscopic) phase separation is caused
by the coexistence of mainly ferromagnetic metallic and charge-ordered
insulating domains at low temperatures with micrometer length scales.
Extensive experimental evidence for such a large-length-scale phase
modulation was found in the Pr$^{3+}$-substituted La$_{5/8}^{3+}$Ca$%
_{3/8}^{2+}$MnO$_{3}$ system where the relative volume of the
electronically- and magnetically-distinct regions varies with Pr
substitution. However, it is important to note that both of the main two
phases have the same charge densities but quite different long-range
strains, resulting from the significantly different crystallographic
structures of the metallic and insulating phases. It is this lattice strain
that produces such large-length-scale phase coexistence at low temperature.
This physical situation resembles what generally happens in martensitic
systems, in which crystallographically-different phases can coexist at low
temperatures to accommodate the significant long-range strain associated
with the martensitic transformation. Even if many magnetotransport
properties of the system can be understood within a percolation model
between two different electronic phases, it is still necessary that
martensitic-transformation-effects should be included to understand
adequately many physical properties of the system. For example, a
martensite-like transformation can naturally bring another charge-disordered
insulating phase with lattice strains different from those of the main two
phases into play. Thus, extremely sensitive and huge magnetoresistance under
an applied magnetic field at low temperature should be also associated with
the special situation that structurally different phases can coexist and
that they show avalanche-effects when they switch to a new phase under
magnetic fields.

The other important phase modulation discussed in this chapter is caused by
nano-scale charge/orbital correlated nano-clusters (or nano-scale
charge/orbital ordering) in the background of the paramagnetic state above
the Curie temperature (or charge ordering temperature). Experimental
evidence from transport, magnetism, optical spectroscopy, and \textit{x}-ray
scattering studies were presented to show that microscopic phase fluctuation
effects can exist over broad doping ranges of the prototypical La$_{1-x}$Ca$%
_{x}$MnO$_{3}$ and Nd$_{1-x}$Sr$_{x}$MnO$_{3}$ systems above the long-range
ordering temperatures. In particular, those results revealed that unusually
large fluctuation of the CE-type charge/orbital ordering exists in the
samples near the special half-doped manganite La$_{0.5}$Ca$_{0.5}$MnO$_{3}$.
Based on present and previous results, the high temperature CE-type
correlations are expected to be dominant over a broad doping ranges of La$%
_{1-x}$Ca$_{x}$MnO$_{3}$ (0.3$\leq $\textit{x}$\leq $0.7) and Nd$_{1-x}$Sr$%
_{x}$MnO$_{3}$ (0.3$\leq $\textit{x}$\leq $0.5). In the future, it will be
interesting to identify the exact nature of nano-correlation as a function
of carrier concentration at high temperature in many ferromagnetic as well
as charge/orbital-ordered manganites. Additionally, we have provided
convincing experimental evidence for the existence of \textit{T}*, the
cluster-forming temperature, in La$_{1-x}$Ca$_{x}$MnO$_{3}$, particularly
with \textit{x}$\geq $0.5. The existence for \textit{T}* in low-hole-doping
manganites is still not clear yet because transport is always
insulating-like up to structural transition temperatures. Future studies are
quite necessary directed toward understanding the phase fluctuation effects
in the low hole-doping regime of La$_{1-x}$Ca$_{x}$MnO$_{3}$. Finally, it is
important to note that many systematic studies of various manganites now
indicate that the insulating nature above the Curie temperature is always
associated with the presence of charge/orbital correlated nano-clusters,
existing only in orthorhombic crystallographic structures. Thus, the
nano-clusters will play a key role in producing CMR near the insulator-metal
transition of the manganites. In other words, sensitive collapse of
nano-scale charge/orbital ordering under a small magnetic field will
essentially produce a huge drop in resistivity, inducing large negative
magnetoresistance just above the Curie temperature.

There remain several interesting unsolved questions. First, most of the
competing interactions in manganites seem to produce fluctuations of
competing order parameters near its thermodynamic critical or bi-critical
points at finite temperatures according to the electronic/magnetic phase
diagram shown in Fig. \ref{Fig2-1}. On the other hand, it remains to be
experimentally clarified whether the giant fluctuations of the half-doped
manganites are a direct consequence of the coexistence of competing order
parameters even at zero temperature as a theoretical model suggests \cite{16}%
. In this context, the role of quantum fluctuations should be investigated
more systematically to determine whether competition of order parameters at
zero temperature can play a key role in producing nanoscale-phase
coexistence/fluctuation in an unusual phase space in perovskite manganites.
Second, the long-range strain-induced phase coexistence might be a generic
feature in the manganites with strong Jahn-Teller effects. However, local
physics based on a Jahn-Teller electron-phonon coupling alone doesn't seem
to be enough to describe the observed unique phase modulation with
micrometer length scale. Theoretical and experimental investigations
directed toward understanding the metal-insulator transition induced by
long-range strain are important. Third, it is still puzzling why
ferromagnetism and metallicity are optimized at the commensurate hole-doping 
\textit{x}=3/8, where enhanced electron-phonon coupling can be expected to
result in localization of carriers \cite{10}. Enhanced metallicity exhibited
for special commensurate doping might be related to a dynamic nano-scale
charge/orbital ordering with stripe correlations. It is now well known that
the high-$T_{c}$ cuprates have dynamic and short-ranged stripe correlations
even in the superconducting regime. While static charge ordering is clearly
compatible with localization of carriers, the exact role of dynamic stripe
correlations on the metallicity or superconductivity is still not understood
in the cuprates. In this context, the role of dynamic stripe correlations on
the carrier mobility should be further investigated in manganites. Finally,
the discovery of two important phase-separation phenomena, micro- and
macro-scopic, increase our understanding of colossal magnetoresistance found
in the perovskite manganites at high and low temperatures, respectively.
Phase-modulation phenomena, and related physics established in manganites,
however, will be broadly applicable to other strongly correlated material
systems, wherein two or multi-competing interactions coexist.

\section{Acknowledgements}

This article is based on the results from close and friendly collaborations
with many coworkers. We sincerely thank our collaborators: M. E. Geshenson,
T. W. Noh, Y. H. Jeong, T. Y. Koo, B. G. Kim, H. J. Lee, M. W. Kim, C. Hess,
V. Podzorov, J. P. Hill, D. Gibbs, Y. J. Kim, and C. S. Nelson. We thank A.
Migliori for a critical reading of the manuscript. We also benefited greatly
from discussions with G. S. Boebinger, M. Jaime, A. J. Millis, C. M. Varma,
G. Kotliar, J. G. Park, N. H. Hur, and J. Yu. We also acknowledge National
Science Foundation for financial support (under the grant number of
NSF-DMR-0103858 and 0080008).

%------------ end of article ------------------->>

%% optional
%\section{Summary}

%% optional
%\begin{acknowledgments}
%...
%\end{acknowledgments}

\QTP{CE
}
%% appendix optional
%\appendix{This is the Appendix Title}
%This is an appendix with a title.

\QTP{CE
}
%\appendix{}
%This is an appendix without a title.

\QTP{CE
}
%
% Bibliography made with BibTeX:
%% apalike is preferred if you have used \kluwerbib, above.
%% Otherwise you may use any .bst style your editor approves.

\QTP{CE
}
%This will allow many Bib\TeX\ bibliographies in one book.
%See the documentation, kapedbk.doc, for more information.

\QTP{CE
}
%\bibliographystyle{apalike}
%\chapbblname{<name of .bbl file>}
%\chapbibliography{<name of .bib file>}

\QTP{CE
}
%or 
%\begin{chapthebibliography}{<widest bib entry>}
%\bibitem[optional]{symbolic name}
%Text of bib item...
%\end{chapthebibliography}

\QTP{CE
}
%%\begin{chapthebibliography}{<widest bib entry>}

\QTP{CE
}
%% \begin{chapthebibliography}{<widest bib entry>}

\QTP{CE
}
%% \end{chapthebibliography}

\end{document}